\documentclass[useAMS,usenatbib]{mn2e}
\usepackage{graphicx}
\usepackage[usenames,dvipsnames]{color}
\usepackage{morefloats}
\usepackage{amsmath}

\title[X-ray polarimetry of Seyfert-2s]
      {Predicting the X-ray polarization of type-2 Seyfert galaxies}
\author[Marin et al.]
      {F.~Marin$^1$\thanks{E-mail: frederic.marin@astro.unistra.fr},
       M.~Dov{\v c}iak$^2$, F. Muleri$^3$, F. F. Kislat$^4$ 
       and H. S. Krawczynski$^4$\\
       $^1$Observatoire Astronomique de Strasbourg, Universit\'e de Strasbourg, CNRS, UMR 7550, 11 rue de l'Universit\'e, 67000 Strasbourg, France\\
       $^2$Astronomical Institute of the Academy of Sciences, Bo{\v c}n\'{\i} II 1401, CZ-14100 Prague, Czech Republic\\
       $^3$INAF/IAPS, Via del Fosso del Cavaliere 100, I-00133 Roma, Italy\\
       $^4$Washington University in Saint Louis, Physics Department and McDonnell Center for the Space Sciences, Saint Louis, MO 63130, USA}

\date{Accepted 2017 September 11;
      Received 2017 September 11;
      in original form 2017 July 25}

\pagerange{\pageref{firstpage}--\pageref{lastpage}}
\pubyear{2017}

\begin{document}

\maketitle

\begin{abstract}
Infrared, optical and ultraviolet spectropolarimetric observations have proven to be ideal tools for the study of the hidden nuclei of type-2 
active galactic nuclei (AGN) and for constraining the composition and morphology of the sub-parsec scale emission components. In this paper, 
we extend the analysis to the polarization of the X-rays from type-2 AGN. Combining two radiative transfer codes, we performed the first 
simulations of photons originating in the gravity dominated vicinity of the black hole and scattering in structures all the way out to the 
parsec-scale torus and polar winds. We demonstrate that, when strong gravity effects are accounted for, the X-ray polarimetric signal of 
Seyfert-2s carries as much information about the central AGN components as spectropolarimetric observations of Seyfert-1s. The spectropolarimetric 
measurements can constrain the spin of the central supermassive black hole even in edge-on AGN, the hydrogen column density along the observer's 
line-of-sight, and the composition of the polar outflows. However, the polarization state of the continuum source is washed out by multiple 
scattering, and should not be measurable unless the initial polarization is exceptionally strong. Finally, we estimate that modern X-ray 
polarimeters, either based on the photo-electric effect or on Compton scattering, will require long observational times on the order of a 
couple of mega-seconds to be able to properly measure the polarization of type-2 AGN.
\end{abstract}

\begin{keywords}
Galaxy: active -- polarization -- radiative transfer -- relativistic processes -- scattering -- X-rays: general.
\end{keywords}

\label{firstpage}

\section{Introduction}
\label{Introduction}
Most, if not all, galaxies contain at least one supermassive black hole (SMBH) at their center, although not all of them show as dramatic 
observational signatures as quasars. In the case of long and steady accretion, the central SMBH can enter a phase where the viscosity of 
the accreting matter can lead to emission outshine the entire host galaxy \citep{Pringle1972,Shakura1973}. This type of object is called 
an active galactic nuclei (AGN) and typically stays in an accretion-efficient state for about 10$^5$~years before returning to a quiescent 
phase \citep{Schawinski2015}. The AGN lifetime is short compared to the total growth time of a galaxy but quasars strongly impact their host 
galaxy during that period. By studying AGN, we can understand how the feedback mechanism, i.e., the material expelled by radiation, winds and 
jets from the vicinity of the potential well, can affect its host galaxy in less than a million years. \citet{Springel2005}, \citet{Nandra2007},
\citet{Schawinski2009}, and other authors have found that quenching star formation from AGN feedback could move the host galaxy from a blue 
(star-forming) to a passive (red sequence) galaxy classification. Additionally, the mass of the central SMBH in nearby galaxies was found to 
correlate with the host bulge luminosity \citep{McLure2000}, velocity dispersion \citep{Ferrarese2000,Gebhardt2000} and mass \citep{Magorrian1998}.
Hence, to understand the evolution of galaxies, exploring the AGN phase is of utmost importance.

According to the standard paradigm, all AGN are relatively similar in terms of physics but several key parameters such as orientation 
\citep{Marin2014,Marin2016}, mass accretion rate \citep{Meier2002,Fanidakis2011} and feedback \citep{Kauffmann2000,Fabian2012} can differ,
resulting in a zoo of active galaxy classes \citep{Antonucci1993,Urry1995}. It is still unclear which of the aforementioned three parameters 
is the main driver in the unification scheme and a detailed multi-wavelength investigation of nearby AGN is mandatory to draw a self-consistent 
picture. 

The X-ray band is particularly well suited for the exploration of the AGN physics and constituents. X-ray radiation is produced close to the 
central SMBH by Inverse Compton scattering of thermally emitted ultraviolet photons in a corona situated above the disk \citep{Haardt1991,Haardt1993}.
This corona of hot electrons becomes the central continuum source of a power-law emission that will illuminate the innermost regions of the 
accretion flow. Observing the X-ray light after absorption, re-emission and scattering inside the compact nuclei will provide crucial information
about the localization, density and composition of gaseous and dusty media along the observer's line-of-sight \citep[e.g.][]{Winter2009,Buchner2015}. 
Imprinted on the X-ray spectra are the signatures of cold or ionized atoms, informing us about the velocities of unresolvable AGN components, 
together with the ionization fraction and intrinsic temperature \citep[e.g.][]{Porquet2000,Tombesi2010,Tombesi2013}. Finally, reprocessing 
will lead to strong constraints on the morphology and composition of the scattering media, hidden in the polarization of the X-ray light
\citep{Matt1993}. The observed X-ray polarization fraction and angle depend on the curved trajectories and the scattering of the photons in 
the vicinity of the black hole \citep{Dovciak2004,Schnittman2009} and on the structure and strength of the magnetic fields in the accretion 
flow \citep{McNamara2009} and thus give additional information to that obtainable through the spectroscopic and timing channels.

It has been 50 years since the pioneering X-ray experiments on cosmic sources \citep{Giacconi1962,Giacconi1964}. Both X-ray spectroscopy 
and timing techniques are now mature and well established but, in this regard, X-ray polarization is far behind. The only dedicated 
high-sensitivity X-ray polarimetry mission was launched in the 70'ties \citep{Novick1972,Weisskopf1976,Weisskopf1978}, and proposed 
follow-up missions based on Bragg and Thomson/Compton scattering were not selected for implementation. The OSO-8 mission achieved only 
one highly significant (19-$\sigma$) detection of the X-ray polarization of the Crab nebula and a handful of 99\%-confidence upper limits 
on additional compact X-ray sources were acquired before the X-ray polarimetric technology became no longer competitive compared to other investigation 
techniques \citep{Weisskopf2010}. We had to wait for the development of new instruments relying on non-rotating photoelectric polarimeter in the 
early 2000s to revive the field \citep{Costa2001}. Combining new gas pixel and Compton scattering detectors with focusing X-ray optics improved
by a factor 100 the sensitivity of X-ray polarimeters with respect to the old generation of instruments, opening a new observational window for 
the high energy sky \citep{Bellazzini2006,Bellazzini2007}. Two balloon-borne instruments, relying on scattering polarimetry, have flown in the 
recent years: X-Calibur (see \citealt{Beilicke2012,Beilicke2014,Krawczynski2016b}) and PoGOLite (see \citealt{Kamae2008,Pearce2012,Chauvin2016}). 
The first X-ray mission to fly a photoelectric imaging polarimeter in space will be launched by NASA in 2020 \citep{Weisskopf2016}. It is thus 
necessary to prepare the ground and to begin to refine the theories and simulations for interpreting the observational results to come.

X-ray polarimetric simulations of AGN has been presented in several key papers that can be divided into two groups: either the simulations were 
focusing on the central SMBH and its accretion disk, excluding any reprocessing regions farther than a thousand of gravitational radii 
\citep{Dovciak2008,Schnittman2009,Schnittman2010,Hoorman2016}, or the models did not account for relativistic effects \citep{Matt1989,Marin2012b,Marin2013,Marin2016b}. 
The propagation of photons through the curved spacetime of rotating black holes, referred to as “strong gravity effects” in the following, significantly 
modify the polarization of the observed radiation. In particular, the polarization angle ($\Psi$) as seen by an observer at infinity is rotated due to aberration 
and light bending  effects \citep[e.g.][]{Connors1977,Pineault1977,Connors1980}. The rotation is larger for smaller radii and higher inclination 
angles. Introducing strong gravity effects is thus important in any X-ray polarimetric modeling of AGN but the computational time becomes significant 
when the code has to account for scattering, absorption and reemission of photons on parsec scales. In the case of type-1 AGN the central
engine is seen from the pole, through the outflowing winds and warm absorber region \citep{Halpern1984}. Forward scattering of high energy radiation 
leads to very small degrees of polarization \citep{Marin2012b} and it is, at first order, possible to use simulations of isolated SMBH with an 
accretion disk to estimate the net X-ray polarization of type-1 AGN. However, in the case of type-2 objects, where the observer's line-of-sight is 
obscured by a dense, cold, circumnuclear medium, photons encounter multiple scattering inside the torus funnel or in the winds before escaping the 
AGN. The resulting polarization is expected to be high (several tens of percents, see \citealt{Marin2016b}). Additionally, the observations of Seyfert-2s 
often indicate the presence of a constant and soft emission component, commonly attributed to the scattering of X-rays by highly ionized gas, but 
which is not clearly separable from the more powerful torus scattering at higher energies. NGC~1068 thought to be dominated by electron scattering 
does not exhibit this emission component \citep{Kinkhabwala2002,Marinucci2016}, but NGC~4945 shows it very clearly \citep{Madejski2000,Puccetti2014}.

This paper focuses on the study of the polarization properties of the 1 -- 100~keV emission of Seyfert-2s. We account for the first time for the effect 
of strong gravity and the photon reprocessing by various structures, such as the equatorial gaseous torus and the conical polar winds. By including strong 
gravity effects close to the black hole, we investigate if the rotation of polarization angle along the photon's null geodesics can be detected by an X-ray 
polarimeter, or if the polarization signal is washed out by multiple scattering. We also include in the code the polarization state of the continuum source 
and explore how the final polarization reaching the observer is affected by these modifications. As previously stated, we focus on type-2 AGN as polar 
scattering is needed to escape the dense circumnuclear environment of the central SMBH, leading to higher polarization degrees than for type-1 objects. 
The higher the polarization degree, the better are the chances for a clean measurement of its X-ray polarization (modulo the X-ray flux of the source). 
We present the models and the code in Sect.~\ref{Modeling}, discuss our outcomes in Sect.~\ref{Results} and provide estimates of the detectability of 
X-ray polarization signals from type-2 AGN in Sect.~\ref{Detectability}. In the light of our results, we conclude in Sect.~\ref{Conclusions} on the feasibility 
of measuring polarimetric signatures imprinted with general relativistic effects and/or physical characteristics of the continuum source (temperature, 
density, composition ...).

\section{Modeling}
\label{Modeling}
We model the X-ray polarization emerging from a complex AGN model by combining the general relativistic {\sc ky} code with the {\sc stokes} 
scattering code \citep{Dovciak2011,Goosmann2007,Marin2012,Marin2015}. The {\sc ky} code simulates the compact object, i.e. tracks photons up 
to a certain radius, and then {\sc stokes} takes over to propagate radiation through the torus and polar winds. We created an AGN model following 
the unified scheme and included the possibility to turn on/off the strong gravity effects and the initial polarization of the coronal 
photons\footnote{For the remainder of this paper, the term ``primary'' emission will refer to the coronal emission. }.

\subsection{The radiative transfer code}
\label{Modeling:code}
The emission and scattering of radiation within the first thousand gravitational radii were simulated using
the updated {\sc ky} code for the relativistic reflection of polarized primary radiation from a cold accretion disk 
(see \citealt{Dovciak2011}). The code assumes that the primary source of X-ray radiation is geometrically small 
and located on the symmetry axis (i.e. lamp-post geometry). The primary emission, considered to be isotropic
and polarized, illuminates the accretion disk below. The disk is assumed to be a Keplerian, geometrically thin, 
optically thick, cold disk with inner edge located at the marginally stable orbit (ISCO) determined by the spin 
of the central super-massive black hole. Re-processing in the disk is calculated with the Monte Carlo 
multi-scattering code NOAR \citep{Dumont2000}, that computes the reflected flux including the iron fluorescent 
K$\alpha$ and K$\beta$ lines. The single-scattering approximation \citep{Chandrasekhar1960} is used for the local 
polarization of the reflected continuum component while the line flux is supposed to be unpolarized. Both the 
reflected flux from the disk and its polarization properties depend on the geometry of scattering, i.e. on the 
incident and emission angles as well as the relative azimuthal angle between incident and emitted light rays. 
Further, the local polarization properties of reflection depends on the polarization degree and angle of the 
incident illumination. A fully relativistic ray-tracing code in vacuum is used for photon paths from the corona 
to the disk and from corona and disk to outer parts of the system where relativistic effects are negligible. 
Note that the polarization direction rotates along the photon trajectory as a consequence of the polarization 
vector being parallel transported along the geodesic between the emission of the photon and the scattering off 
the disk and thus the illuminating flux has a distribution of incident polarization angles. As a consequence of all 
these effects, the local polarization properties of the re-processed radiation depends on the position on the disk 
where it is reflected from (both radial as well as azimuthal). The relativistic effects further change the 
polarization angle as the light travels from the disk to the more distant components of the system, especially 
for the light coming from the innermost parts of the disk. Therefore, the polarization properties of the radiation 
from the inner parts of the accretion flow result from adding the polarized emission from the corona and from all 
parts of the accretion disk and it will depend on the relative location of the inner accretion flow and the particular 
distant components.

{\sc stokes} is a Monte Carlo radiative transfer code developed to study the broadband, scattering-induced, polarization 
signatures of AGN \citep{Goosmann2007,Marin2012,Marin2015}. The code can handle a large variety of geometrical structures
arranged around the continuum source and accounts for all the reprocessing mechanisms from the near-infrared to the hard 
X-ray bands. Multiple scattering allows the radiative coupling between all the AGN components and, once radiation has 
escaped the model, it is recorded by a spherical web of virtual detectors, allowing to compute the intensity and polarization
spectra for all polar and azimuthal angles at once. Special and general relativity are not included in {\sc stokes}, hence 
the use of the input data from {\sc ky}. The scattering code samples photons according to the flux distribution from {\sc ky}, 
including the polarization state of radiation and its direction of propagation. In this paper, we neglect magnetic field effects 
and limit the modeling to radio-quiet Seyfert galaxies. A manual for the utilization of {\sc stokes} can be found on-line 
(http://www.stokes-program.info), together with a free version of the code working in the optical and ultraviolet bands.

\subsection{The AGN model}
\label{Modeling:AGN}

\begin{figure}
    \includegraphics[trim = 0mm 70mm 0mm 30mm, clip, width=8.5cm]{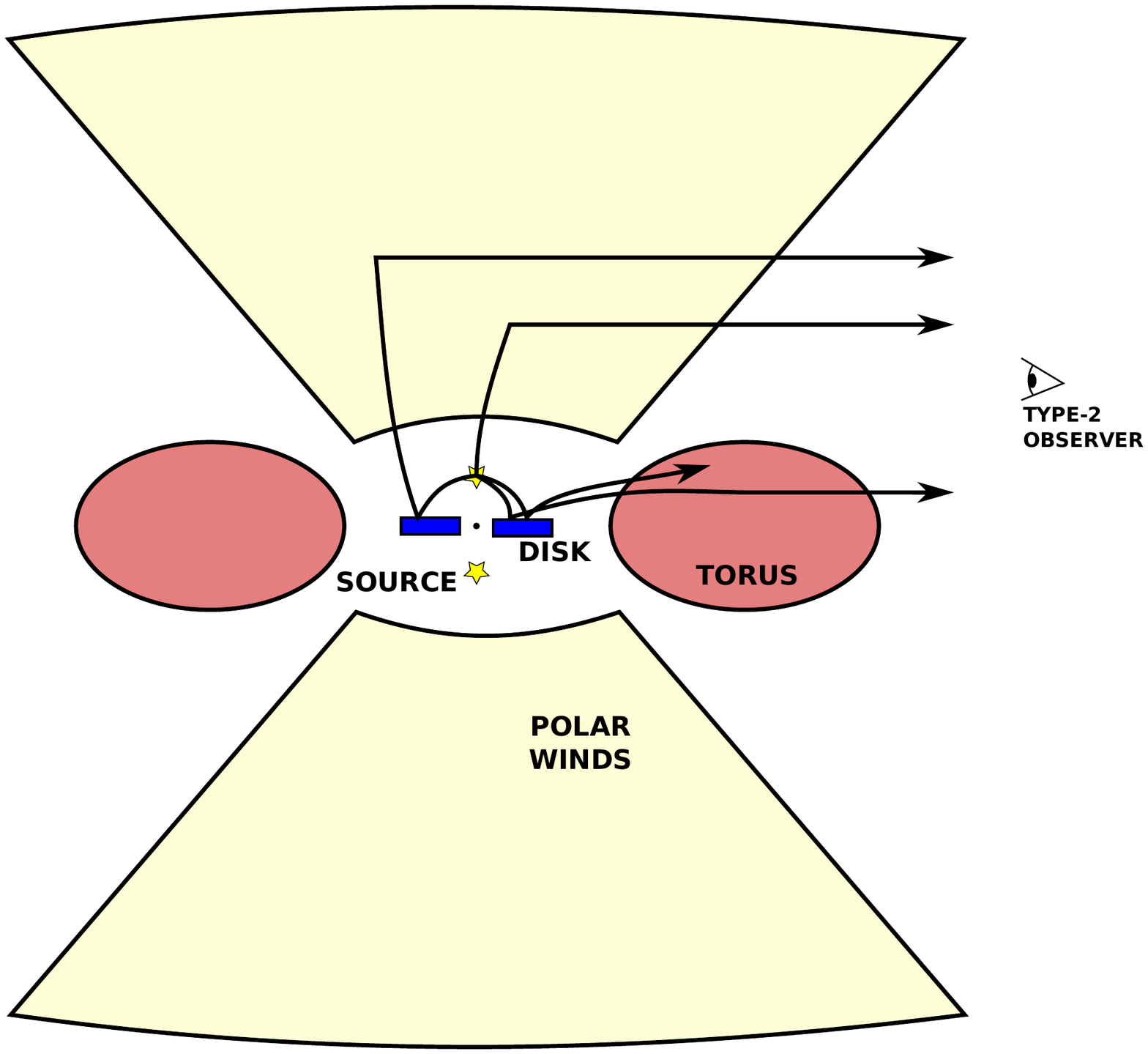}
    \caption{Artist representation of the AGN model. Scales
	    have been exaggerated for better visualization of 
	    the inner components. The point-like coronas 
	    are represented with yellow stars, the cold 
	    accretion disk is in blue, the gaseous torus 
	    in red and the polar outflows in primrose yellow. 
	    The photon trajectories are bend close to the 
	    central SMBH. Photons can scatter inside the 
	    polar winds to reach the observer or, depending 
	    upon the energy of radiation and Compton-thickness
	    of the equatorial material, can pass through the gas.
	    }
    \label{Fig:Scheme}
\end{figure}

We followed the geometrical depiction of AGN from \citet{Antonucci1993} to construct our baseline model (see Fig.~\ref{Fig:Scheme}). This model consists of 
a central SMBH surrounded by a geometrically thin accretion disk. The disk is irradiated by two compact coronas situated above and below the disk emitting 
the X-ray photons irradiating the accretion disk. Along the equatorial plane, at a parsec-scale distance is a compact gaseous torus that blocks radiation 
and collimates a conical wind that extends towards the polar region. 

The black hole mass distribution observed in large radio-quiet Seyfert catalogs is quite narrow, with an average mass of $\sim$ 10$^8$ solar masses 
\citep{Woo2002}, so we fixed the mass of our SMBH to the same value. We allowed the spin of the black hole to be either 0 (Schwarzschild case) or
1 (Kerr black hole). The dimensionless spin parameter gives the angular momentum of the black hole and impacts the location of the innermost stable 
circular orbit and thus the inner edge of the accretion disk. The more rapidly the black hole spins, the closer the accretion disk extends towards 
the black hole, and the stronger are the general relativistic effects (see, e.g., \citealt{Dovciak2011}).

Our black hole is surrounded by a geometrically thin (radius $\gg$ height), optically thick (electron optical depth $>$ 1) accretion disk filled 
with neutral matter with solar abundances and composition. X-ray photons are emitted by a compact source with a height of 3 gravitational radii and situated 
in a lamp-post geometry (i.e. on the disk symmetry axis). The photon index $\Gamma$ is equal to 2, where N(E) $\propto$ E$^{-\Gamma}$ and $\alpha$ = -($\Gamma$-1).
This power law photon index has been chosen accordingly to the observed indexes measured in different AGN samples, with typical values lying between 
1.5 and 2.5 \citep[e.g.][]{Nandra1994,Page2005}. We set the continuum source to radiate photons in the 1 -- 100~keV band, with either an unpolarized 
corona emission, or a polarized primary with a 2\% linear polarization. The polarized primary can have two different configurations: either perpendicular or
parallel. Polarization is described as parallel when its electric vector is aligned with the projected symmetry axis of the model (polarization position 
angle $\Psi$ = 90$^\circ$). In the case of orthogonality, the polarization is described as perpendicular ($\Psi$ = 0$^\circ$). We estimate the polarization 
from the Comptonization of the photons in the corona with the classical result for scattering-dominated atmospheres presented in \citet{Chandrasekhar1960}.

At larger distances from the central engine is an obscuring circumnuclear cold matter torus. The torus extends from 0.01~pc from the center of the model 
to 5~pc. The inner radius has been set accordingly to the mass and luminosity of the central SMBH \citep{Suganuma2006} and the outer radius is consistent 
with the maximum extension of the circumnuclear dusty torus as observed/modeled in the infrared \citep{Fritz2005,Siebenmorgen2015}. Its half-opening angle 
is set to 60$^\circ$ from the polar axis, so that the line-of-sight of an observer situated along the equatorial plane (a type-2 view) is obscured by the 
gaseous medium. We assume three different amounts of hydrogen column densities along the observer's line-of-sight: 10$^{23}$, 10$^{24}$, or 10$^{25}$~at/cm$^2$. 
By doing so, we model both Compton-thin and Compton-thick AGN classes (with the transition between Compton-thin and Compton-thick Seyferts starting at 
10$^{24}$~at/cm$^2$, see \citealt{Risaliti2005}).

Finally, along the pole, we added the possibility to have an outflowing wind, collimated by the torus funnel. The winds extends 60$^\circ$ from the polar 
axis, and it is either composed of neutral gas in a Compton-thin regime, or filled with electrons only (radial optical depth much lower than unity). 
The wind base is located at a radial distance of 0.1~pc from the center of the model and extends up to 25~pc, before mixing with the interstellar 
medium. In total, there are three kind of AGN: models without polar winds, models with low-density, cold, gaseous winds and models with a highly ionized 
outflows.

We thus developed a baseline model with a variety of input parametrizations in order to explore different kinds of radio-quiet AGN: from Compton-thin 
to Compton-thick type-2s, AGN with a maximally or non-rotating black hole, AGN lacking outflowing signatures, AGN with winds dominated by electron or 
neutral gas, and AGN with unpolarized or partially polarized corona emission..

\subsection{Polarization of the continuum source}
\label{Modeling:Polarization}

\begin{figure}
    \includegraphics[trim = 0mm 0mm 0mm 0mm, clip, width=8.5cm]{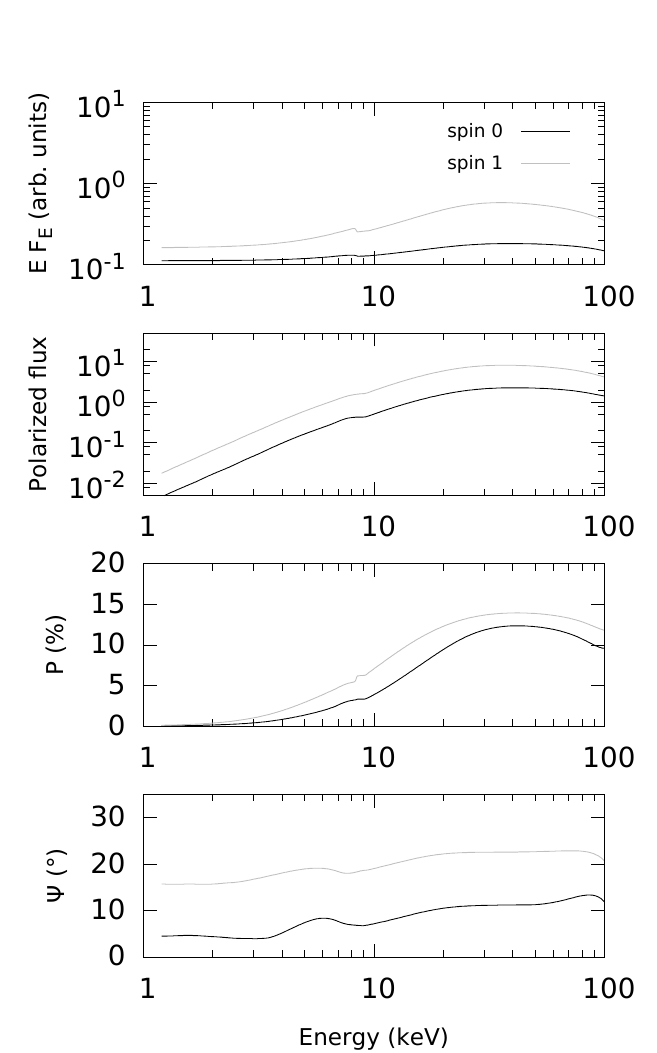}
    \caption{X-ray flux (F$_{\rm E}$ is energy flux at energy E), polarized flux (E~F$_{\rm E}$ 
	    times the polarization fraction), polarization degree $P$ and polarization position angle 
	    $\Psi$ seen by an observer at infinity, resulting from an elevated point-like corona that
	    irradiates an accretion disk inclined by 70$^\circ$. The source is unpolarized. The 
	    variation in $P$ and $\Psi$ are due to general relativistic effects that will induce
	    a parallel transport of the polarization angle along geodesics, plus the scattering 
	    and reemission of photons from the cold accretion matter. Two flavors of black holes
	    are shown: a non-spinning Schwarzschild black hole (black line) and a maximally spinning 
	    Kerr black hole (gray line).}
    \label{Fig:Initial_NOTpolarization}
\end{figure}

\citet{Schnittman2010} and \citet{Schnittman2013} derived the polarization of the corona self-consistently for the case of stellar-mass black holes but 
the intrinsic polarization of the primary radiation was never properly calculated for AGN. Such polarization is expected to occur if the primary spectrum 
emitted by the lamp-post is indeed due to inverse Compton scattering of ultraviolet photons thermally emitted by the accretion disk. However the exact 
degree and polarization position angle resulting from this kind of interaction in the vicinity of a potential well was never investigated for the quasar 
class. Here we upgraded the {\sc ky} code (Dov{\v c}iak et al., in prep.) to account for a non-null polarization of the corona emission.

\subsubsection{Benchmark case: unpolarized primary}
\label{Modeling:Polarization:Unpolarized}

In Fig.~\ref{Fig:Initial_NOTpolarization}, we plot the X-ray flux, polarized flux, polarization degree $P$ and polarization position angle $\Psi$ seen by 
an observer at infinity, resulting from our baseline model without the torus and polar winds. We fixed the inclination of the observer to 70$^\circ$ 
(type-2 view) and explored two flavors of SMBH: a non-spinning Schwarzschild black hole and a maximally spinning Kerr black hole. We first examine a model 
where the source emits unpolarized photons. As we can see from Fig.~\ref{Fig:Initial_NOTpolarization} (top panel), the X-ray spectra of the two black hole 
systems are different in terms of intensity, the Kerr black hole re-emitting more photons that the non-spinning one. This is due to the radius of the innermost 
stable circular orbit that is six times smaller in the former case, allowing radiation to scatter from the disk towards the observer, rather than crossing the
event horizon in the case of a Schwarzschild black hole. The polarized flux is the multiplication of the intensity times the polarization degree. Since both 
the flux and $P$ are higher for a spinning black hole, the polarized flux of a Kerr SMBH is also higher. Finally, the light bending aberration effects drive 
$\Psi$ to a different value for the two black hole flavors, as the radius of the photon orbit decreases when the black hole spin increasing.

\subsubsection{Polarized primary}
\label{Modeling:Polarization:Polarized}

We now switch on the polarization of the corona emission and investigate two configurations: either the initial polarization position angle is equal to 90$^\circ$ 
(parallel with respect to the disk axis) or $\Psi$ = 0$^\circ$ (perpendicular polarization). We compute the net polarization at infinity, that is a combination
of direct light and photons that have scattered onto the disk surface, and present the outcomes of the simulations for two observer's inclinations: 20$^\circ$ 
and 70$^\circ$.

\begin{figure*}
    \includegraphics[trim = 0mm 0mm 0mm 0mm, clip, width=8.5cm]{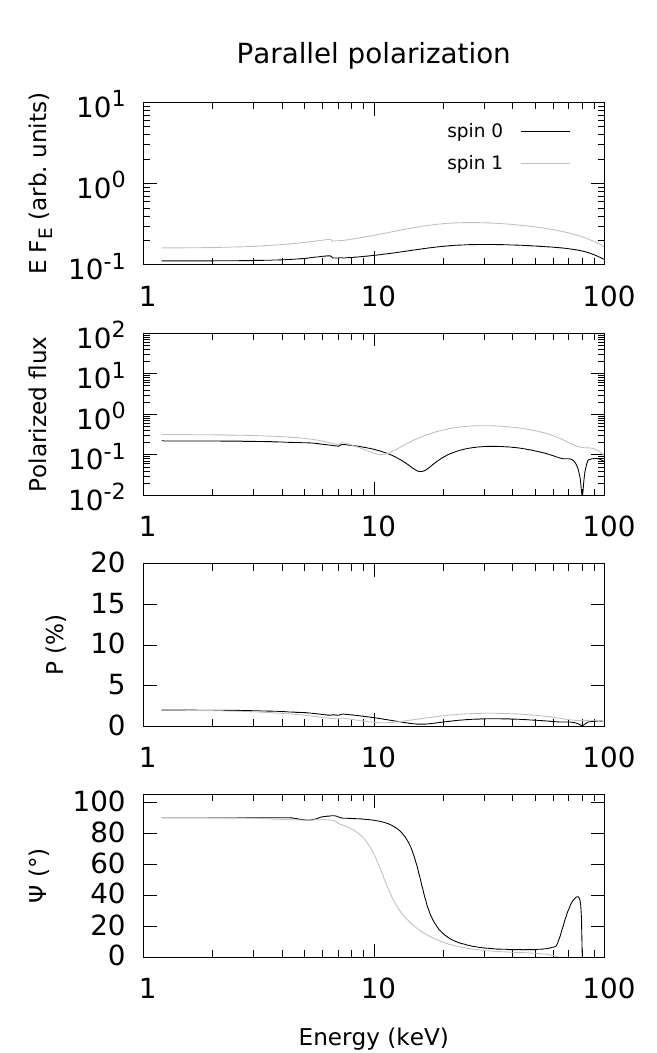}  
    \includegraphics[trim = 0mm 0mm 0mm 0mm, clip, width=8.5cm]{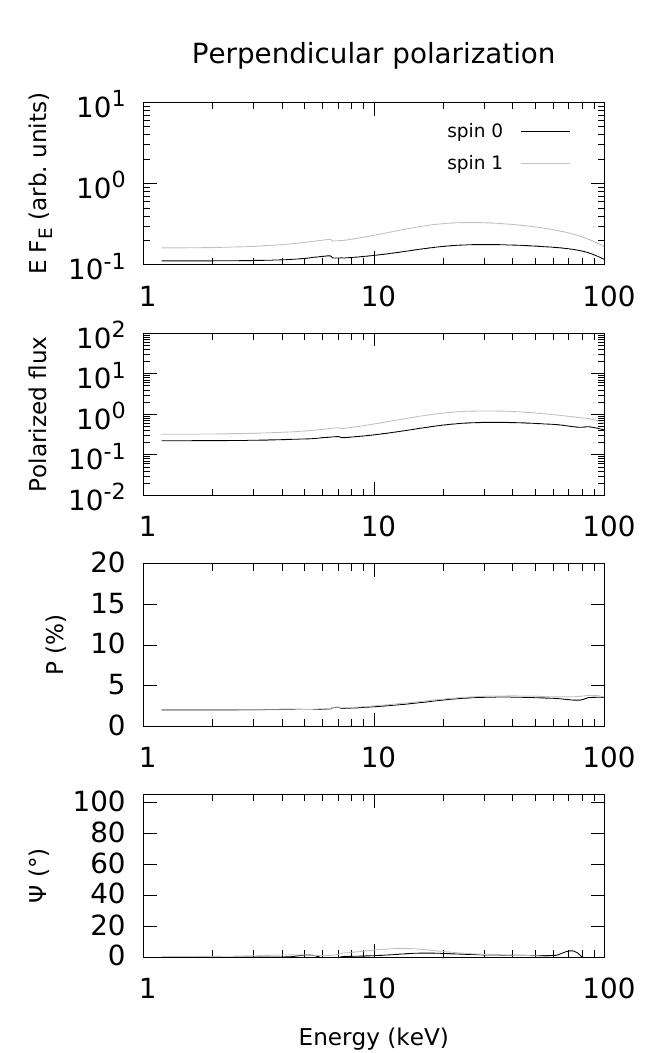}
    \caption{X-ray flux (F$_{\rm E}$ is energy flux at energy E), polarized flux (E~F$_{\rm E}$ 
	    times the polarization fraction), polarization degree $P$ and polarization position angle 
	    $\Psi$ seen by an observer at infinity, resulting from an elevated point-like corona that 
	    irradiates an accretion disk inclined by 20$^\circ$. The initial polarization is set to 
	    2\% with a parallel or a perpendicular polarization angle (left and right column, 
	    respectively). The variation in $P$ and $\Psi$ are due to general relativistic effects 
	    that will induce a parallel transport of the polarization angle along geodesics, plus
	    the scattering and reemission of photons from the cold accretion matter. Two flavors
	    of black holes are shown: a non-spinning Schwarzschild black hole (black line) and a
	    maximally spinning Kerr black hole (gray line).}
    \label{Fig:Initial_polarization_20deg}
\end{figure*}

Fig.~\ref{Fig:Initial_polarization_20deg} shows our results for a type-1 viewing angle (i.e. inclination of 20$^\circ$). The left column has an initial 2\%
parallel polarization, and the right column has a 2\% perpendicular polarization angle. The intensity spectra are exactly the same for both configuration 
of the primary polarization, indicating that spectroscopy would be insensitive to two different emission mechanisms that would mimic a compact source 
dominated by Compton scattering processes. Polarization, however, carries more information about the physical state of the emitting corona. If the 
polarization degree is twice lower than in the case of an unpolarized corona emission, with $P <$~5\% in both cases (and for both spin parametrization), it 
is primarily due to inclination (here 20$^\circ$, and 70$^\circ$ in the previous case). Polarization is sensitive to the geometry of the flow near the 
inner-disk boundary and decreases for larger inclinations (see \citealt{Dovciak2004}). A second, less critical, depolarization effect is due to the initial 
polarization angle of radiation. We focus on the Kerr black hole for the following explanation, keeping in mind that the Schwarzschild case is similar. 
Assuming unpolarized corona emission, the net signal emerging from the accreting Kerr black hole has a net polarization direction angle of 15 -- 20$^\circ$ 
from 1 to 100~keV, see Fig.~\ref{Fig:Initial_NOTpolarization}. However, if the corona emission is polarized, $P$ will only increase if the polarization 
vectors have a similar $\Psi$; otherwise the two polarization will cancel each other, resulting in a lower $P$. From Fig.~\ref{Fig:Initial_polarization_20deg}
(bottom panel), we can see that $\Psi$ is almost zero for all the energy bins in the case of a corona emission with perpendicular polarization (i.e. 0$^\circ$), and 
changes from 90$^\circ$ to 0$^\circ$ in the case of a parallel initial polarization (i.e. 90$^\circ$). This reinforces the decrease of $P$ with respect 
to the unpolarized case. It is interesting to note that, if the corona emission imposes its initial polarization state to the signal seen by an observer at infinity in the low energy 
band (E $\le$ 10~keV), at higher energies scattering from the disk will force the polarization angle to another value (shaped by special and general 
relativistic effects). This is due to the fact that, at low energies, photons are preferentially absorbed by the disk (with the potential reemission of 
fluorescent lines) and photons reaching the observers are mostly the ones that escaped the model without scattering, hence a $P$ close to the initial value. 
At higher energies, scattering from the disk dominates and $P$ can rise. Finally, for both configuration of the initial polarization, one can see two
variations of $\Psi$ at soft ($\sim$ 6--7~keV) and at high (above 60~keV) energies: in some cases relativistic effects may cause sudden changes in flux with 
energy to be visible in the polarization signal. This is due to the energy shift of the locally emitted spectrum due to Doppler effect and gravitational redshift. 
Total flux and polarization at a given energy have contributions from nearby energy bins according to energy shifts that vary across the accretion disk. Thus if 
there is a drop in flux, the contributions from those parts of the disk where the relativistic energy shift pushes lower flux part to certain energy will contribute 
less to the overall polarization properties. This can have several outcomes for the total polarization properties at this energy when compared to the energies 
farther away from the flux drop:

\begin{itemize}
\item increase of polarization: the lower-flux-contribution parts of the disk provide varied polarization angles, 
and thus their depolarizing contribution to given energy is decreased due to the drop in flux,
\item decrease of polarization: the lower-flux-contribution parts of the disk provide substantial contribution 
with similar polarization angles as the resultant one, and thus their polarization contribution is missing at 
given energy due to the drop in flux,
\item fluctuation in polarization properties: a combined effect of the two above where lower flux polarization 
contributions may depend on the energy shift,
\item no effect: the lower-flux-contribution parts of the disk do not contribute substantially to the overall 
polarization, neither by depolarizing nor strengthening the polarization.
\end{itemize}

In some cases we can see these (subtle) effects around iron edge energy (7.3~keV) or at high energies where the flux decreases quickly above the Compton hump 
(above 60~keV). The effect appears clearly in case of a disk inclination of 20$^\circ$ (Fig.~\ref{Fig:Initial_polarization_20deg}) where for extremely spinning
black hole there is an increase in polarization above 80~keV (case number 1 from the above list) and for non-rotating black hole there is a fluctuation in 
polarization properties at these energies (case 3 above). One should also note that the sharp cut-off for the re-processed flux above 100~keV might also 
artificially impact the last energy bins. However, as it will be demonstrated in Sect.~\ref{Results}, this will have no influence on the modeling results.

~\

\begin{figure*}
    \includegraphics[trim = 0mm 0mm 0mm 0mm, clip, width=8.5cm]{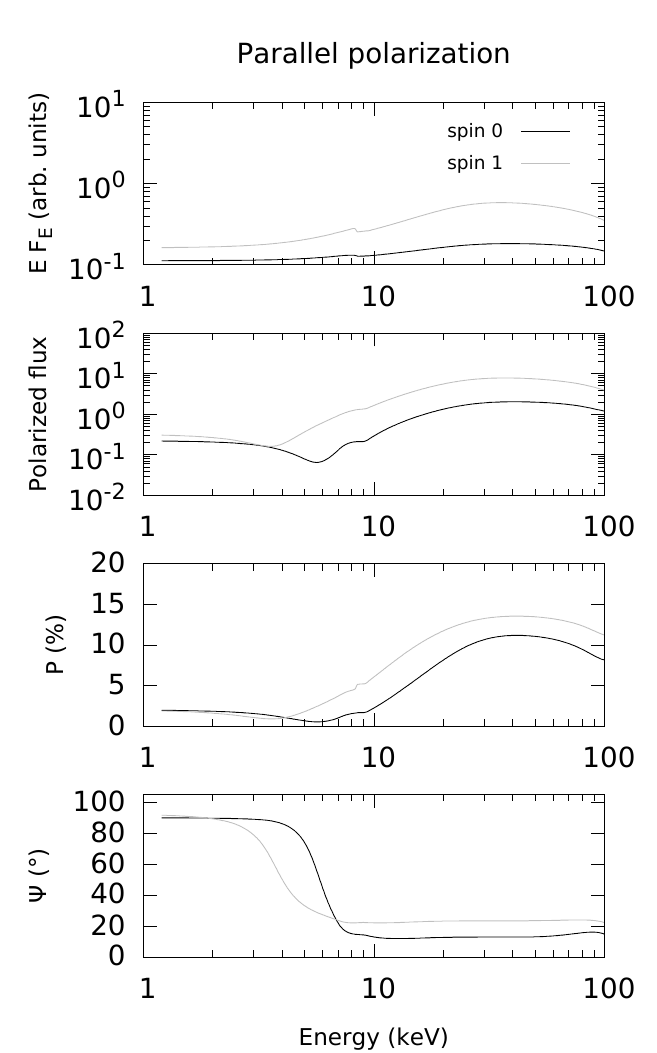}  
    \includegraphics[trim = 0mm 0mm 0mm 0mm, clip, width=8.5cm]{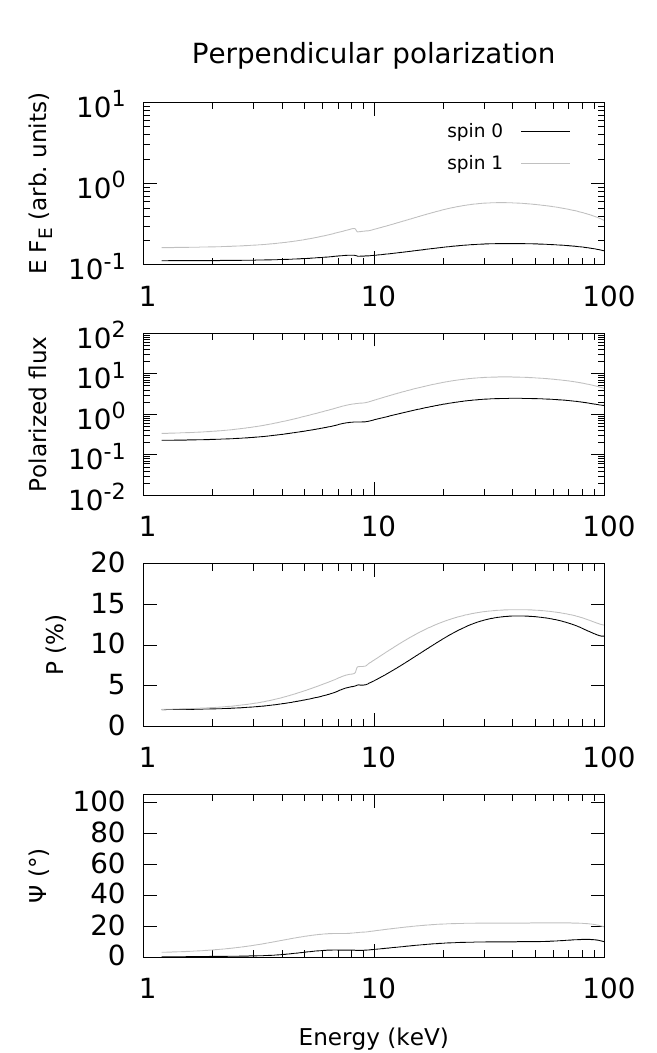}
    \caption{Same as Fig.~\ref{Fig:Initial_polarization_20deg} but 
	      for an accretion disk inclined by 70$^\circ$.}
    \label{Fig:Initial_polarization_70deg}
\end{figure*}

In Fig.~\ref{Fig:Initial_polarization_70deg}, the intensity and polarization spectra are shown for an inclination of 70$^\circ$. Again, there is no 
difference between models with parallel, perpendicular or even no polarization of the corona emission in terms of spectroscopy. However, the polarization degree
and polarization angle are rather different. Comparing simulations at similar observer's inclinations, the model without primary polarization shows a much
lower polarization degree in the low energy band, where absorption processes dominates. Since most of the escaping photons have traveled from the source to 
the observer without scattering, they carry the information about the initial polarization. This is why $P$ is about 2\% below 4~keV, and has the same $\Psi$
than the source. However, as seen from a lower inclination, the polarization position angle rotates as soon as scattering from the disk becomes efficient.
The energy at which the rotation occurs depends on the spin parameter, related to the location of the innermost stable circular orbit. Since the accretion 
disk gets closer to the potential well for maximally spinning black holes, the polarization angle variation occurs sooner in terms of energies. The difference 
between the two flavors of black holes is very easy to spot between the soft and high energy band, as predicted by \citet{Schnittman2009,Schnittman2010}, 
but we see here that the polarization of the corona emission can change both the value of $P$ and the energy at which the $\Psi$ rotation will happen.

In conclusion, for an isolated system of a SMBH plus an accretion disk, including the polarization of the corona emission will have a drastic impact onto the degree 
of polarization in the low energy band, and can cause the polarization position angle to have a large amplitude variation at an energy that depends on the 
spin parameter. The polarization angle of the initial photons can lead $\Psi$ observed at infinity to have an orthogonal rotation, facilitating the 
discrimination between different theoretical models of X-ray reprocessing in the vicinity of a compact object.

\section{Results}
\label{Results}
We ran the code {\sc stokes}, using the relativistic results of {\sc ky} as an input parametrization of the light reprocessed by the accretion disk in 
the strong gravity regime. We explored a set of 81 models (with and without strong gravity effects, unpolarized or polarized corona emission, two sets 
of different initial polarization, three different hydrogen column densities of the torus and three parametrizations of the outflows). Each model took 
approximatively 192~hours on the platforms for intensive computing at the meso-center of the University of Strasbourg. The total amount of CPU time 
allocated for this research was 15552~hours, corresponding to 21~months, but reduced down to 3~months of real time continuous computation thanks to 
computer parallelism.

Since our principal interest is to look at the polarimetric signatures of type-2 Seyferts, for the remainder of this paper we fixed the observer's 
viewing angle to 70$^\circ$. All the spectra of this intensive program are presented in the Appendix section to improve the readability of the paper.
A plot synthesizing our results is presented in Fig.~\ref{Fig:TF_PO_PA_difference}. In the following subsections, we will describe the main outcomes 
from our simulations, organized as follow: first our results for an unpolarized corona emission, then for a 2\% parallel polarized primary and later 
for a 2\% perpendicular polarized primary. 

\subsection{Results for an unpolarized primary without strong gravity effects}
\label{Results:UNPOL_PL}
We display our results for a type-2 AGN model without strong gravity effects and for an unpolarized source in three figures in the appendix. The first, 
Fig.~\ref{Fig:Unpolarized_primary_PL_Absorbing_nh21_wind}, presents an AGN where the polar winds are dominated by cold material. There are three 
sub-figures: top-left is for a torus hydrogen column density of 10$^{23}$~at/cm$^2$ along the observer's line-of-sight, top-right 10$^{24}$~at/cm$^2$,
and bottom 10$^{25}$~at/cm$^2$. This configuration will be the same for the remainder of the figures in appendix. We see that the amount of absorbing 
material shapes the intensity spectra such as expected from typical modeling of type-2 AGN, see, e.g., \citet{Ueda2015}, with the flux at lower energies 
being more suppressed by photoelectric absorption than at higher energies. It is fairly easy to differentiate Compton-thin and Compton-thick Seyfert-2s 
from spectroscopy, but polarization adds specific characteristics. First, the polarization degree is very high ($\sim$ 30\%) in the soft energy band. 
If the hydrogen column density of the torus is too high for the low energy photon to pass through, radiation is absorbed unless it scatters from 
the polar winds towards the observer. This periscope-like path is responsible for the high polarization, as orthogonal Thomson/Compton scattering 
produces maximum $P$. The composition of the wind is imprinted in the unpolarized fluorescent emission lines seen in the polarization spectra and 
the decrease of $P$ indicates where photons start to leak from the equatorial plane, carrying a $\Psi$ value of 90$^\circ$. This signature also 
corresponds to an orthogonal rotation of the polarization position angle. It follows that the exact amount of hydrogen column density along the observer's 
line-of-sight can be probed with great precision by the energy at which the orthogonal switch of $\Psi$ happens. Note that this, of course, is strongly 
orientation-dependent. At higher energies, the polarization angle remains fixed and the increase of $P$ with n$_{\rm H}$ is due to enhanced multiple 
scattering in the Compton hump.

If we compare our previous results to a model with a highly ionized, electron-filled, wind (Fig.~\ref{Fig:Unpolarized_primary_PL_Ionized_wind}), the 
main difference resides in the absence of signatures of fluorescent lines in the polarization spectrum for the later case. The difference can be seen 
in the intensity spectra too, where the low energy part of the spectrum is much less absorbed. It indicates that, regardless the ionization stage of 
the wind, if polar scattering is happening, we expect a large polarization degree at soft X-ray energies (up to 30\% in the case of no additional 
dilution). Seyfert-2s with clear electron scattering in their soft X-ray spectra are thus excellent potential targets for an X-ray polarimeter.

The case of a Seyfert-2 AGN without polar winds (a sub-class of thermal AGN characterized by a very weak or absent amount of intrinsic warm 
absorption, see, e.g., \citealt{Patrick2011}) is presented in Fig.~\ref{Fig:Unpolarized_primary_PL_No_wind}. The lack of polar scatterers leads 
the intensity spectra to be very dim in the soft energy band where absorption dominates. This paucity of photons translates in polarization spectra 
with very poor statistics despite the large amount of computational time. The degree of polarization is lower ($<$ 20\%) and decreases with energy 
until the transition due to Compton opacity. The polarization position angle also rotates from 0$^\circ$ to 90$^\circ$ at this peculiar point and  
$\Psi$ starts to stabilize. In comparison with the two other cases, an type-2 AGN without outflows would require much more observing time to get a 
significant X-ray polarization spectrum.

\subsection{Results for an unpolarized primary with strong gravity effects}
\label{Results:UNPOL_GR}

We now turn on the strong gravity effects but keep an unpolarized corona emission. The results, presented in Fig.~\ref{Fig:Unpolarized_primary_GR_Absorbing_nh21_wind},
Fig.~\ref{Fig:Unpolarized_primary_GR_Ionized_wind} and Fig.~\ref{Fig:Unpolarized_primary_GR_No_wind}, are now subdivided according to the dimensionless
spin parameter (0 or 1).

For the three configurations (Compton-thin wind, ionized winds or ``bare'' AGN), we find no differences in their intensity spectra between a non-rotating
and a maximally rotating black hole. A spectroscopic investigation of type-2 AGN is therefore unable to distinguish between the two black hole flavors, as 
scattering onto the torus funnel and on the winds almost completely mask the signatures of general relativity \citep{Streblyanska2004}. The resulting narrow 
feature at 6.4~keV is originating from reemission from the torus region and the extended, relativistic, red wing of the iron fluorescent emission is lost. 
A spectropolarimetric observation, on the other hand, provides a lot of additional information. In comparison to our previous models without strong gravity 
effects where the polarization position angle could only take only 2 values (0$^\circ$ and 90$^\circ$), adding general relativity leads to a smooth rotation 
of $\Psi$ between the soft and hard X-ray bands. The presence of winds has strong impacts. First the variation of $\Psi$ with energy is clearly different between 
a cold and a highly ionized wind. The lack of polar absorption in the second case leads the polarization angle to follow the results of unpolarized light 
scattering off the disk (see Fig.~\ref{Fig:Initial_NOTpolarization}), while absorption by the Compton-thin wind is more efficient to suppress the signatures of 
the initial polarization. X-ray polarimetry can clearly probe the composition of the outflows in this case. Second, when scattering occurs in the winds, the 
polarization angle of the soft X-ray radiation is naturally fixed to 0$^\circ$, and at the Compton opacity transition, strong gravity effects become visible.
The energy-dependent variations of the polarization position angle are directly related to the energy-dependent albedo and scattering phase function of the 
disk material, and help to distinguish between two different spins \citep{Dovciak2008,Dovciak2011}. This is particularly visible for the Compton-thin cases, 
where the small opacity of the circumnuclear torus helps to detect the effects of strong gravity near the central SMBH. The situation is less trivial for 
Compton-thick type-2 Seyferts, where the difference is just a matter of a couple of degrees. This difference is completely washed out when the torus is Compton 
thick and collimated winds are absent (see Fig.~\ref{Fig:Unpolarized_primary_GR_No_wind}). 

Thereby, similarly to what we foresee for spectroscopic \citep{Nandra1997,Yaqoob2004,Nandra2007} and polarimetric \citep{Dovciak2008,Schnittman2009,Schnittman2010,Marin2013} 
observations of type-1 AGN, measuring the X-ray polarization position angle of type-2 Seyfert galaxies will definitively tell us if strong gravity effects are important 
close to the central compact source, or if the signatures traditionally attributed to general relativity are in fact caused by pure absorption and Compton 
scattering by a distant cloudy medium \citep{Inoue2003,Miller2008,Miller2009,Miller2013}.

\subsection{Results for a 2\% parallel polarized primary without strong gravity effects}
\label{Results:POL_PARA_PL}
We now fix the polarization of the corona emission to a value of 2\% (linear polarization). The polarization position angle is set to 90$^\circ$, i.e. parallel 
to the projected symmetry axis of the disk. We first investigate the resulting polarization from a polarized source without strong gravity effects.

As we can see from Fig.~\ref{Fig:Polarized_primary_PL_para_Absorbing_nh21_wind}, Fig.~\ref{Fig:Polarized_primary_PL_para_Ionized_wind} and 
Fig.~\ref{Fig:Polarized_primary_PL_para_No_wind}, the spectroscopic and polarimetric spectra are quite similar to the case of an AGN without general 
relativistic effects nor polarized primary, with two exceptions. First the degree of polarization $P$ is higher by about 2\% (corresponding to the 
initial $P$) in the hard energy band, where photons have crossed the equatorial material without suffering from heavy absorption. In the soft band, 
scattering off the wind material overwhelms the intrinsic primary polarization and it is impossible to identify the influence of the polarized primary.
Second, the polarization position angle switch from 0$^\circ$ (scattering off the polar structure) to 90$^\circ$ (equatorial scattering) at a slightly 
different energy. For a type-2 AGN model with Compton-thin winds and a hydrogen column density of 10$^{23}$~at/cm$^2$ along the observer's line-of-sight,
in the case of an unpolarized corona emission the transition happens at E = 2.1~keV (Fig.~\ref{Fig:Unpolarized_primary_GR_Absorbing_nh21_wind}, top-left case, 
$\Psi$ panel), while in the case of a parallelly polarized corona emission, the rotation happens at E = 1.7~keV (Fig.~\ref{Fig:Polarized_primary_PL_para_Absorbing_nh21_wind}, 
top-left case, $\Psi$ panel). This difference, marginally detectable, is the result of the input polarization matching the scattering-induced polarization 
from the model. Since the polarization vectors have the same $\Psi$, the rotation of the polarization angle is facilitated and can happen at lower energies.

We thus find that using a 2\% parallel polarized primary radiation enhances the observed polarization degree at high energies and slightly alter 
the energy at which the orthogonal rotation of the polarization angle happens. Otherwise, the results are very similar to the ones obtained for 
an unpolarized corona emission in the same conditions (without strong gravity effects).

\subsection{Results for a 2\% perpendicular polarized primary without strong gravity effects}
\label{Results:POL_PERP_PL}
In this subsection, the polarized corona emission has also a 2\% linear polarization but its initial $\Psi$ is set to 0$^\circ$, i.e. perpendicular to the projected
symmetry axis of the disk. Results are plotted in Fig.~\ref{Fig:Polarized_primary_PL_perp_Absorbing_nh21_wind}, Fig.~\ref{Fig:Polarized_primary_PL_perp_Ionized_wind}
and Fig.~\ref{Fig:Polarized_primary_PL_perp_No_wind}.

Similarly to the previous case (2\% parallel), the spectroscopic channels are exactly the same as the unpolarized corona emission models. There is almost no 
differences in terms of polarization albeit the two remarks from Sect.~\ref{Results:POL_PARA_PL}, except that this time $P$ is smaller than in the 
unpolarized cases at high energies and that the rotation of $\Psi$ may vanish for some specific models. The first change is due to the orthogonality 
of the polarization vectors (from the primary source and from equatorial scattering), leading to a depolarization effect. The second is also linked 
with the transmission of photons with $\Psi$ = 0$^\circ$ through the circumnuclear gaseous medium when the absorption column density is too low 
($<$ 10$^{24}$~at/cm$^2$), forcing the net polarization angle to the same value. Finally, we observe a slightly higher value of $P$ in the soft energy 
band (when the Compton opacity is $>$~1) for all the models with outflows. Since the polarization position angle due to polar scattering is the same as 
the initial polarization angle, $P$ is strengthened by almost 2\%.

Ultimately, using a polarized primary in a type-2 AGN model without special or general relativity has minimal impact onto the resulting $P$ and $\Psi$
(if $P_{\rm init}$ is small). Depending on the configuration of the emission source, both polarization indicators can vary. If the initial polarization 
angle is parallel, the observed $P$ will be slightly higher in the hard X-ray band, while if $\Psi_{\rm init}$ is perpendicular, it is the polarization 
of the soft X-ray band that will increase. However, in the soft band, scattering off the wind and torus completely dominates over the input polarization. 
If strong gravity effects are not important close to the central SMBH, it is unlikely that we will be able to retrieve the initial polarization of the 
continuum source when looking at type-2 AGN.

\subsection{Results for a 2\% parallel polarized primary including strong gravity effects}
\label{Results:POL_PARA_GR}
For the last two subsections, we include both strong gravity effects and the polarization of the corona emission. Similarly to the previous case, we start with 
a 2\% parallel polarized primary and plot the results in Fig.~\ref{Fig:Polarized_primary_GR_perp_Absorbing_nh21_wind}, Fig.~\ref{Fig:Polarized_primary_GR_perp_Ionized_wind}
and Fig.~\ref{Fig:Polarized_primary_GR_perp_No_wind}.

The most striking result between models with and without strong gravity effects (and including a polarized corona emission) is visible in the energy-dependent 
polarization angle. $\Psi$ smoothly rotates from large values to small angles at high energies, where photons can cross the equatorial region. The 
inclusion of a polarized corona emission drastically alters the value of $\Psi$ for the models with a polar wind, while it has only a modest influence on 
the ``bare'' AGN model. In particular, the rotation of the polarization angle is postponed to higher energies and the polarization degree observed 
at infinity is stronger than in the unpolarized primary case. Relativistic signatures can be seen in the polarimetric signal of photons with 
energy superior to the energy of $\Psi$ transition, except in the case of a torus with hydrogen column density $\ge$ 10$^{25}$~at/cm$^2$. Photo-electric 
absorption is too strong for the remaining photons that scattered along equatorial plane to impose their polarization state over photons that have scattered 
off the winds or off the torus. An AGN dominated by highly ionized winds will also tend to smooth out the scattering-induced changes, revealing with 
higher precision the strong gravity effects.

\subsection{Results for a 2\% perpendicular polarized primary including strong gravity effects}
\label{Results:POL_PERP_GR}
We find similar conclusions in the case of a 2\% perpendicular polarized corona emission including strong gravity effects, see 
Fig.~\ref{Fig:Polarized_primary_GR_para_Absorbing_nh21_wind}, Fig.~\ref{Fig:Polarized_primary_GR_para_Ionized_wind} and Fig.~\ref{Fig:Polarized_primary_GR_para_No_wind}.
Most of the differences are related to the energy-dependent polarization angle that is quite distinct between the two initial polarization states.
A primary emission with parallel polarization will create larger $\Psi$ rotations than a corona emission with perpendicular polarization, allowing 
for a clear distinction between the two models. The degree of polarization follows the levels dictated by the source, with the exception 
of the high, scattering-induced polarization in the very soft band, where polar scattering dominates. Similarly to Sect.~\ref{Results:POL_PERP_PL},
$P$ is higher in that energy band due to the similarity of the polarization angles between the source and the polar scattering mechanism.

\begin{figure}
    \includegraphics[trim = 0mm 0mm 0mm 0mm, clip, width=8.5cm]{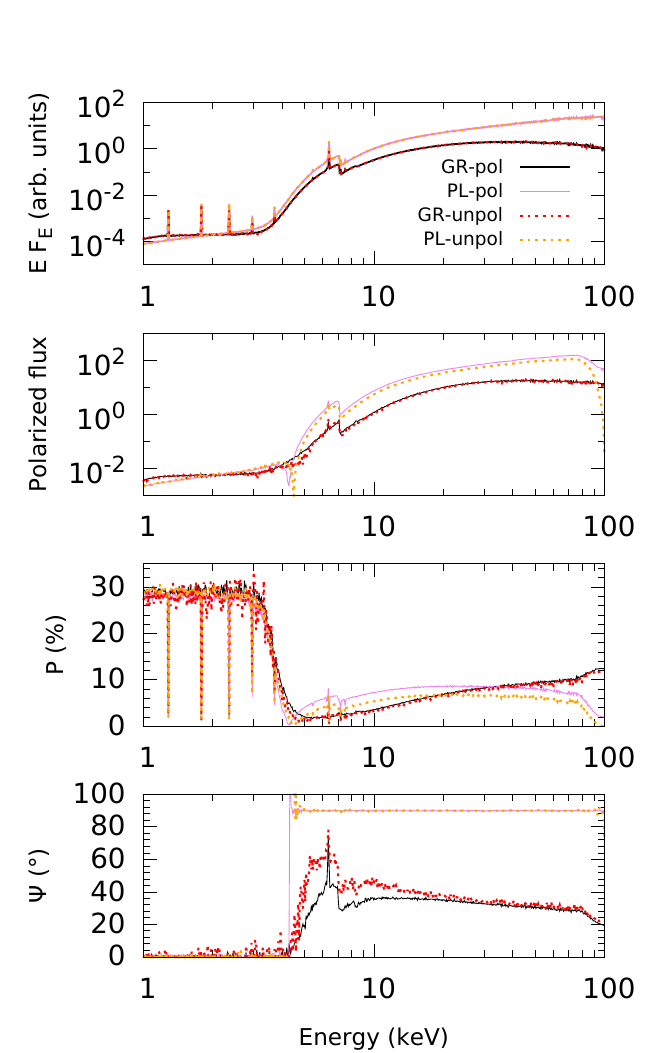}
    \caption{X-ray flux (F$_{\rm E}$ is energy flux at energy E), polarized flux (E~F$_{\rm E}$ 
	    times the polarization fraction), polarization degree $P$ and polarization position angle 
	    $\Psi$ seen by an observer at infinity for a Seyfert-2 AGN. The nucleus inclination is set 
	    to 70$^\circ$. The model consists of two continuum sources above and below the disk irradiating
	    a cold material, plus a circumnuclear molecular torus (n$_{\rm H_{torus}}$ = 10$^{24}$~at.cm$^{-2}$) 
	    and a pair of collimated, Compton-thin, absorbing polar outflows. The 
	    four spectra show the observed fluxes and polarization for different 
	    sources (perpendicularly polarized or unpolarized power-laws with/without 
	    general relativistic effects). Only one flavor of SMBH is shown (spin = 1).}
    \label{Fig:TF_PO_PA_difference}
\end{figure}

~\

Our principal conclusions can be summarized in Fig.~\ref{Fig:TF_PO_PA_difference}. We plotted the results of a model with a Compton-thin absorbing 
polar outflow plus a circumnuclear molecular torus with n$_{\rm H_{torus}}$ = 10$^{24}$~at.cm$^{-2}$. We included the results of a model where strong 
gravity effects are off, and results for a model where relativity is accounted for. We overplotted the outcomes for different models investigated 
in this paper, including perpendicularly polarized or unpolarized power-laws, leading to a total four models. We leave aside the spectroscopic and 
polarization spectra of a model with parallel input polarization for clarity purposes, as the conclusions are very similar to the conclusions 
for the perpendicular polarization case. 

All the models give very similar results in terms of spectroscopy, particularly below 10~keV. At higher energies, strong gravity effects tend to 
suppress the flux from the Compton hump with respect to the models with Newtonian physics. This is a behavior expected from observations and simulations, 
see, e.g., \citet{Risaliti2013} for an application to the galaxy NGC~1365, but it remains difficult to properly estimate all the model constituents
to reproduce observations. A more secure option is to rely on X-ray polarimetry, as it can be seen in the third and fourth panels of Fig.~\ref{Fig:TF_PO_PA_difference}.
A model without a polarized corona emission, nor strong gravity effects, will present a large degree of polarization in the soft X-ray band if polar 
scattering takes place but its $P$ will be among the smallest at high energies. A rotation of 90$^\circ$ is expected between the two bands, 
with the polarization angle being either 0$^\circ$ or 90$^\circ$. Including a polarized primary will only result in increasing $P$ at high energies,
making any distinction between the two models difficult, but allowing for an easier detection. A soon as the strong gravity effects are included,
the polarization degree changes in the hard X-ray band and follows the polarization of the corona emission if photons are sufficiently energetic to pass
through the circumnuclear gas without being absorbed. The polarization position angle keeps a trace of its initial polarization, showing smooth 
variations of $\Psi$ with energy. Those variations strongly depend on the polarization of the primary emission, but they never rotate by 90$^\circ$
such as in the Newtonian case.

\section{Detectability}
\label{Detectability}
We saw that including an initial polarization and strong gravity effects can have major repercussions onto the polarimetric signal from type-2 AGN. 
From our simulation, it is clear that even at edge-on inclinations it might be possible to detect special and general relativity effects imprinted 
in the X-ray polarimetric spectra if the amount of obscuring hydrogen column density along the observer's line-of-sight is not equal or larger than
10$^{25}$~at.cm$^{-2}$. Hence, in the following section, we will investigate the detectability of a typical type-2 AGN such as presented in 
Fig.~\ref{Fig:Scheme} and Fig.~\ref{Fig:TF_PO_PA_difference}, using three different space missions. To illustrate this example, we set the X-ray 
flux of the model to equal the broadband X-ray flux of NGC~1068 \citep{Matt2004,Pounds2006,Cardamone2007}, an archetypal Seyfert-2 galaxy. The Galactic 
column density towards this AGN was estimated to be close to 2.99~$\times$~10$^{20}$~cm$^{-2}$ \citep{Murphy1996}, and we accounted for this value 
in our estimations of the observed X-ray polarization.

\begin{figure*}
    \includegraphics[trim = 0mm 0mm 0mm 0mm, clip, width=17.5cm]{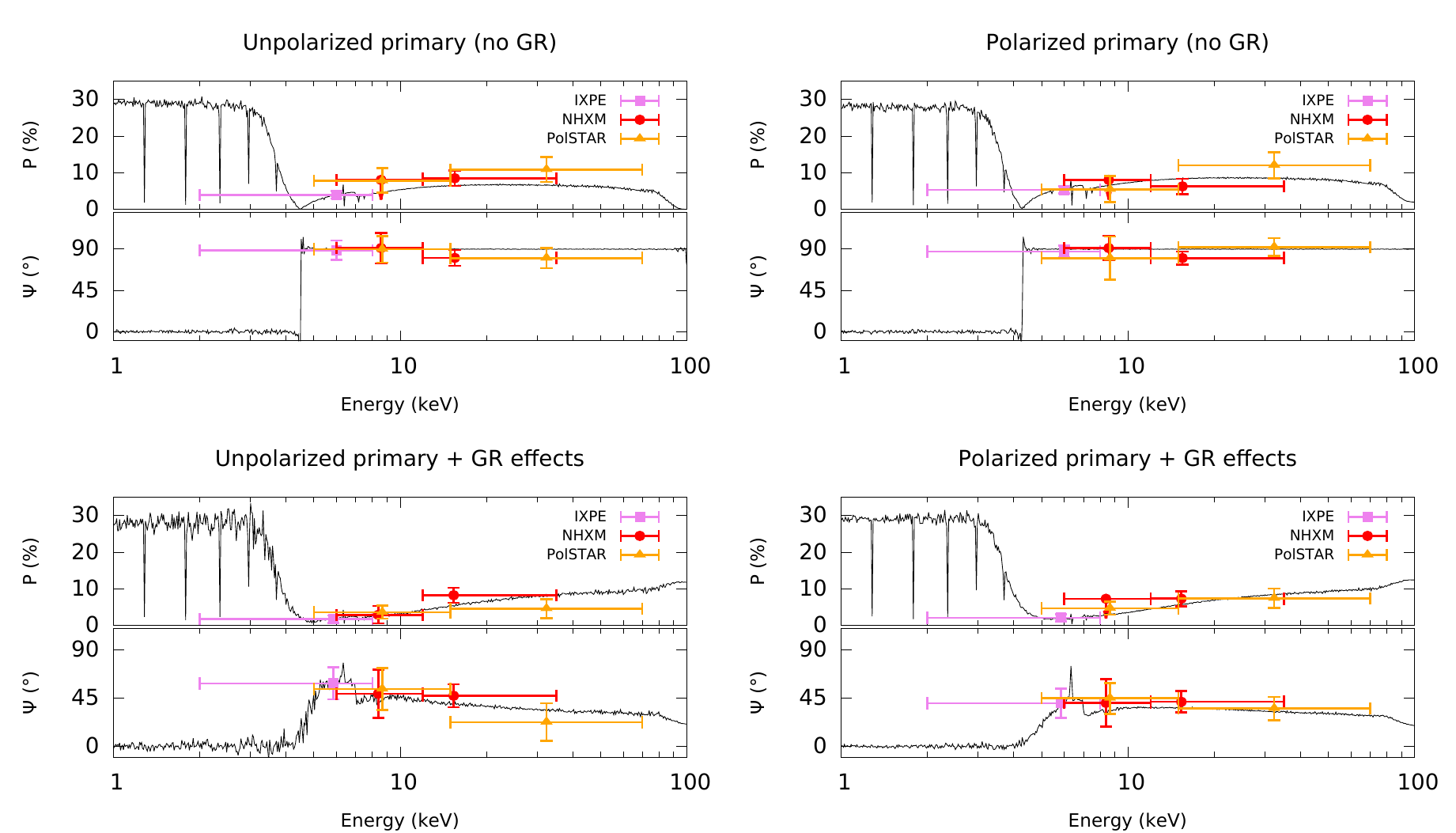}
    \caption{Simulated polarimetric observation for a generic Seyfert-2 
	    galaxy. We used the energy spectrum and the polarization 
	    degrees and angles from the simulations presented in 
	    Fig.~\ref{Fig:TF_PO_PA_difference}. The four spectra 
	    show the polarization for different parametrizations 
	    of the central source (top-left: unpolarized power-law without 
	    general relativity; bottom-left: unpolarized power-law with general 
	    relativity; top-right: perpendicularly polarized corona emission without 
	    general relativity; bottom-right: perpendicularly polarized corona emission
	    with general relativity). The violet squares correspond to a
	    20~Ms observation with $IXPE$, the red circles correspond 
	    to a 2~Ms observation with $NHXM$, and the orange triangles
	    show a simulated 20~Ms observation with $PolSTAR$.}
    \label{Fig:Detectability}
\end{figure*}

\subsection{With the Imaging X-ray Polarimetry Explorer}
\label{Detectability:IXPE}
Our first prediction concerns the Imaging X-ray Polarimetry Explorer (IXPE), a NASA-SMEX mission that was selected on January the 3$^{\rm rd}$, 2017,
for launch in late 2020. The IXPE spacecraft will study black holes and other high energy astronomical phenomena thanks to its three telescopes 
dedicated to X-ray polarimetry \citep{Weisskopf2016}. The energy range of sensitivity, between 2 and 8~keV, will allow to observe representative 
objects belonging to basically all the classes of high energy sources. For several extended sources, IXPE will perform imaging polarimetry for the
first time. The capability to measure also the time of arrival and the energy of the absorbed photons will allow to perform time and and 
spectrally-resolved measurements.

Detectability of the effects presented in the previous sections was evaluated by means of Monte Carlo simulations, performed with a the same code 
used in \citet{Dovciak2011} or in \citet{Taverna2014}. The source spectra and the instrument effective area are used to compute the counting rate 
on the instrument, whereas the source polarization and the amplitude of the instrumental response to polarization, expressed through the modulation 
factor, are used to derive the amplitude of the expected signal. The code eventually returns a modulation curve which is representative of the 
output of the real instrument; this is processed like real data and an estimate of the measured polarization, together with its error, is obtained.

We present in Fig.~\ref{Fig:Detectability} the detectability of the IXPE mission. The Minimum Detectable Polarization (MDP), which is the degree 
of polarization corresponding to the amplitude of modulation that has only a 1\% probability of being detected by chance, is about 3\% in this case.
Errors are at 1-$\sigma$ confidence level only if we want to measure just one parameter (the degree or the angle of polarization, but not both), see
\citet{Strohmayer2013}. If we aim to measure both polarization parameters, the 1-$\sigma$ errors are about 50\% larger. To achieve a detection 
of the polarimetric signal in the 2--8~keV, we estimate that a 20~Ms observation with IXPE is necessary. Despite a large polarization degree in 
the soft X-ray band, the relative dimness of type-2 AGN compared to type-1 objects (unobscured by the equatorial torus) drives longer observational 
requirements. For this reason, we integrated the X-ray polarization over the whole energy band of IXPE in order to minimize the observational 
time, resulting in only one measurement. Since the polarization of our models shows energy-dependent polarization properties, it would be 
beneficial to have a complementary observation in a harder X-ray band. This could be achieved thanks to the following polarimetric mission
concepts.

\subsection{With the New Hard X-ray Mission}
\label{Detectability:NHXM}
The New Hard X-ray Mission (NHXM) was a concept mission based on new technologies in mirror and detector manufacturing, aiming to 
achieve imaging X-ray spectroscopy and polarimetry in a broad-band energy range \citep{Tagliaferri2012}. A set of three X-ray optics
based on multi-layer technology were coupled with focal plane detectors to achieve imaging and spectroscopy between 0.5 and 80~keV; 
a forth optics was dedicated to imaging X-ray polarimetry with the alternate use of two Gas Pixel Detectors optimized in the 2--10~keV 
and 6--35~keV energy range. The latter, called Medium Energy Polarimetry (MEP), is particularly indicated to study the effects presented above, 
because the sensitivity is in the energy range where the polarization signal is higher. To evaluate the detectability with NHXM, we used 
the same approach followed for IXPE.

Thanks to its MEP, the NHXM mission would have been able to measure the energy-dependent X-ray polarization of Seyfert-2 galaxies. The MDP 
in the soft band is of the order of 8\% while it is about 6\% in the higher energy band. The amount of time necessary to achieve those measurement 
would have been ten times shorter than for the IXPE mission, about 2~Ms here (see Fig.~\ref{Fig:Detectability}). The acquisition of two 
data point would have been sufficient to distinguish between the models with and without strong gravity effects. However, a longer exposure 
time would have been necessary to distinguish between a polarized and an unpolarized primary emission.

\subsection{With the Polarization Spectroscopic Telescope Array}
\label{Detectability:polSTAR}
The Polarization Spectroscopic Telescope Array (PolSTAR) is a mission designed to measure 3 -- 50~keV polarization of compact objects with a 
scattering polarimeter, which was proposed in response to NASA's 2014 SMEX announcement of opportunity \citep{Krawczynski2016}. PolSTAR is 
built on technology developed for NuSTAR, namely its X-ray optics, extensible telescope boom, optical bench, and CdZnTe detectors. In PolSTAR, 
the X-rays are focused onto a cylindrical LiH scattering element surrounded by 16 CZT detectors to measure the scattered X-rays. The technique 
makes use of the fact that photons scatter preferentially perpendicular to their polarization direction. PolSTAR has a relatively uniform 
modulation factor of $\mu\sim0.5$ throughout its entire energy bandwidth. It achieves its maximum polarization sensitivity between
7 -- 14~keV, with an MDP of 1\% for an 860~ksec observation of a 20~mCrab source. 

In order to simulate the response of PolSTAR for our AGN model, we normalized the absorbed model flux to the observed ASCA 2 -- 8~keV flux of 
5.19 $\times$ 10$^{-12~}$erg~cm$^{-2}$~s$^{-1}$. We then numerically integrated the reduced Stokes parameters of the model, $\mathcal{Q}_{r}$ 
and $\mathcal{U}_{r}$, see \citet{Kislat2015}, over the simulated energy bins (5 -- 15~keV and 15 -- 70~keV). Finally, we calculated the expected 
number of signal and background events, $N_{s}$ and $N_{BG}$, in each bin. The observed polarization fraction and angle were then drawn from the 
distribution \citep{Vinokur1965,Weisskopf2006,Krawczynski2011a}:
 
\begin{equation*}
	P(p,\psi|p_{0},\psi_{0}) = \frac{N_{obs}^{2}\mu^{2}p}{4\pi(N_{obs}+N_{BG})}\exp\left[A\right],
\end{equation*}

with

\begin{equation*}
	A = -\frac{N_{obs}^{2}\mu^{2}}{4(N_{obs}+N_{BG})}\left(p_{0}^{2}+p^{2}-2pp_{0}\cos\left(2(\psi-\psi_{0})\right)\right),
\end{equation*}

where $p_{0}=\sqrt{\mathcal{Q}_{r}+\mathcal{U}_{r}}$ and $\psi_{0}=\frac{1}{2}\arctan\frac{\mathcal{U}_{r}}{\mathcal{Q}_{r}}$ are the true polarization 
fraction and angle, and $N_{obs}$ is the simulated number of signal photons due to $N_{s}$. The expected number of background events, $N_{BG}$, is 
based on the observed NuSTAR background, scaled to account for the larger detector area.

To measure the soft and hard X-ray polarization of our model of type-2 AGN and distinguish between a unpolarized/polarized source dominated (or not) 
by strong gravity effects, about 20~Ms is required (see Fig~\ref{Fig:Detectability}). This is ten times higher than for the NHXM mission, but it is 
scaled with the physical size of the detectors (where NHXM was intended to be a medium-sized satellite). A broadband X-ray polarimeter able to measure 
the polarization in the Compton hump gives a clear picture of the importance of strong gravity effects, with the polarization position angle showing a 
larger rotation between the soft and hard X-ray bins of PolSTAR with respect to the other two missions.

~\

Ultimately, a measurement of the X-ray polarization of type-2 AGN is within the capabilities of the aforementioned three instruments but long 
observing times are required. This is due to the obscuration of the central source by the optically thick molecular gas around the equatorial plane.
Despite the high degree of polarization expected in the soft X-ray band, the starvation of photons hampers an easy detection of the polarimetric 
signatures of obscured objects, and we showed here that a least a couple of megaseconds are required (with the exact amount of time needed being 
model-dependent).

\section{Conclusions}
\label{Conclusions}
In this paper, we explored in great details the X-ray polarization emerging from complex type-2 AGN modeling. For the first time, we coupled the 
strong gravity effects near the horizon of the central supermassive black hole to the distant scattering and absorbing media that shape the 
observed fluxes of Seyfert galaxies. To do so, we used the {\sc ky} code that computes the parallel transport of polarization along the photon 
null geodesics close to a potential well and, from a certain radius where relativistic effects are no longer important, the radiative transfer code
{\sc stokes} takes over to propagate radiation through the torus and polar winds. It results in the first coherent modeling of X-ray polarization 
from type-2 AGN. We explored a large variety of AGN structures, including or excluding winds (either ionized or filled with neutral, cold matter),
and varying the hydrogen column density along the observer's line-of-sight intercepted by the puffed-up torus. The polarization state of the 
continuum source was investigated and we analyzed how the initial polarization modify the final polarization observed at infinity.

We found that Seyfert-2s with clear electron scattering in their soft X-ray spectra (such as NGC~4945, see \citealt{Madejski2000,Puccetti2014}) 
are excellent potential targets for a future polarimetric detection as, regardless of the ionization stage of the wind, we expect a large 
polarization degree at soft X-ray energies. On the other hand, Compton-thick, windless AGN such as the equivalent of bare type-1 AGN \citep{Patrick2011}
might be more problematic to observe due to the strong absorption of photons below 10~keV. If strong gravity effects are not dominant close to 
the central engine, then the polarization position angle can take only 2 values (0$^\circ$ and 90$^\circ$). However, when special and general 
relativity are accounted, then the polarization position angle can rotate smoothly between the soft and hard X-ray bands. This is an important 
result as, similarly to what we foresee for type-1s, looking at the polarization angle of type-2s will definitively tell us if strong gravity 
effects are important close to the central compact source. 

If special and general relativity are not shaping the X-ray spectrum of AGN close to the potential well, modifying the initial polarization 
of the continuum will not affect the final polarization degree. Scattering off the winds and torus completely dominates the input polarization
and it is impossible to retrieve the initial polarization of the continuum source when looking at type-2s. However, by looking at the energy 
at which the polarization position angle rotates from 0$^\circ$ to 90$^\circ$, it becomes feasible to derive the hydrogen column density along 
the line-of-sight (with a small degeneracy on the observer's inclination). 

Adding strong gravity effects completely changes the picture. First the polarization angle becomes energy-dependent and differs between a Schwarzschild 
and a Kerr black hole. In the soft X-ray band, photo-ionization dominates and most of the information about the spin of the central source is 
lost due to the overwhelming importance of polar scattering in the wind. At higher energies (or for Compton-thin type-2s), photons can travel 
through the equatorial gaseous medium and it becomes feasible to observe the energy-dependent variation of the polarization angle, together with 
a different polarization degree than what is expected from the purely Newtonian case. Unfortunately, a polarized or an unpolarized primary radiation 
has almost no effect onto the final polarization spectrum of a type-2 AGN, independently of the inclusion of strong gravity effects or not. This 
is a very different conclusion from what we expect for type-1s. 

We have shown that the future generation of X-ray polarimeters will be able to measure the degree and angle of polarization for type-2 objects, 
albeit long integration times (over a mega-second, even for bright type-2s). With the development of photo-electric and scattering polarimeters,
and the ever-improving resolution and sensitivity of modern satellites, it will be soon possible to unveil the physical effects and the organization 
of matter even in the most obscured astronomical sources thanks to X-ray polarimetry.

\section*{Acknowledgments}
The authors are grateful to Ski Antonucci for his remarkable suggestions about the X-ray spectra of type-2 AGN, and to
Ren\'e W. Goosmann for his help with the code and the English grammar. FM would like to thank the meso-center of Strasbourg, 
partner of the CPER 2015 - 2020 project "Alsacalcul", for allocating us the time to run our simulations. FFK and HSK acknowledge 
funding from NASA through Grants NNX10AJ56G, NNX12AD51G and NNX14AD19G.

\label{lastpage}
\clearpage

\appendix

\setcounter{figure}{0}
\renewcommand{\thefigure}{A\arabic{figure}}

\begin{figure*}
    \includegraphics[trim = 0mm 0mm 0mm 0mm, clip, width=6.5cm]{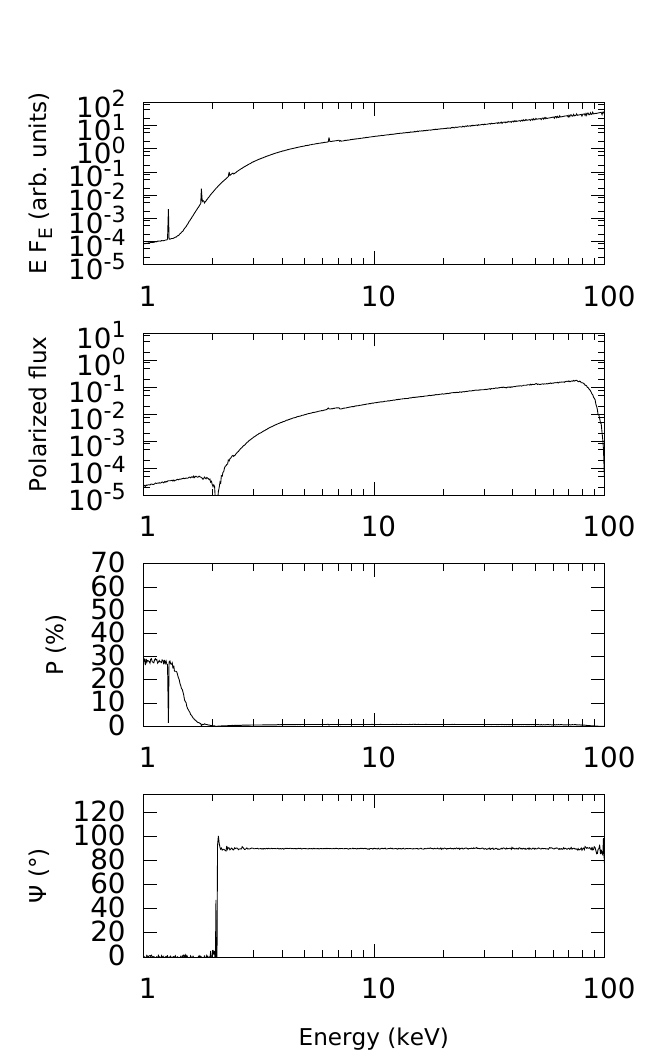}
    \hspace{5pt}\vrule\hspace{5pt}%
    \includegraphics[trim = 0mm 0mm 0mm 0mm, clip, width=6.5cm]{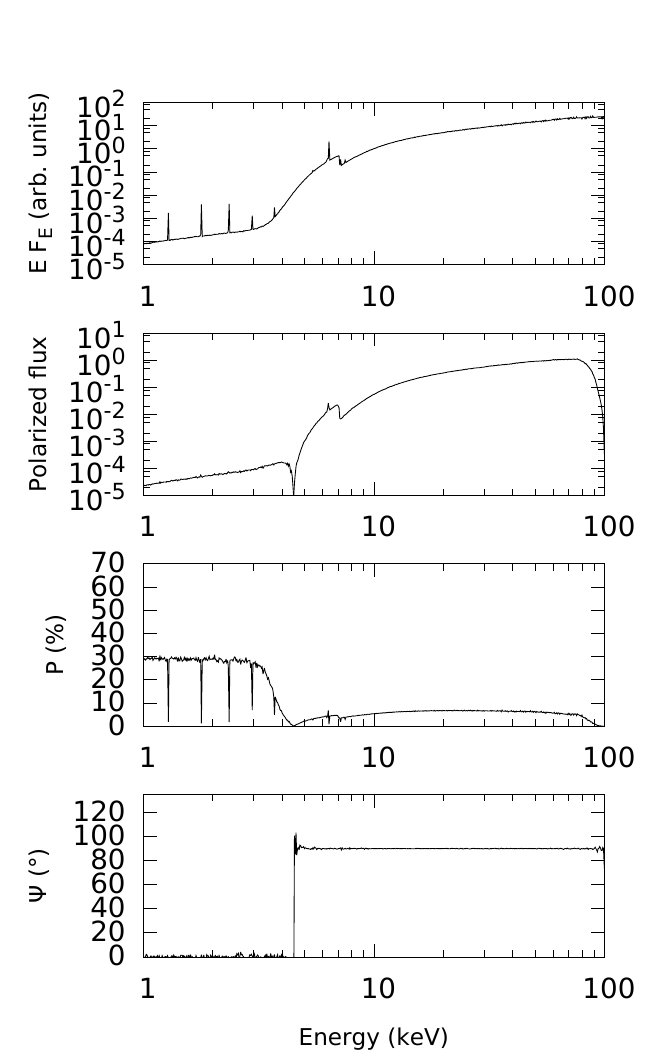}
    \includegraphics[trim = 0mm 0mm 0mm 0mm, clip, width=6.5cm]{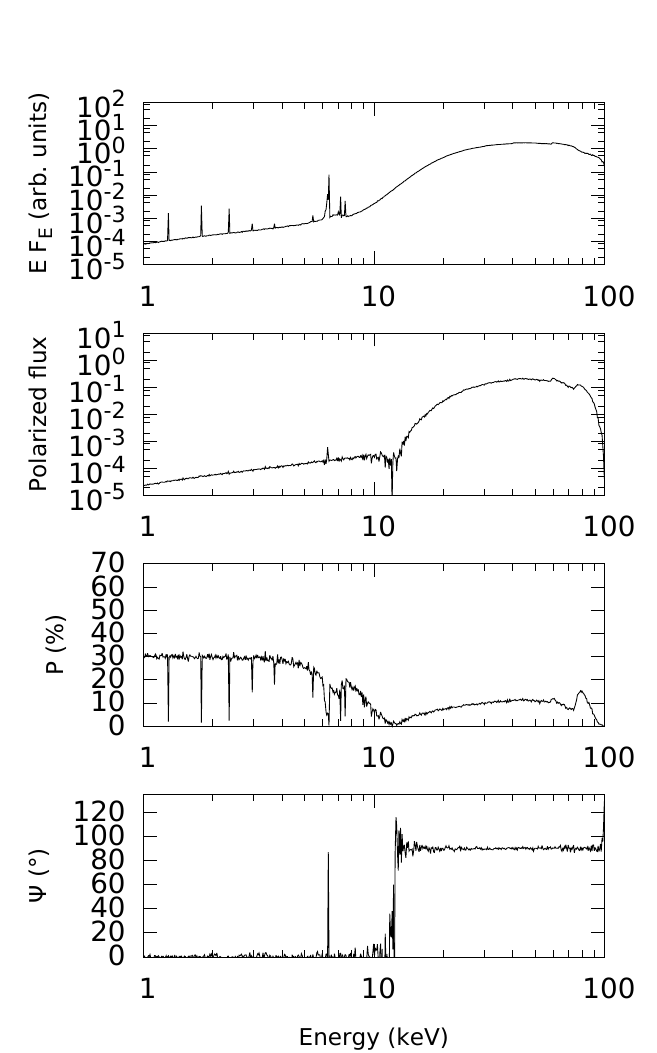}    
    \caption{X-ray flux (F$_{\rm E}$ is energy flux at energy E), polarized flux, polarization 
	    degree and polarization position angle for a type-2 AGN with Compton-thin 
	    absorbing polar winds (n$_{\rm H_{wind}}$ = 10$^{21}$~at.cm$^{-2}$).
	    Top-left: n$_{\rm H_{torus}}$ = 10$^{23}$~at.cm$^{-2}$; top-right:
	    10$^{24}$~at.cm$^{-2}$; bottom: 10$^{25}$~at.cm$^{-2}$. See text 
	    for additional details about the model components. The input 
	    spectrum is unpolarized. Strong gravity effects are not included.}
    \label{Fig:Unpolarized_primary_PL_Absorbing_nh21_wind}
\end{figure*}

\begin{figure*}
    \includegraphics[trim = 0mm 0mm 0mm 0mm, clip, width=6.5cm]{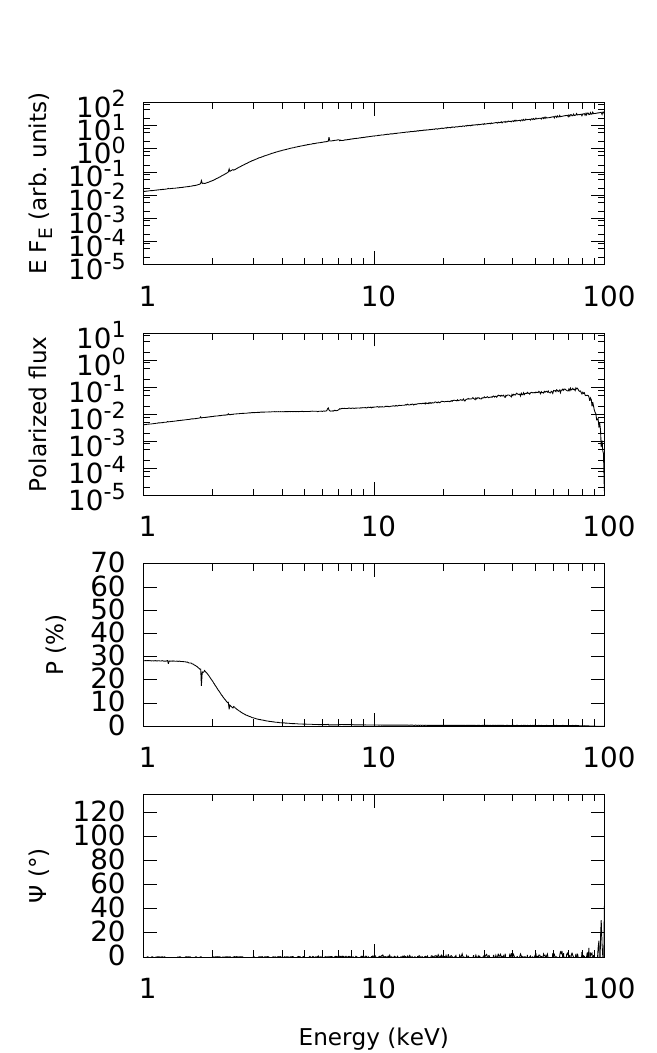}
    \hspace{5pt}\vrule\hspace{5pt}%
    \includegraphics[trim = 0mm 0mm 0mm 0mm, clip, width=6.5cm]{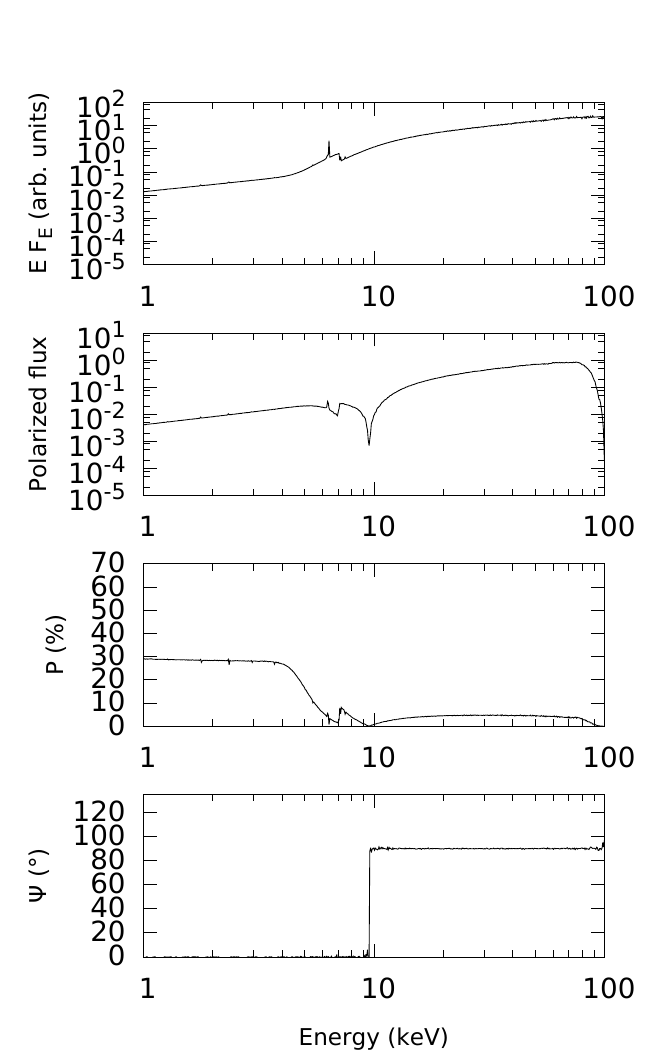}
    \includegraphics[trim = 0mm 0mm 0mm 0mm, clip, width=6.5cm]{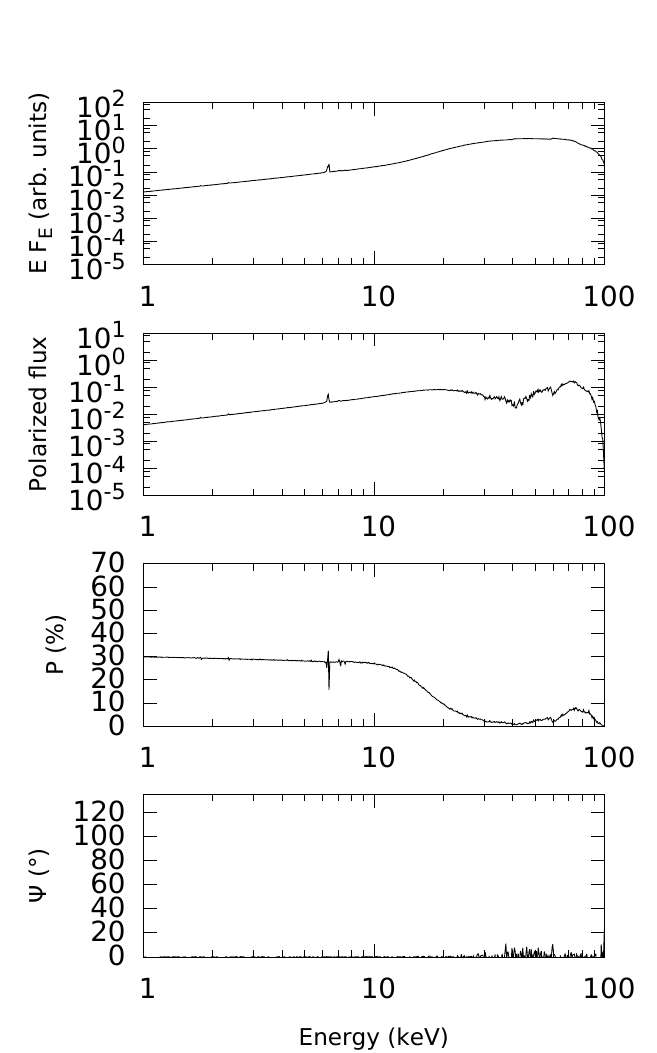}    
    \caption{X-ray flux (F$_{\rm E}$ is energy flux at energy E), polarized flux, polarization 
	    degree and polarization position angle for a type-2 AGN with fully-ionized 
	    polar winds. Top-left: n$_{\rm H_{torus}}$ = 10$^{23}$~at.cm$^{-2}$; top-right:
	    10$^{24}$~at.cm$^{-2}$; bottom: 10$^{25}$~at.cm$^{-2}$. See text 
	    for additional details about the model components. The input 
	    spectrum is unpolarized. Strong gravity effects are not included.}
    \label{Fig:Unpolarized_primary_PL_Ionized_wind}
\end{figure*}

\begin{figure*}
    \includegraphics[trim = 0mm 0mm 0mm 0mm, clip, width=6.5cm]{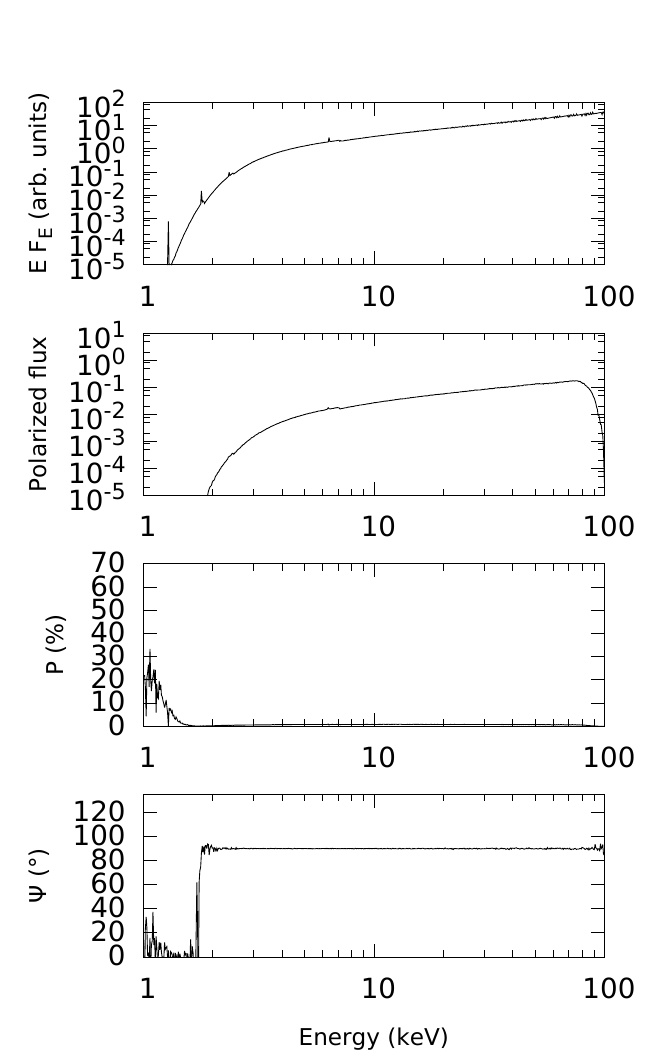}
    \hspace{5pt}\vrule\hspace{5pt}%
    \includegraphics[trim = 0mm 0mm 0mm 0mm, clip, width=6.5cm]{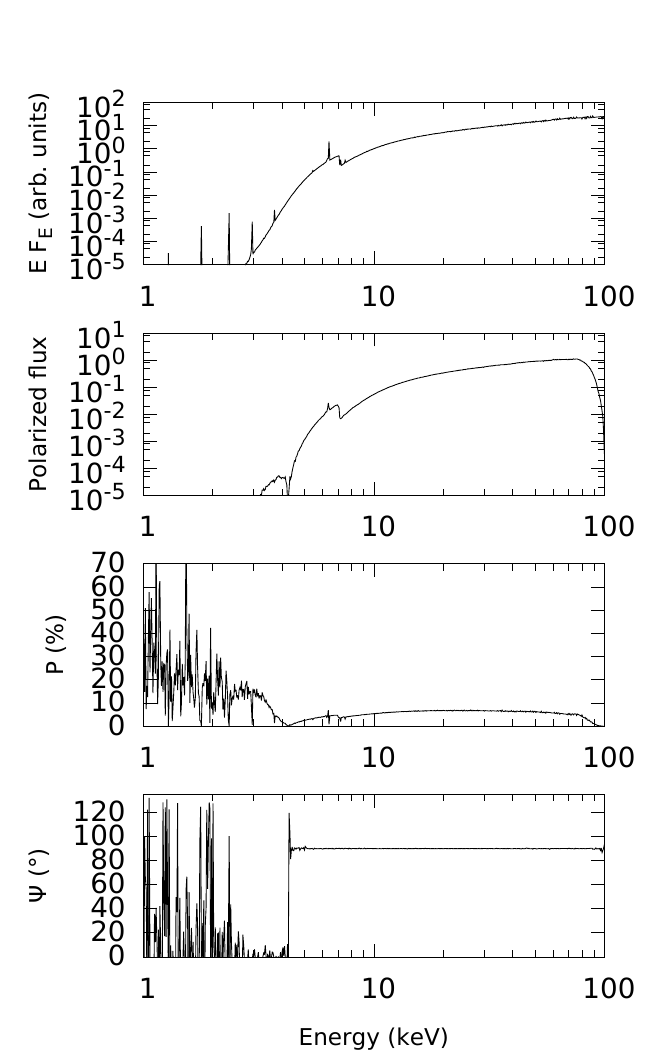}
    \includegraphics[trim = 0mm 0mm 0mm 0mm, clip, width=6.5cm]{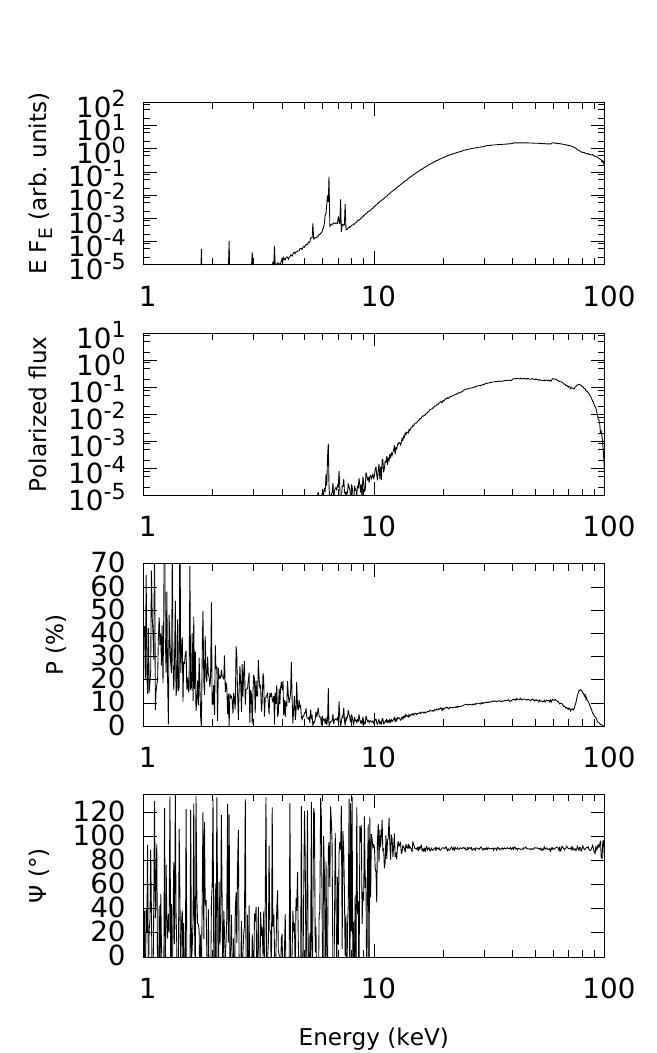}    
    \caption{X-ray flux (F$_{\rm E}$ is energy flux at energy E), polarized flux, polarization 
	    degree and polarization position angle for a type-2 AGN without polar winds.
	    Top-left: n$_{\rm H_{torus}}$ = 10$^{23}$~at.cm$^{-2}$; top-right:
	    10$^{24}$~at.cm$^{-2}$; bottom: 10$^{25}$~at.cm$^{-2}$. See text 
	    for additional details about the model components. The input 
	    spectrum is unpolarized. Strong gravity effects are not included.}
    \label{Fig:Unpolarized_primary_PL_No_wind}
\end{figure*}

\clearpage

\setcounter{figure}{0}
\renewcommand{\thefigure}{B\arabic{figure}}

\begin{figure*}
    \includegraphics[trim = 0mm 0mm 0mm 0mm, clip, width=8.5cm]{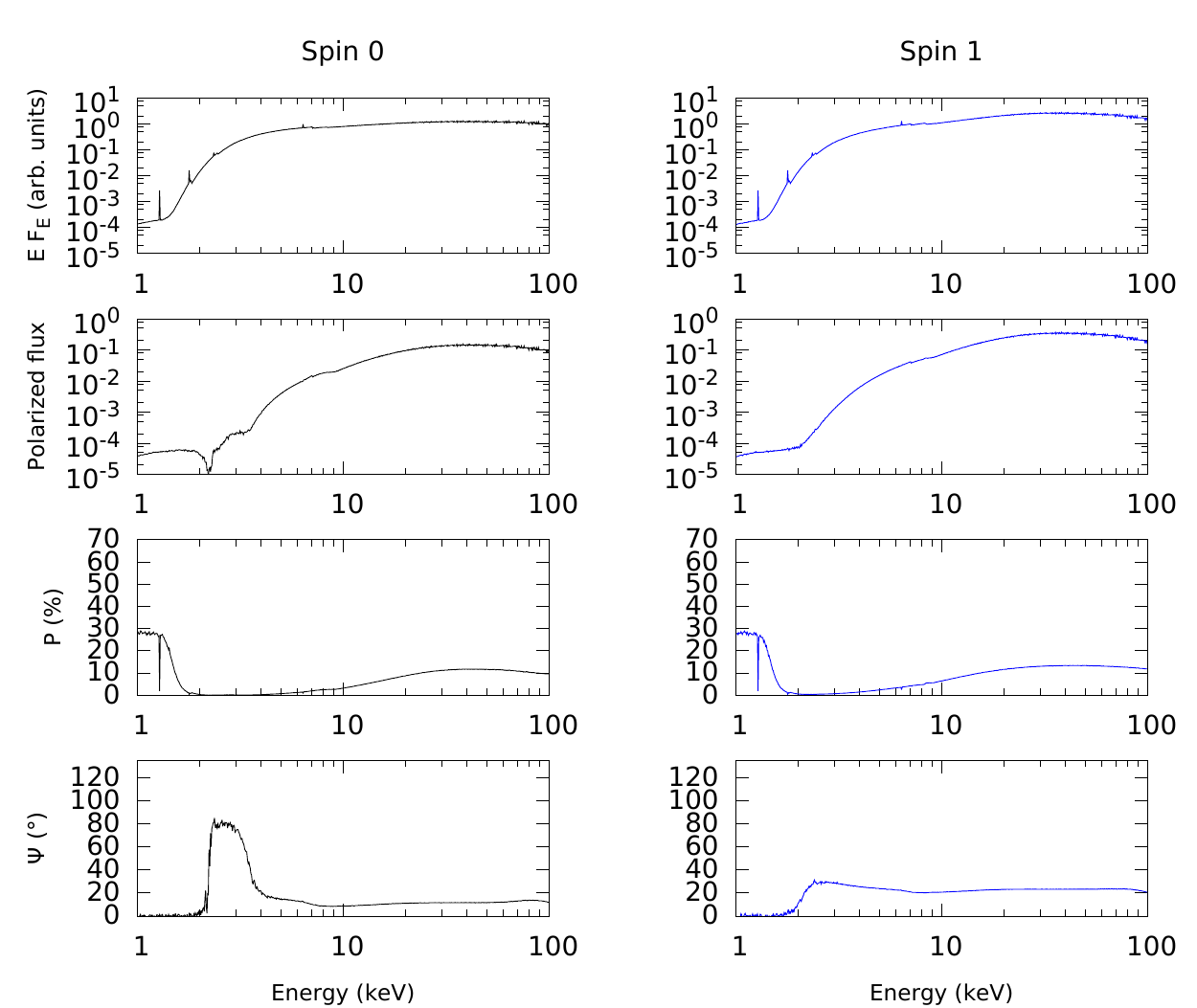}
    \hspace{5pt}\vrule\hspace{5pt}%
    \includegraphics[trim = 0mm 0mm 0mm 0mm, clip, width=8.5cm]{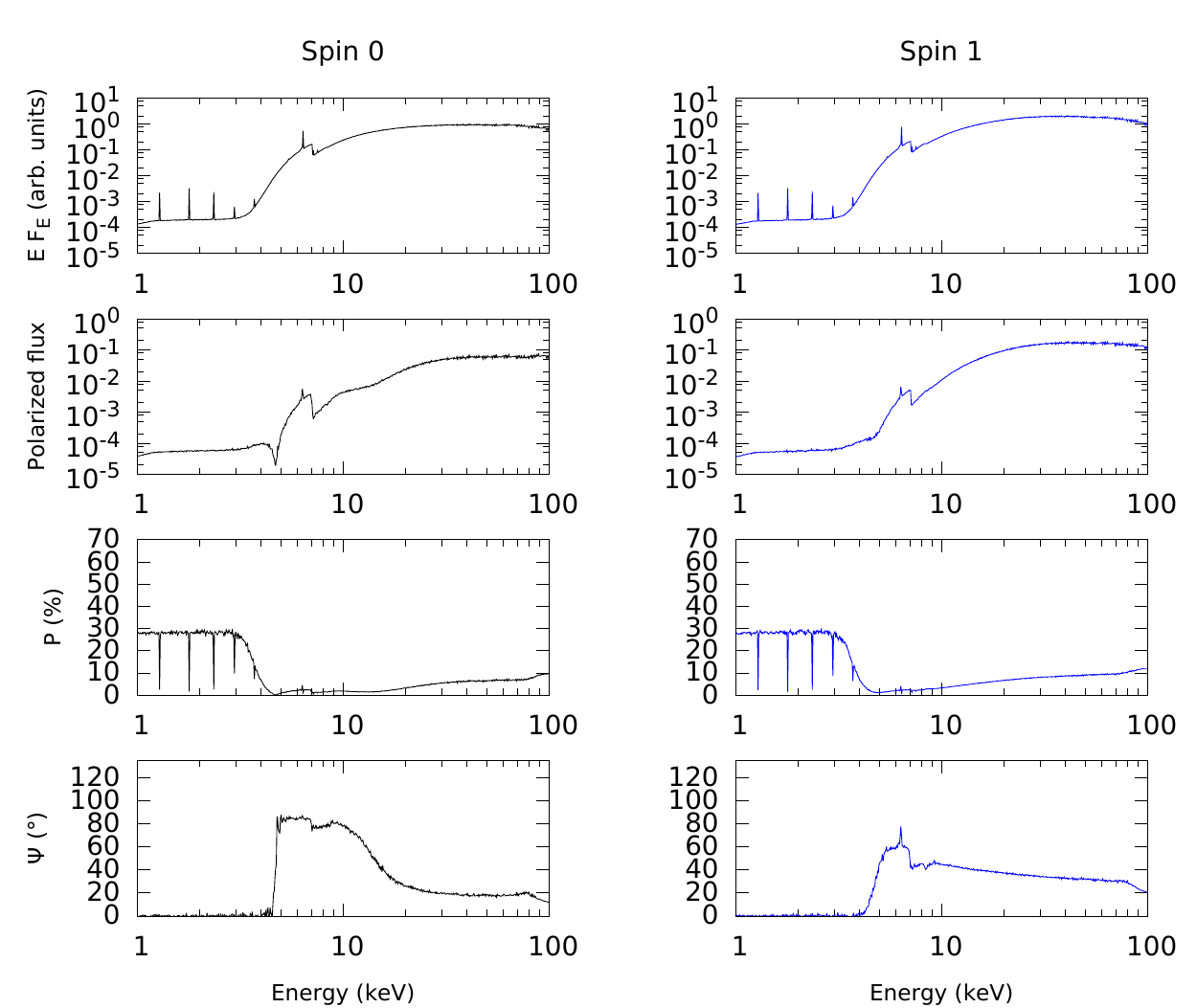}
    \includegraphics[trim = 0mm 0mm 0mm 0mm, clip, width=8.5cm]{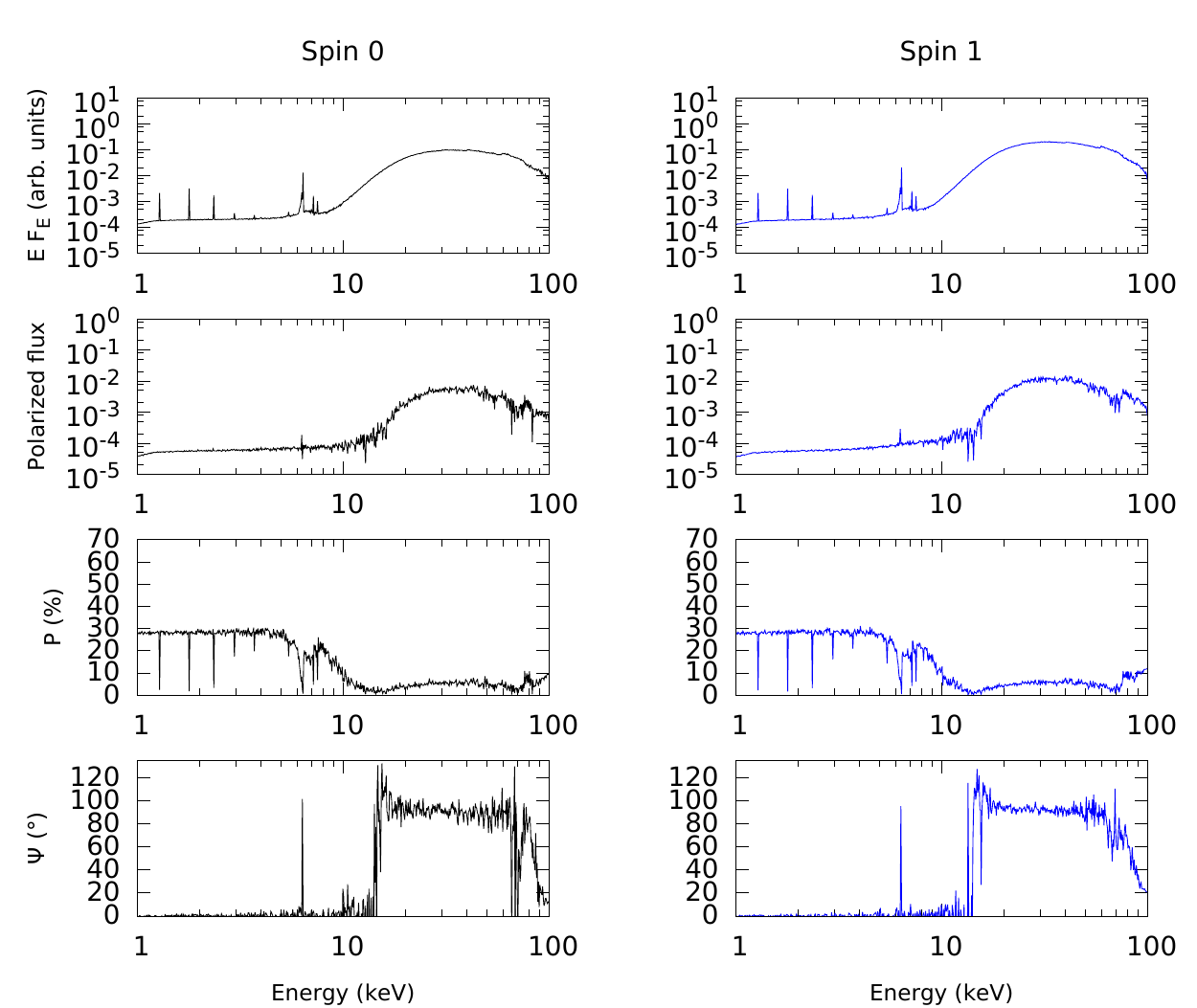}    
    \caption{X-ray flux (F$_{\rm E}$ is energy flux at energy E), polarized flux, polarization 
	    degree and polarization position angle for a type-2 AGN with Compton-thin 
	    absorbing polar winds (n$_{\rm H_{wind}}$ = 10$^{21}$~at.cm$^{-2}$).
	    Top-left: n$_{\rm H_{torus}}$ = 10$^{23}$~at.cm$^{-2}$; top-right:
	    10$^{24}$~at.cm$^{-2}$; bottom: 10$^{25}$~at.cm$^{-2}$. See text 
	    for additional details about the model components. The input 
	    spectrum is unpolarized. Strong gravity effects are included.}
    \label{Fig:Unpolarized_primary_GR_Absorbing_nh21_wind}
\end{figure*}

\begin{figure*}
    \includegraphics[trim = 0mm 0mm 0mm 0mm, clip, width=8.5cm]{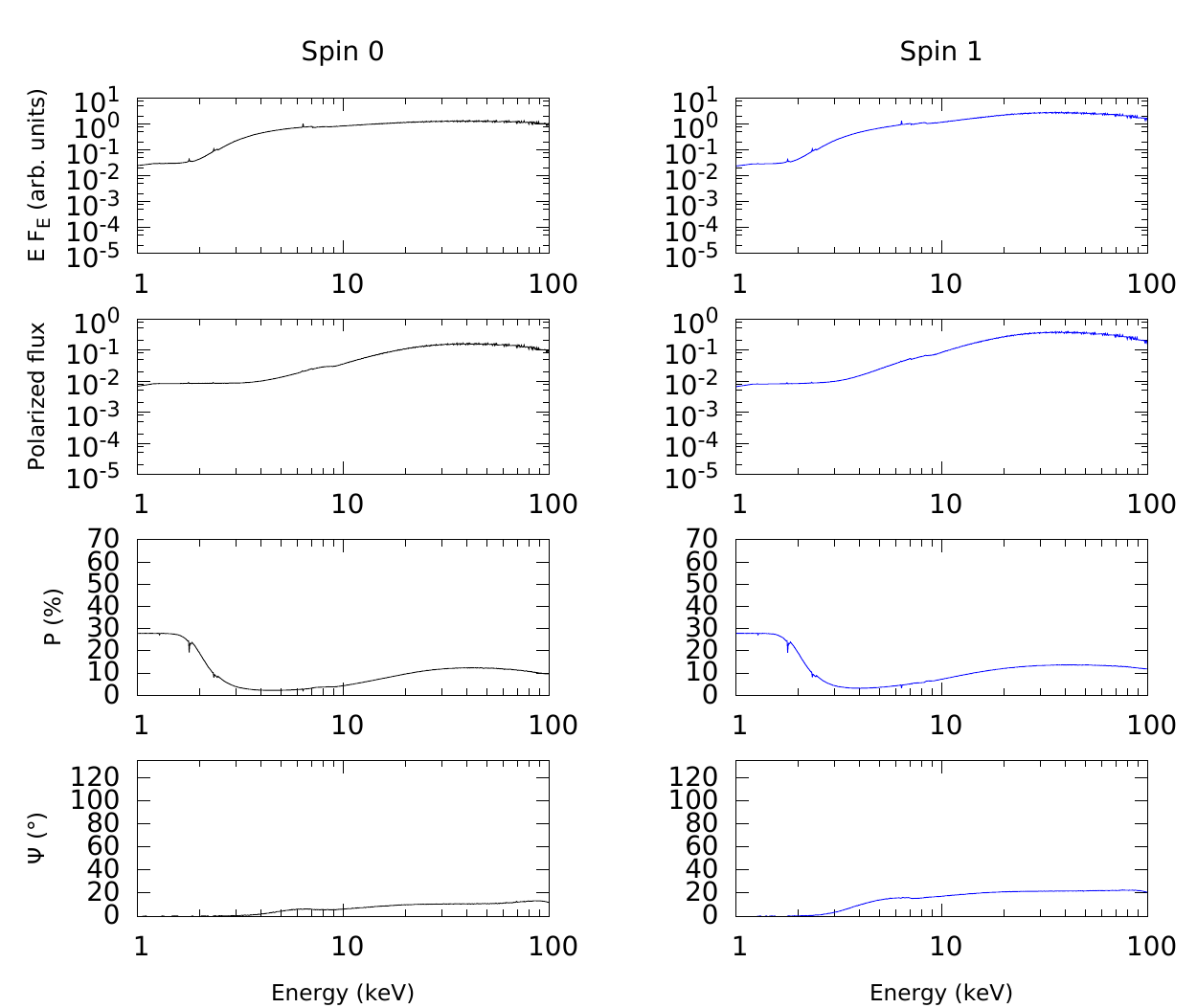}
    \hspace{5pt}\vrule\hspace{5pt}%
    \includegraphics[trim = 0mm 0mm 0mm 0mm, clip, width=8.5cm]{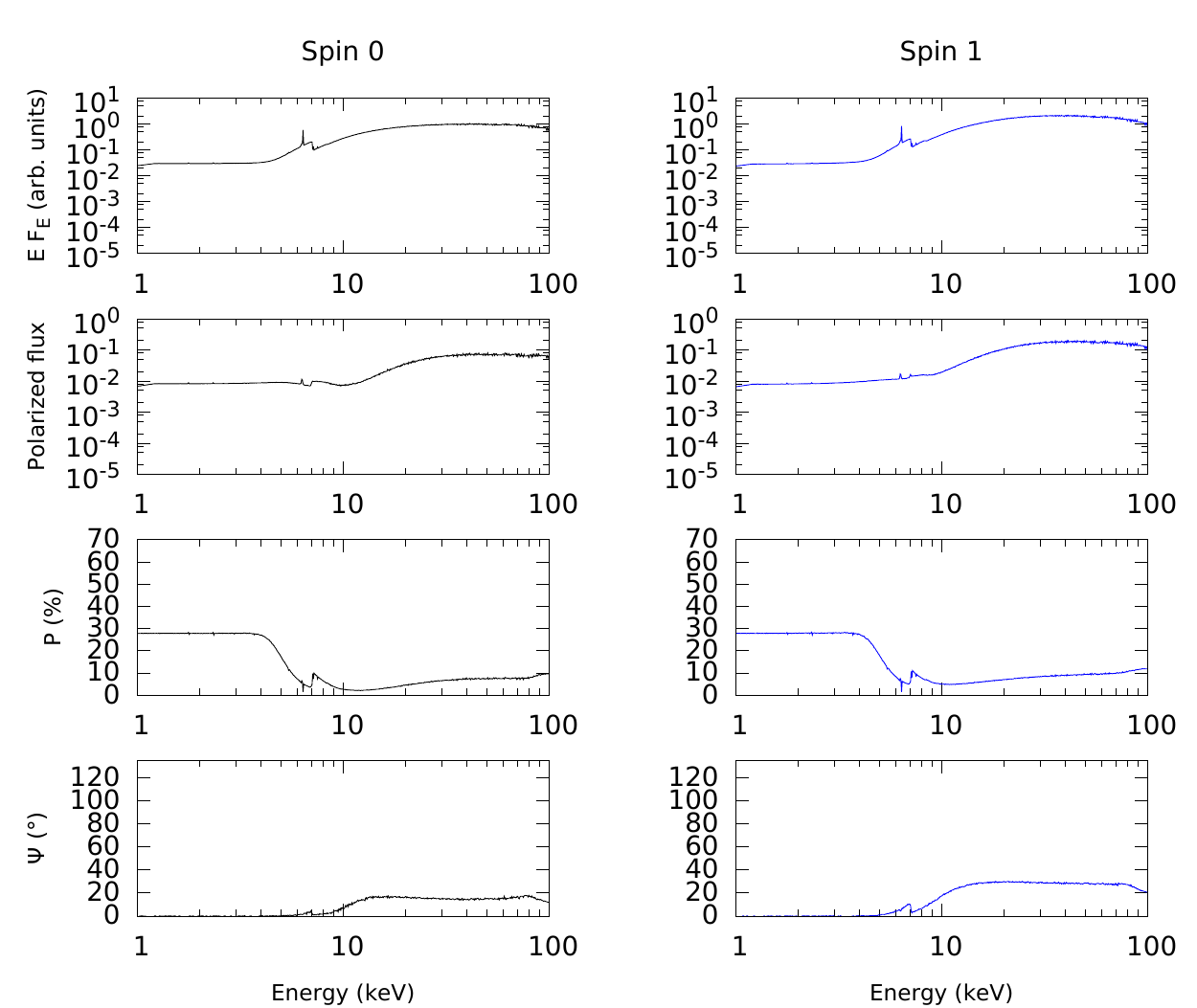}
    \includegraphics[trim = 0mm 0mm 0mm 0mm, clip, width=8.5cm]{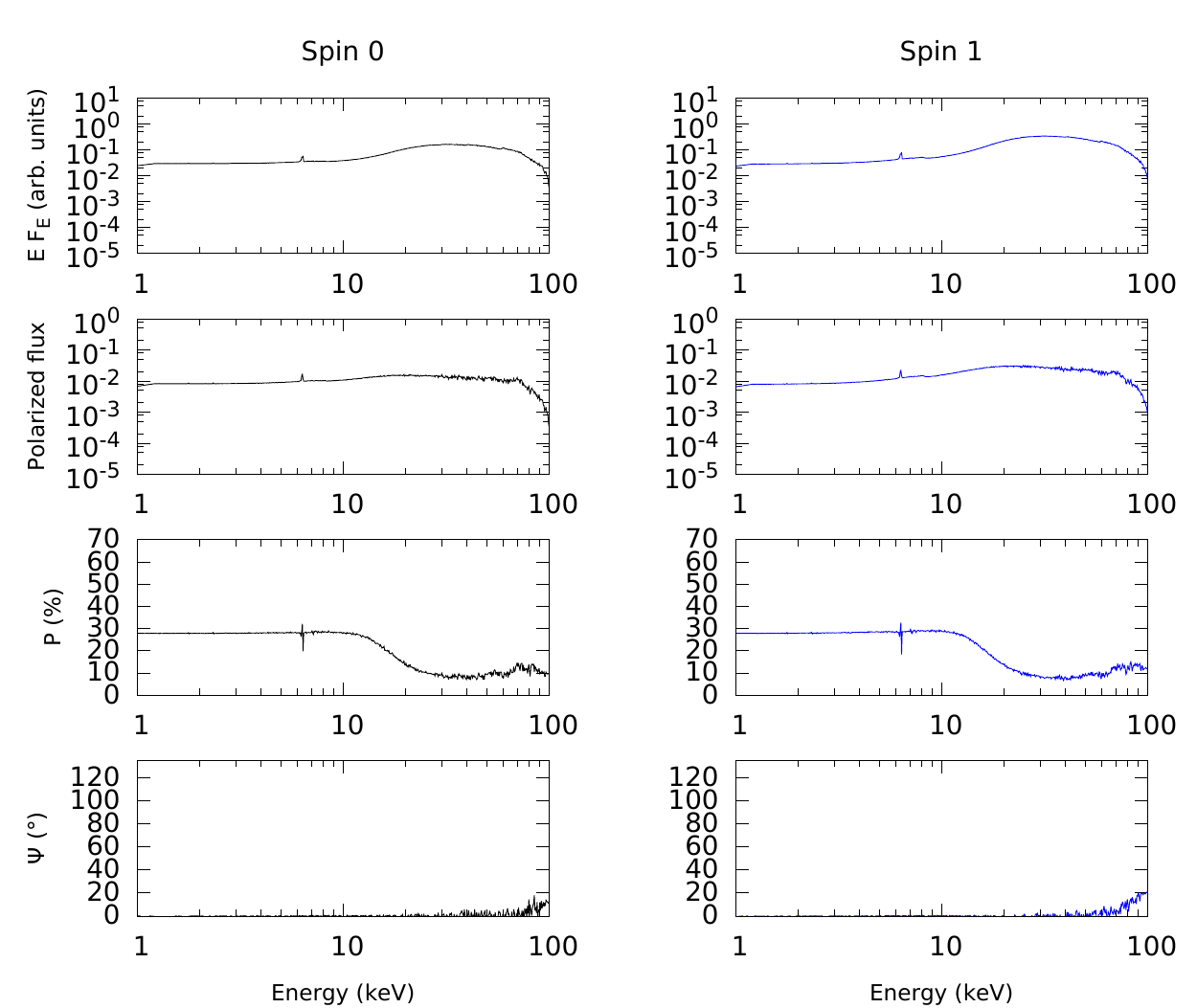}    
    \caption{X-ray flux (F$_{\rm E}$ is energy flux at energy E), polarized flux, polarization 
	    degree and polarization position angle for a type-2 AGN with fully-ionized 
	    polar winds. Top-left: n$_{\rm H_{torus}}$ = 10$^{23}$~at.cm$^{-2}$; top-right:
	    10$^{24}$~at.cm$^{-2}$; bottom: 10$^{25}$~at.cm$^{-2}$. See text 
	    for additional details about the model components. The input 
	    spectrum is unpolarized. Strong gravity effects are included.}
    \label{Fig:Unpolarized_primary_GR_Ionized_wind}
\end{figure*}

\begin{figure*}
    \includegraphics[trim = 0mm 0mm 0mm 0mm, clip, width=8.5cm]{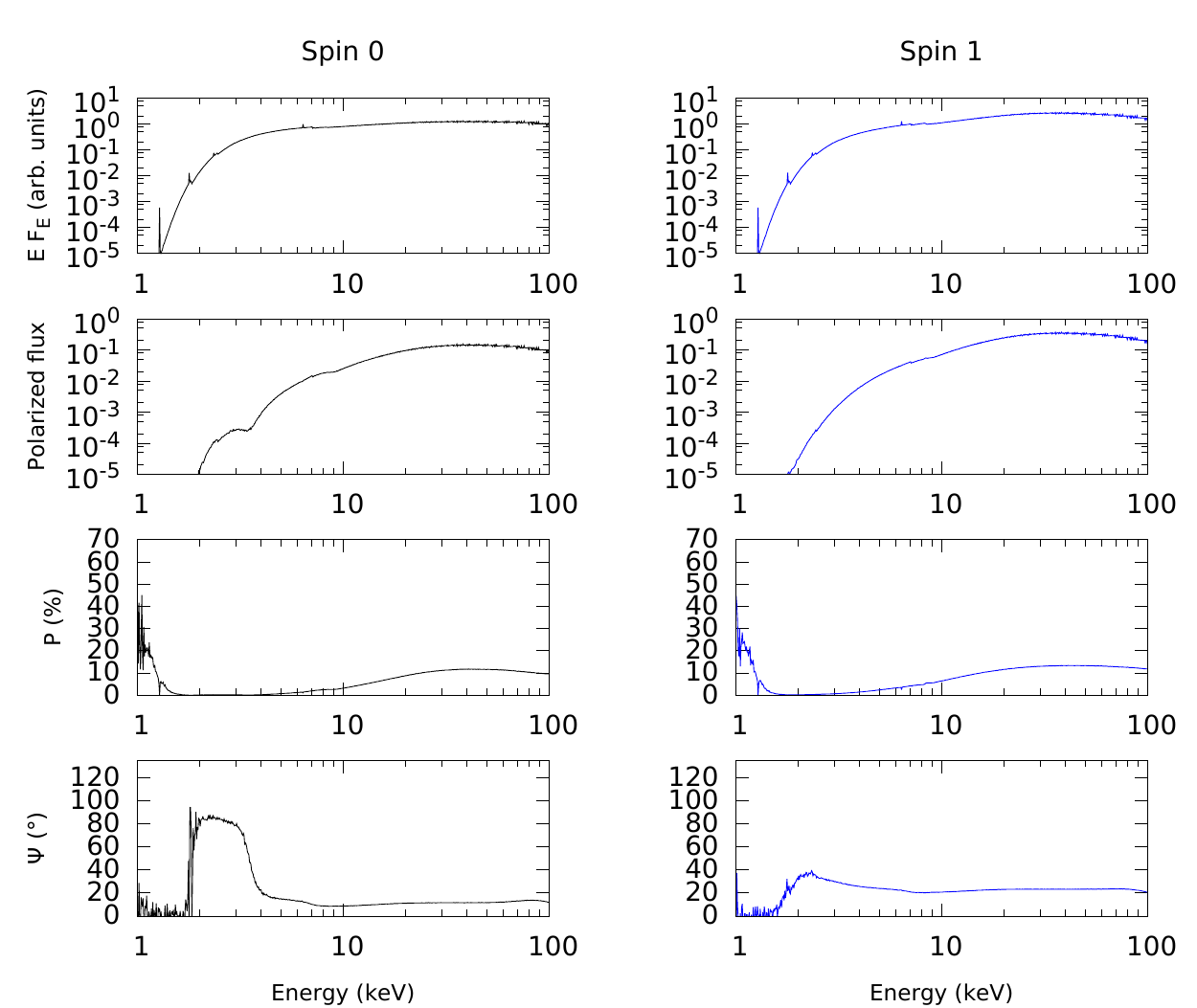}
    \hspace{5pt}\vrule\hspace{5pt}%
    \includegraphics[trim = 0mm 0mm 0mm 0mm, clip, width=8.5cm]{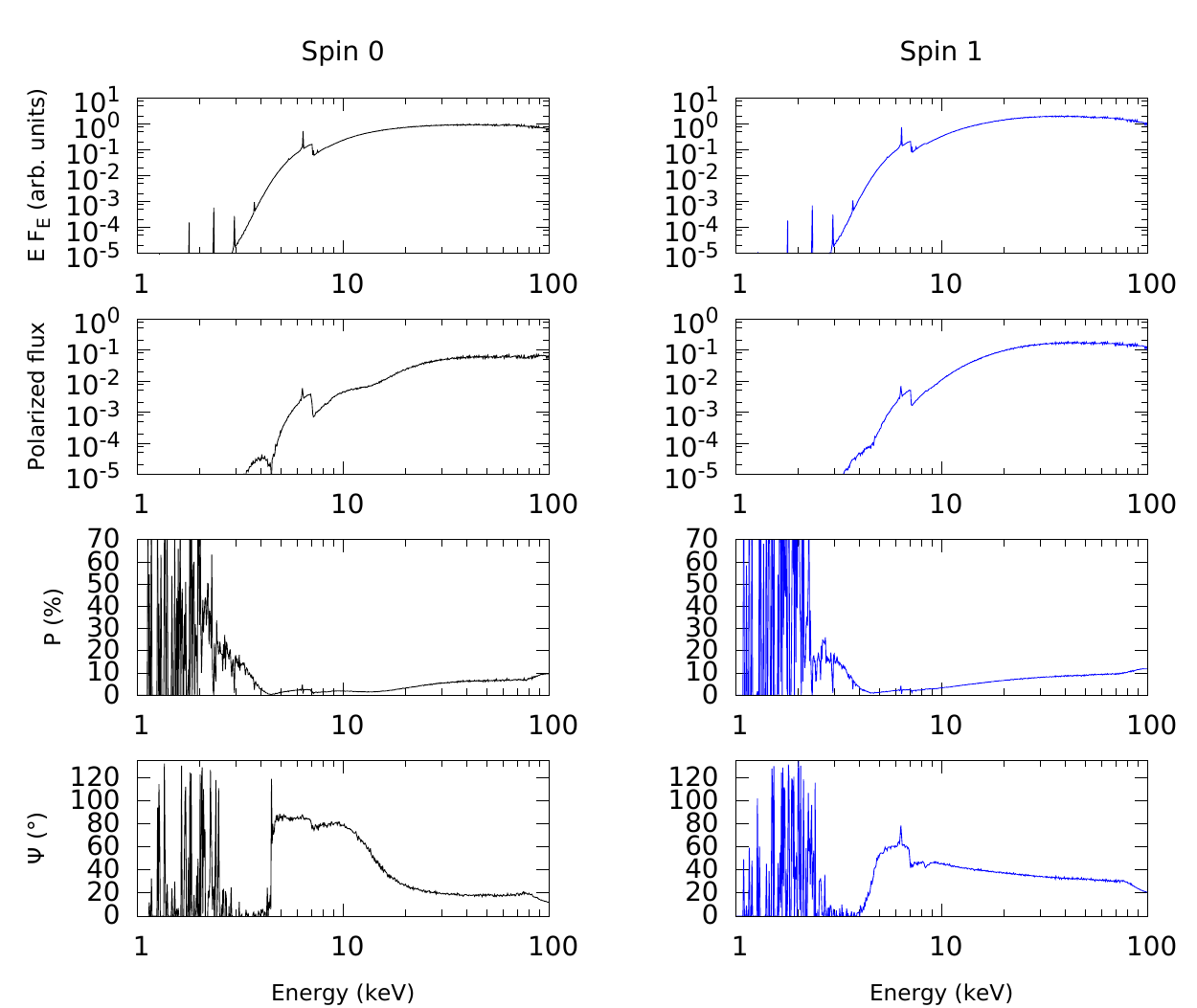}
    \includegraphics[trim = 0mm 0mm 0mm 0mm, clip, width=8.5cm]{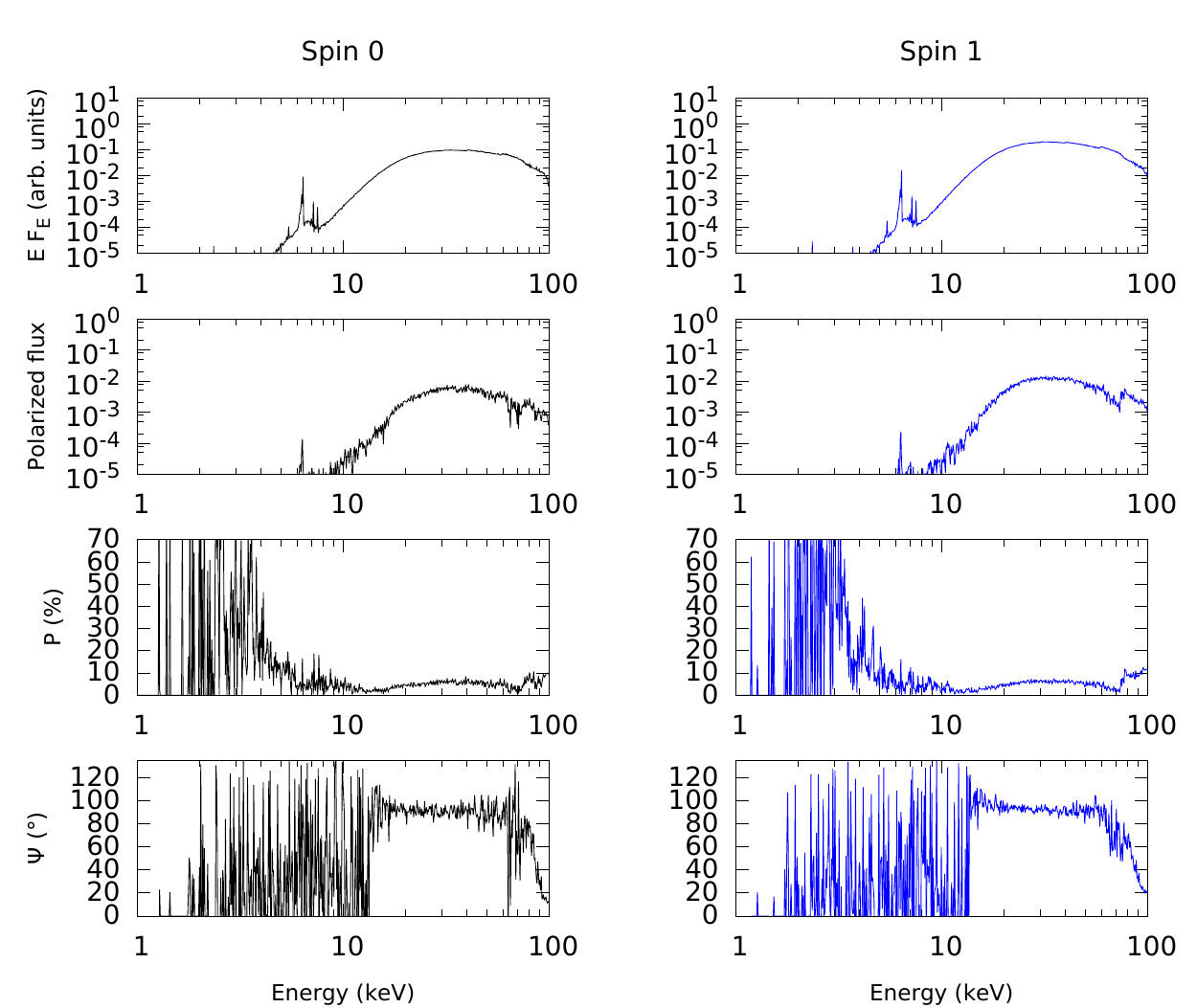}    
    \caption{X-ray flux (F$_{\rm E}$ is energy flux at energy E), polarized flux, polarization 
	    degree and polarization position angle for a type-2 AGN without polar winds.
	    Top-left: n$_{\rm H_{torus}}$ = 10$^{23}$~at.cm$^{-2}$; top-right:
	    10$^{24}$~at.cm$^{-2}$; bottom: 10$^{25}$~at.cm$^{-2}$. See text 
	    for additional details about the model components. The input 
	    spectrum is unpolarized. Strong gravity effects are included.}
    \label{Fig:Unpolarized_primary_GR_No_wind}
\end{figure*}

\clearpage

\setcounter{figure}{0}
\renewcommand{\thefigure}{C\arabic{figure}}

\begin{figure*}
    \includegraphics[trim = 0mm 0mm 0mm 0mm, clip, width=6.5cm]{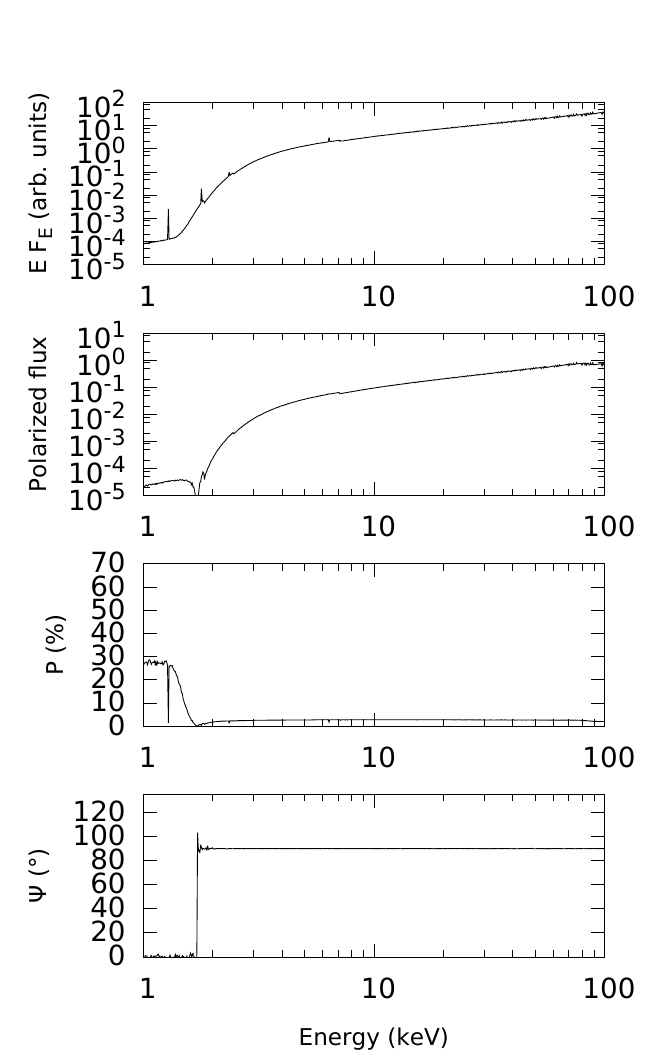}
    \hspace{5pt}\vrule\hspace{5pt}%
    \includegraphics[trim = 0mm 0mm 0mm 0mm, clip, width=6.5cm]{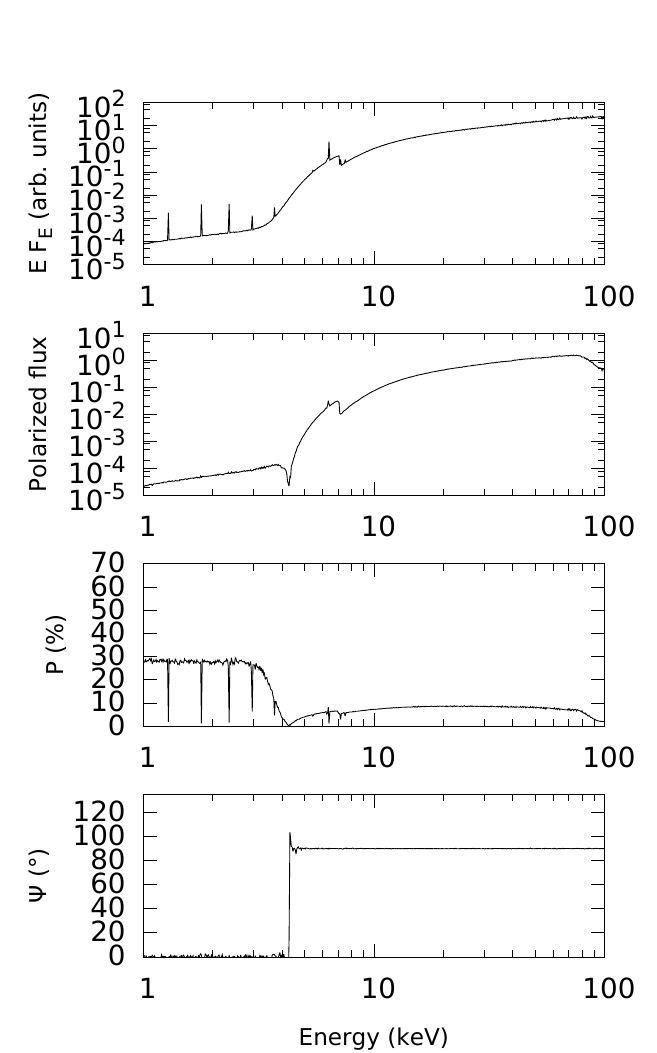}
    \includegraphics[trim = 0mm 0mm 0mm 0mm, clip, width=6.5cm]{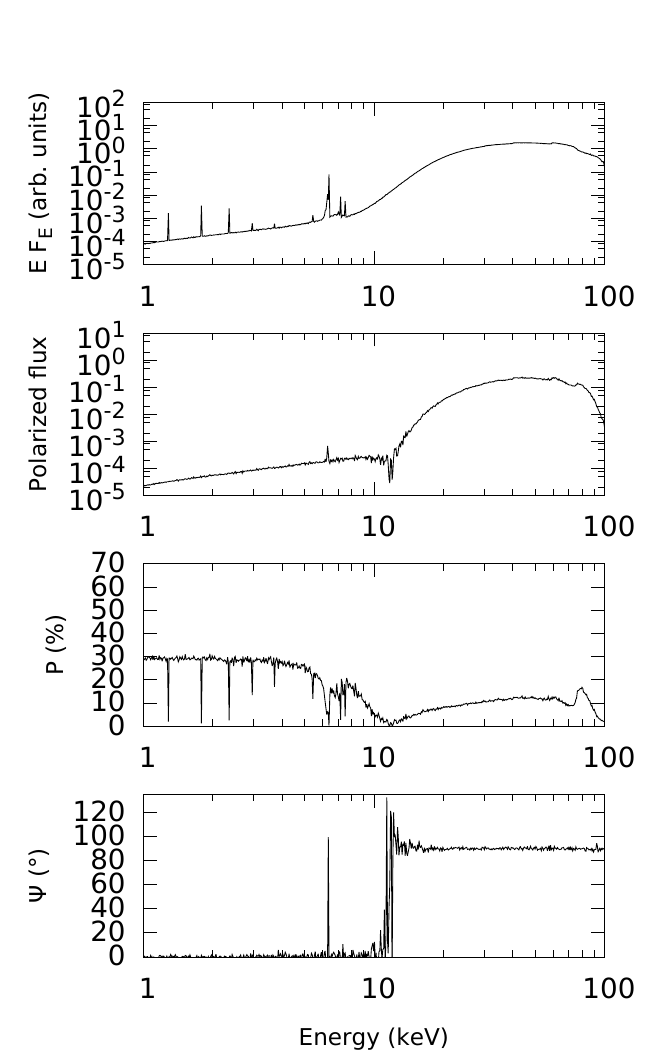}    
    \caption{X-ray flux (F$_{\rm E}$ is energy flux at energy E), polarized flux, polarization 
	    degree and polarization position angle for a type-2 AGN with Compton-thin 
	    absorbing polar winds (n$_{\rm H_{wind}}$ = 10$^{21}$~at.cm$^{-2}$).
	    Top-left: n$_{\rm H_{torus}}$ = 10$^{23}$~at.cm$^{-2}$; top-right:
	    10$^{24}$~at.cm$^{-2}$; bottom: 10$^{25}$~at.cm$^{-2}$. See text 
	    for additional details about the model components. The input 
	    spectrum is polarized (2\% parallel polarization).
	    Strong gravity effects are not included.}
    \label{Fig:Polarized_primary_PL_para_Absorbing_nh21_wind}
\end{figure*}

\begin{figure*}
    \includegraphics[trim = 0mm 0mm 0mm 0mm, clip, width=6.5cm]{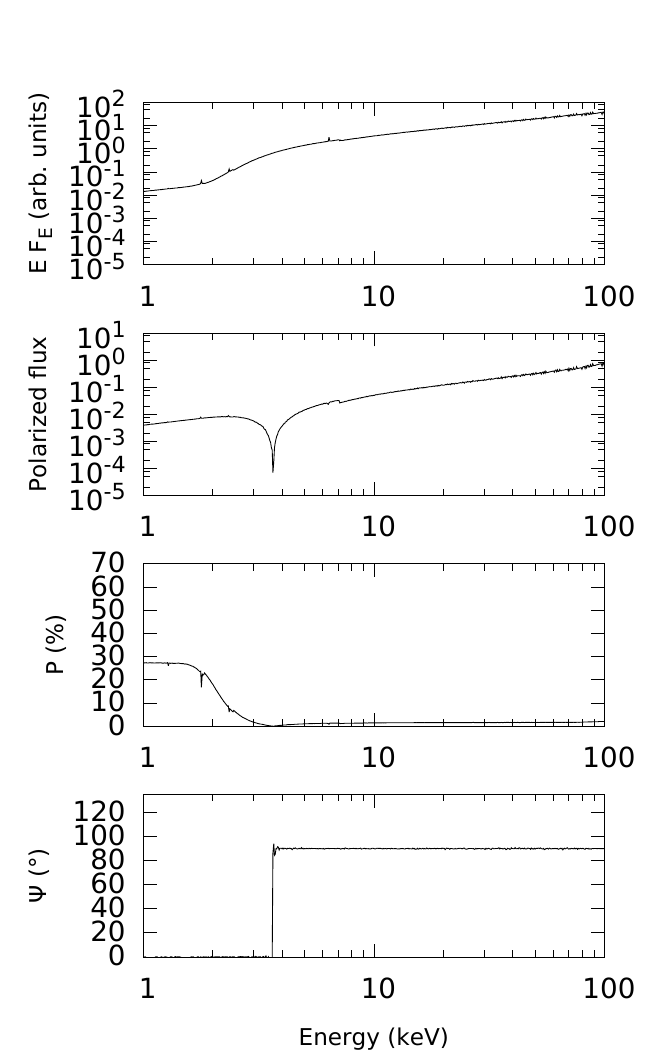}
    \hspace{5pt}\vrule\hspace{5pt}%
    \includegraphics[trim = 0mm 0mm 0mm 0mm, clip, width=6.5cm]{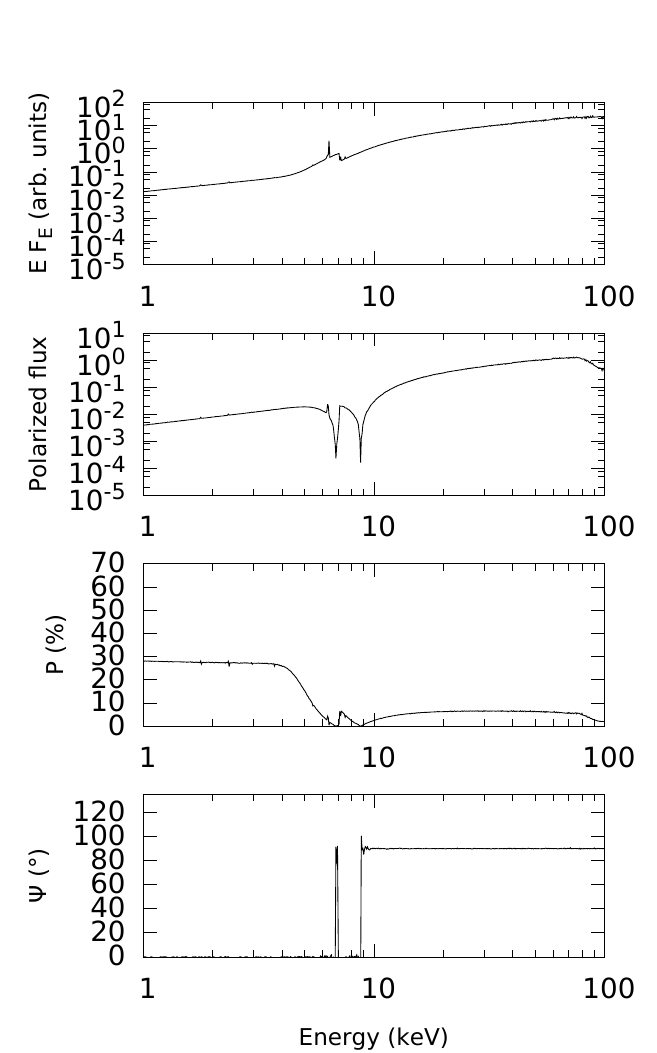}
    \includegraphics[trim = 0mm 0mm 0mm 0mm, clip, width=6.5cm]{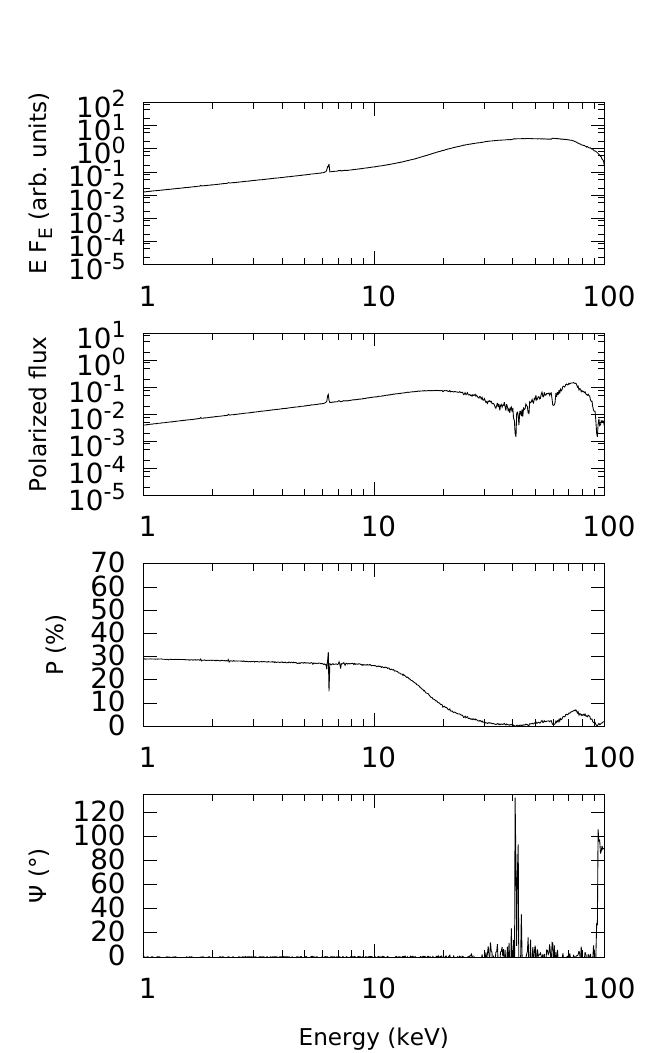}    
    \caption{X-ray flux (F$_{\rm E}$ is energy flux at energy E), polarized flux, polarization 
	    degree and polarization position angle for a type-2 AGN with fully-ionized 
	    polar winds. Top-left: n$_{\rm H_{torus}}$ = 10$^{23}$~at.cm$^{-2}$; top-right:
	    10$^{24}$~at.cm$^{-2}$; bottom: 10$^{25}$~at.cm$^{-2}$. See text 
	    for additional details about the model components. The input 
	    spectrum is polarized (2\% parallel polarization).
	    Strong gravity effects are not included.}
    \label{Fig:Polarized_primary_PL_para_Ionized_wind}
\end{figure*}

\begin{figure*}
    \includegraphics[trim = 0mm 0mm 0mm 0mm, clip, width=6.5cm]{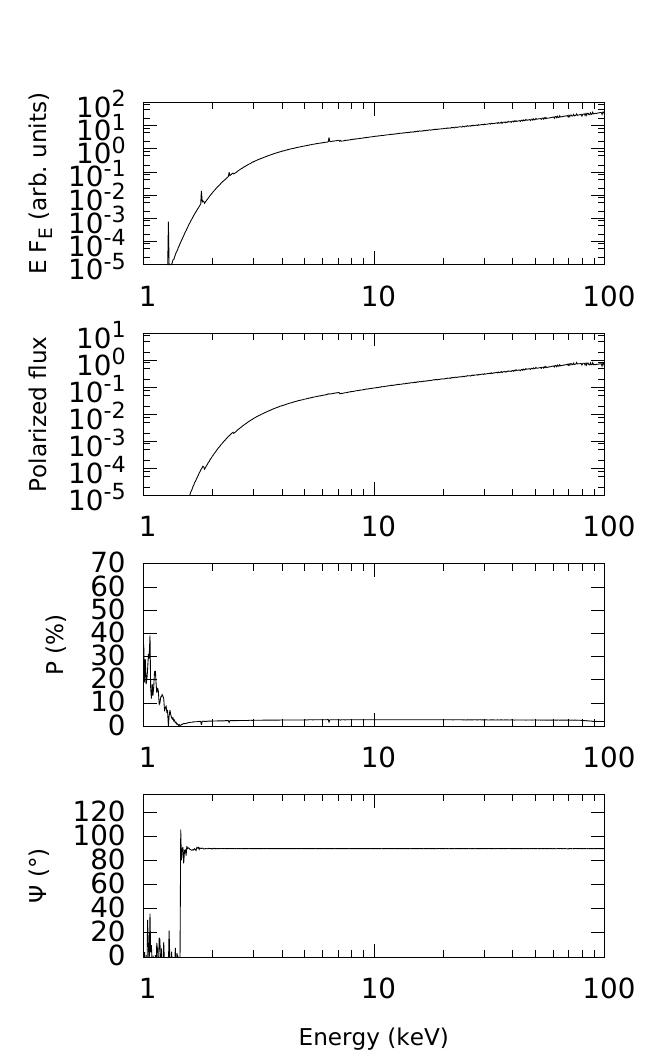}
    \hspace{5pt}\vrule\hspace{5pt}%
    \includegraphics[trim = 0mm 0mm 0mm 0mm, clip, width=6.5cm]{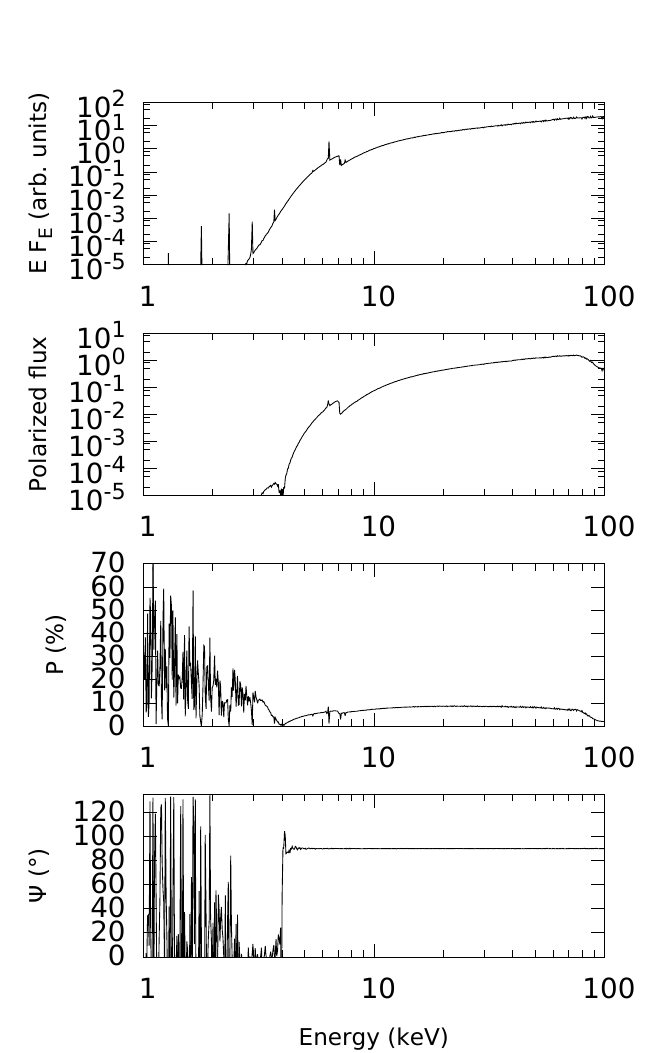}
    \includegraphics[trim = 0mm 0mm 0mm 0mm, clip, width=6.5cm]{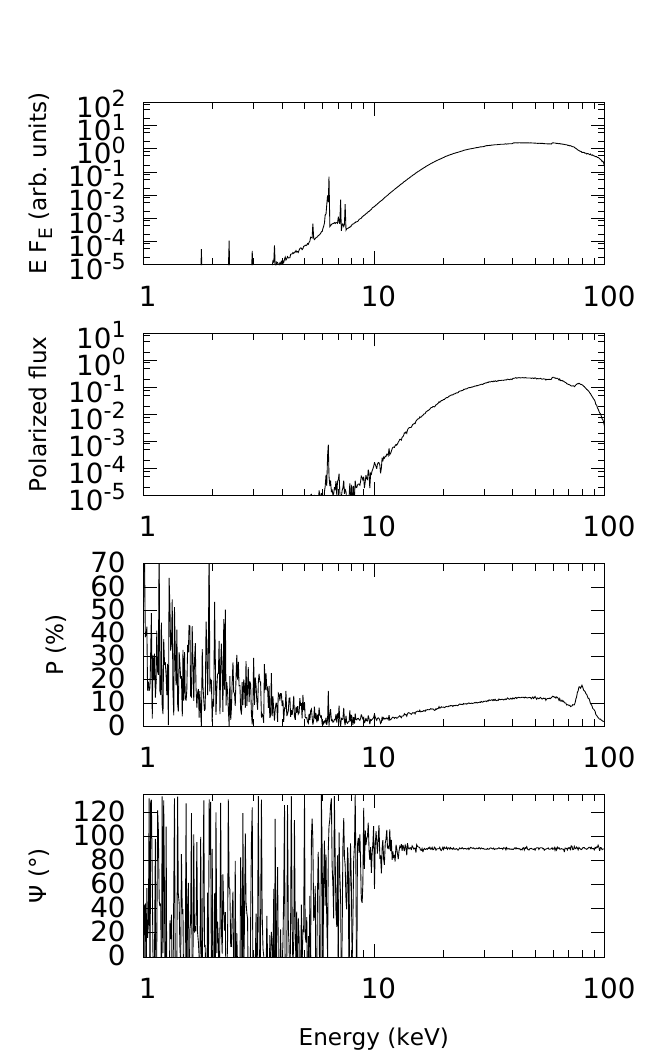}    
    \caption{X-ray flux (F$_{\rm E}$ is energy flux at energy E), polarized flux, polarization 
	    degree and polarization position angle for a type-2 AGN without polar 
	    winds. Top-left: n$_{\rm H_{torus}}$ = 10$^{23}$~at.cm$^{-2}$; top-right:
	    10$^{24}$~at.cm$^{-2}$; bottom: 10$^{25}$~at.cm$^{-2}$. See text 
	    for additional details about the model components. The input 
	    spectrum is polarized (2\% parallel polarization).
	    Strong gravity effects are not included.}
    \label{Fig:Polarized_primary_PL_para_No_wind}
\end{figure*}

\clearpage

\setcounter{figure}{0}
\renewcommand{\thefigure}{D\arabic{figure}}

\begin{figure*}
    \includegraphics[trim = 0mm 0mm 0mm 0mm, clip, width=6.5cm]{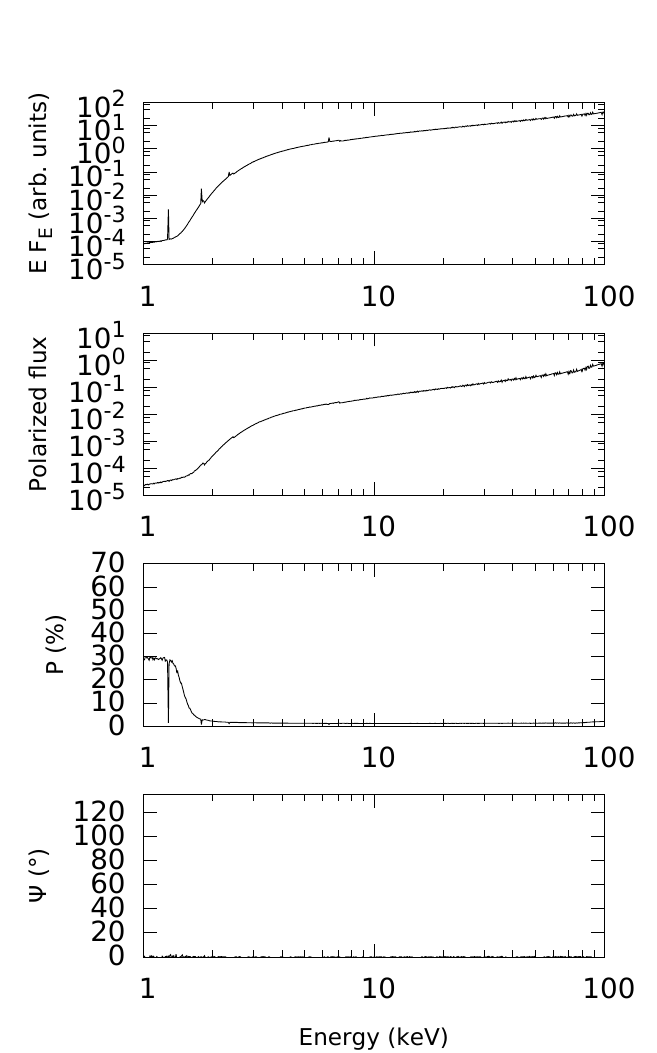}
    \hspace{5pt}\vrule\hspace{5pt}%
    \includegraphics[trim = 0mm 0mm 0mm 0mm, clip, width=6.5cm]{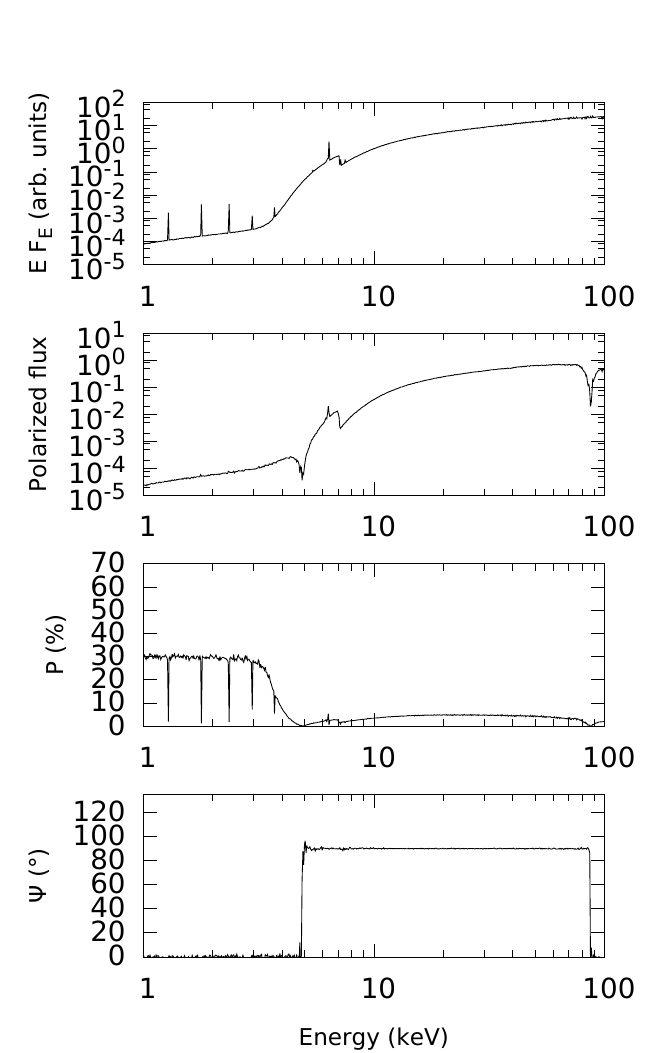}
    \includegraphics[trim = 0mm 0mm 0mm 0mm, clip, width=6.5cm]{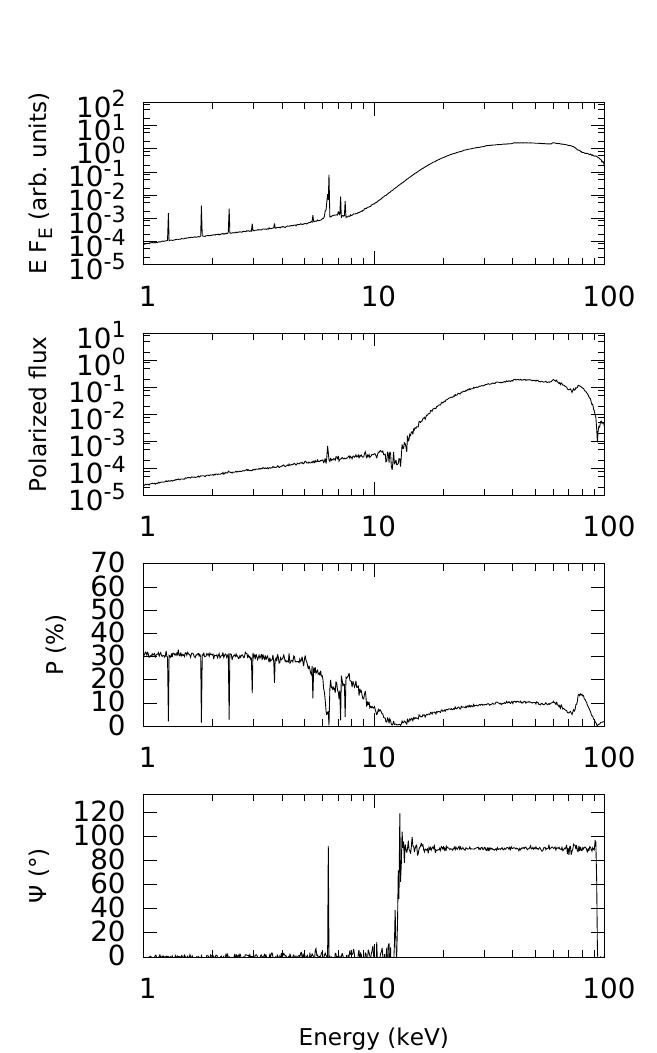}    
    \caption{X-ray flux (F$_{\rm E}$ is energy flux at energy E), polarized flux, polarization 
	    degree and polarization position angle for a type-2 AGN with Compton-thin 
	    absorbing polar winds (n$_{\rm H_{wind}}$ = 10$^{21}$~at.cm$^{-2}$).
	    Top-left: n$_{\rm H_{torus}}$ = 10$^{23}$~at.cm$^{-2}$; top-right:
	    10$^{24}$~at.cm$^{-2}$; bottom: 10$^{25}$~at.cm$^{-2}$. See text 
	    for additional details about the model components. The input 
	    spectrum is polarized (2\% perpendicular polarization).
	    Strong gravity effects are not included.}
    \label{Fig:Polarized_primary_PL_perp_Absorbing_nh21_wind}
\end{figure*}

\begin{figure*}
    \includegraphics[trim = 0mm 0mm 0mm 0mm, clip, width=6.5cm]{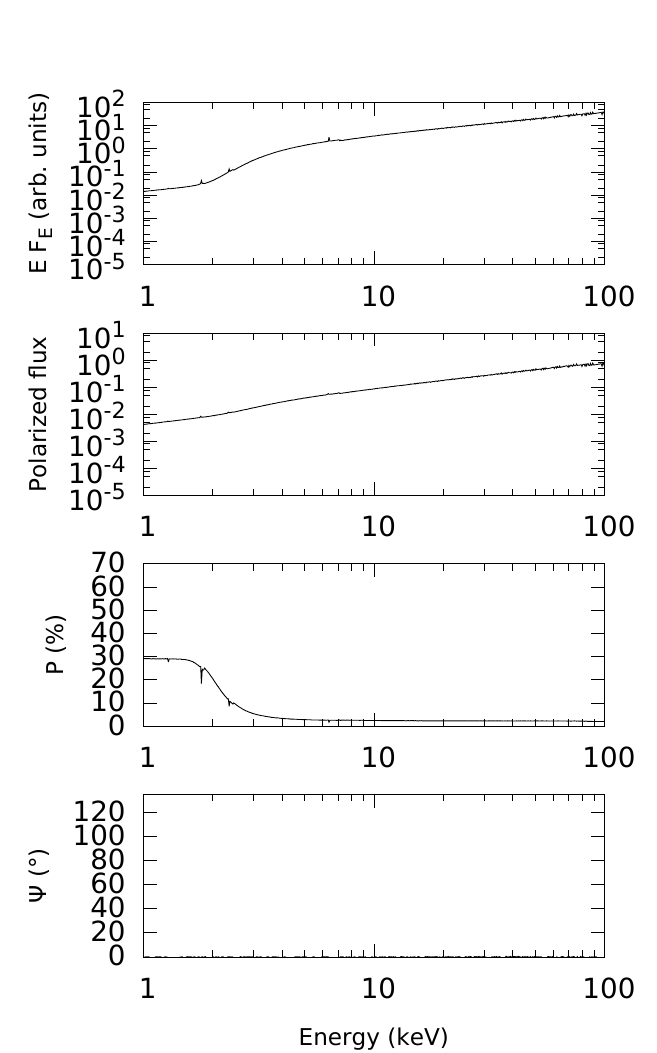}
    \hspace{5pt}\vrule\hspace{5pt}%
    \includegraphics[trim = 0mm 0mm 0mm 0mm, clip, width=6.5cm]{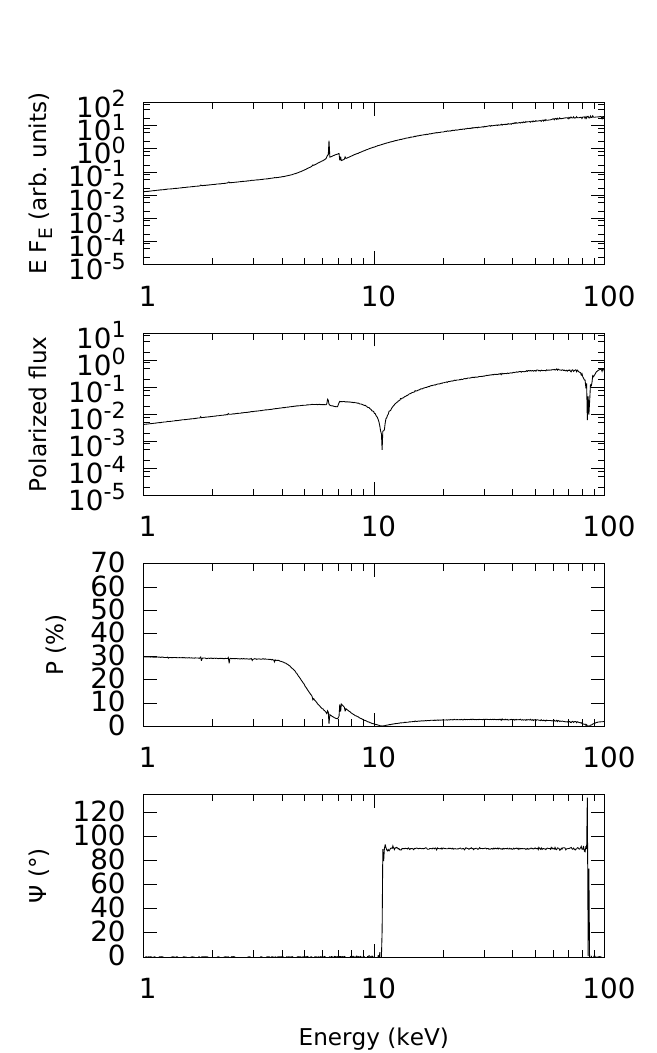}
    \includegraphics[trim = 0mm 0mm 0mm 0mm, clip, width=6.5cm]{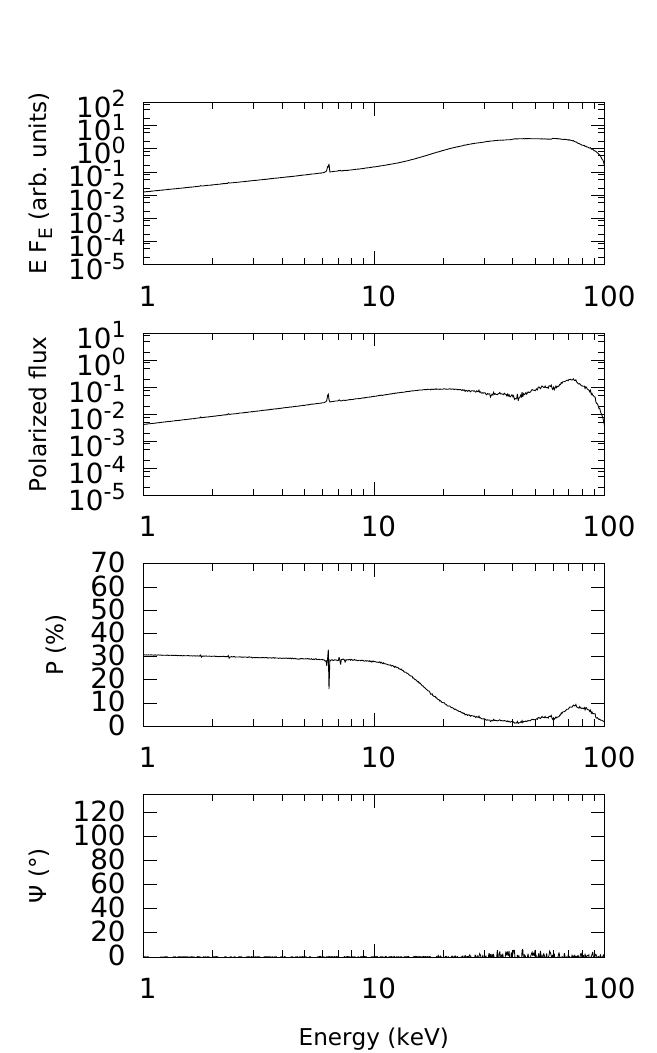}    
    \caption{X-ray flux (F$_{\rm E}$ is energy flux at energy E), polarized flux, polarization 
	    degree and polarization position angle for a type-2 AGN with fully-ionized 
	    polar winds. Top-left: n$_{\rm H_{torus}}$ = 10$^{23}$~at.cm$^{-2}$; top-right:
	    10$^{24}$~at.cm$^{-2}$; bottom: 10$^{25}$~at.cm$^{-2}$. See text 
	    for additional details about the model components. The input 
	    spectrum is polarized (2\% perpendicular polarization).
	    Strong gravity effects are not included.}
    \label{Fig:Polarized_primary_PL_perp_Ionized_wind}
\end{figure*}

\begin{figure*}
    \includegraphics[trim = 0mm 0mm 0mm 0mm, clip, width=6.5cm]{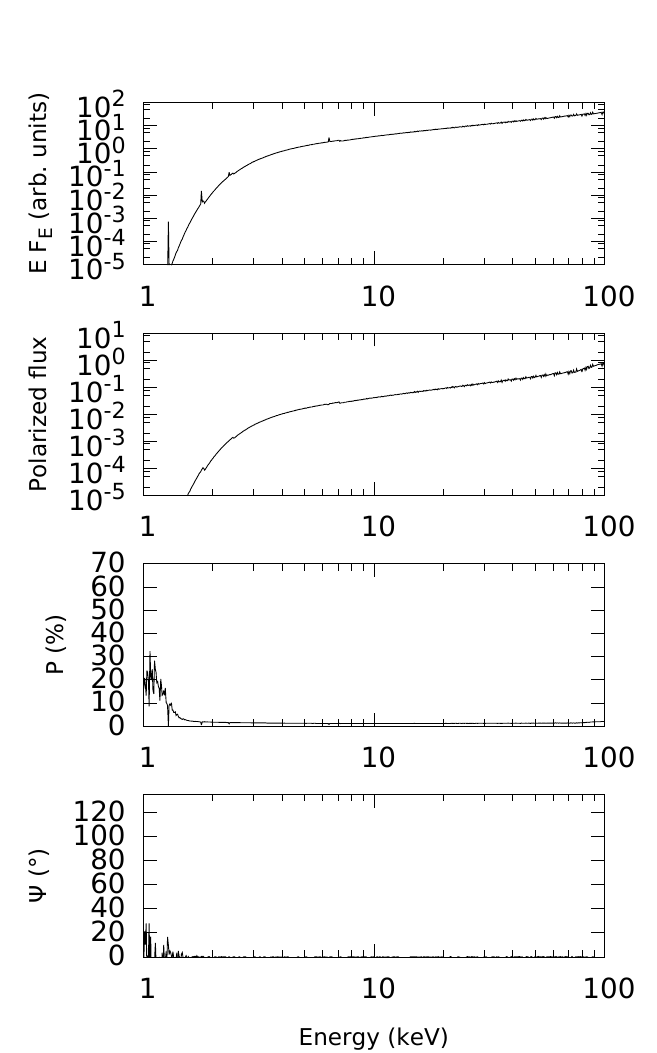}
    \hspace{5pt}\vrule\hspace{5pt}%
    \includegraphics[trim = 0mm 0mm 0mm 0mm, clip, width=6.5cm]{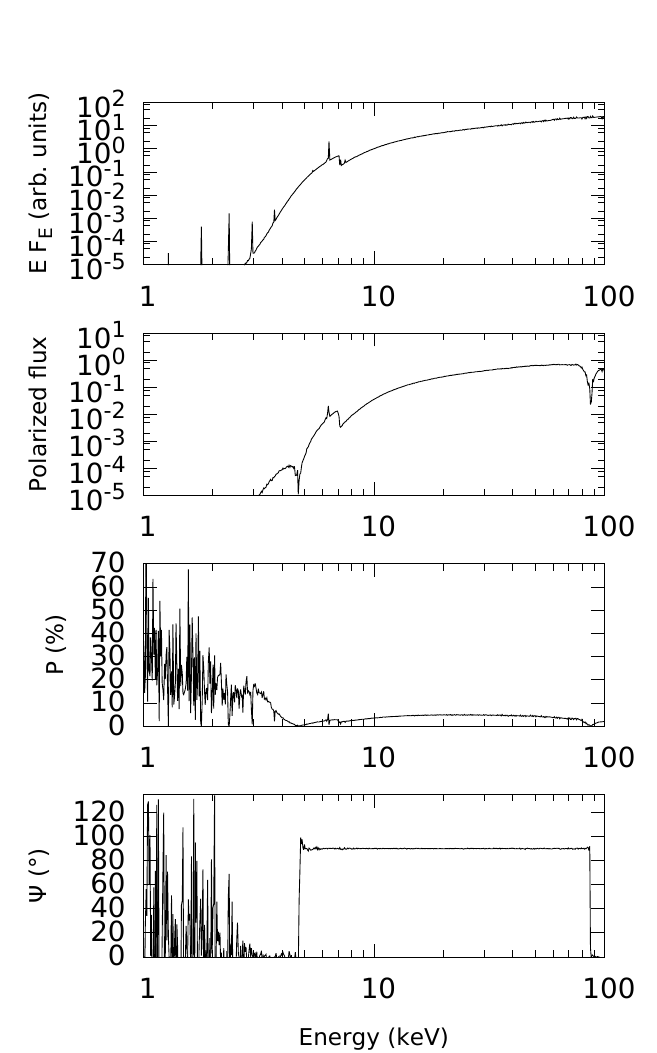}
    \includegraphics[trim = 0mm 0mm 0mm 0mm, clip, width=6.5cm]{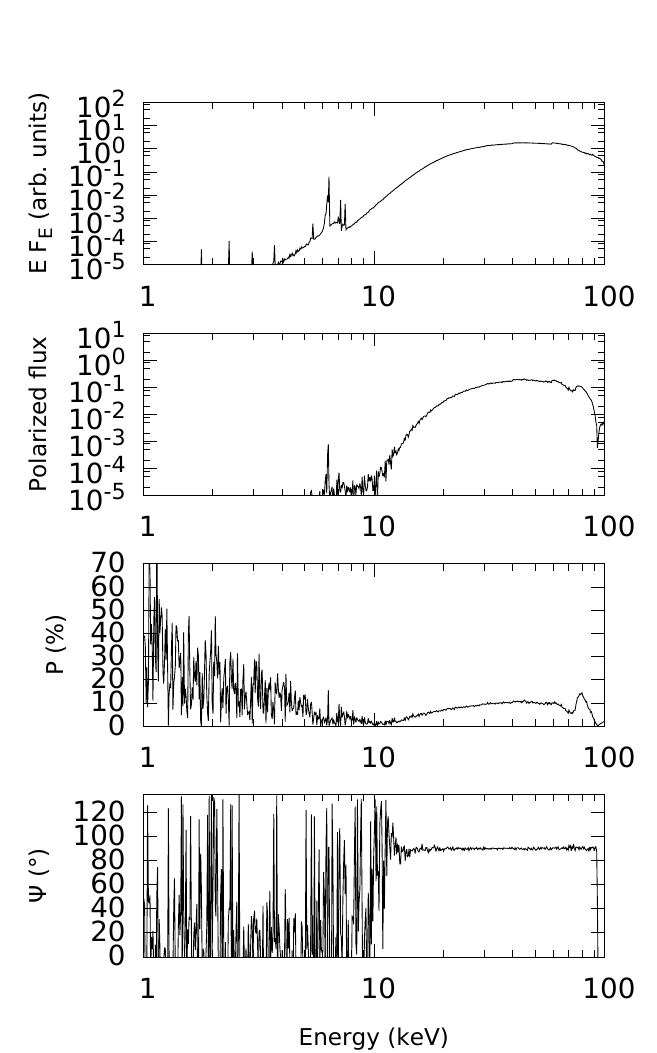}    
    \caption{X-ray flux (F$_{\rm E}$ is energy flux at energy E), polarized flux, polarization 
	    degree and polarization position angle for a type-2 AGN without polar winds.
	    Top-left: n$_{\rm H_{torus}}$ = 10$^{23}$~at.cm$^{-2}$; top-right:
	    10$^{24}$~at.cm$^{-2}$; bottom: 10$^{25}$~at.cm$^{-2}$. See text 
	    for additional details about the model components. The input 
	    spectrum is polarized (2\% perpendicular polarization).
	    Strong gravity effects are not included.}
    \label{Fig:Polarized_primary_PL_perp_No_wind}
\end{figure*}

\clearpage

\setcounter{figure}{0}
\renewcommand{\thefigure}{E\arabic{figure}}

\begin{figure*}
    \includegraphics[trim = 0mm 0mm 0mm 0mm, clip, width=8.5cm]{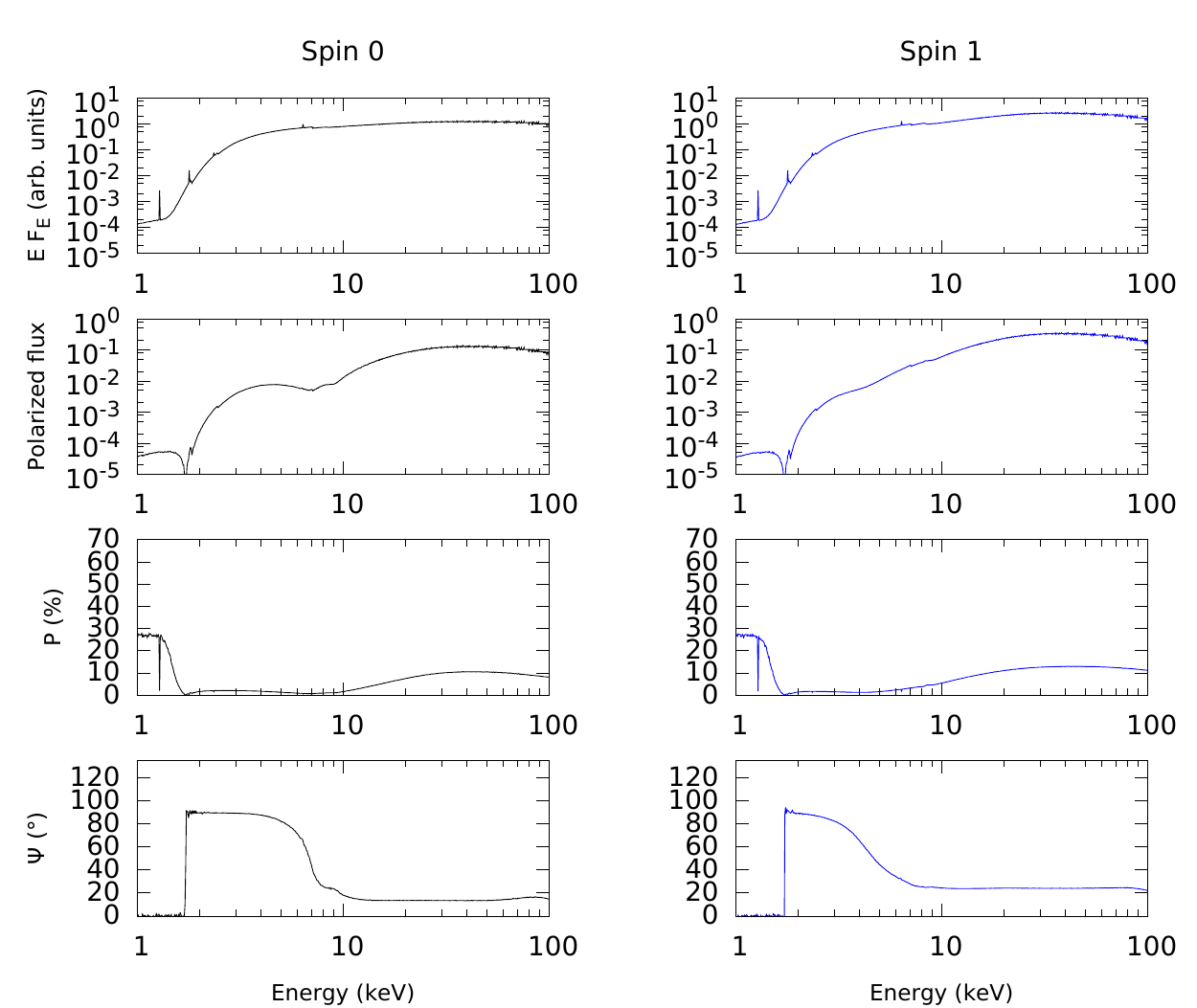}
    \hspace{5pt}\vrule\hspace{5pt}%
    \includegraphics[trim = 0mm 0mm 0mm 0mm, clip, width=8.5cm]{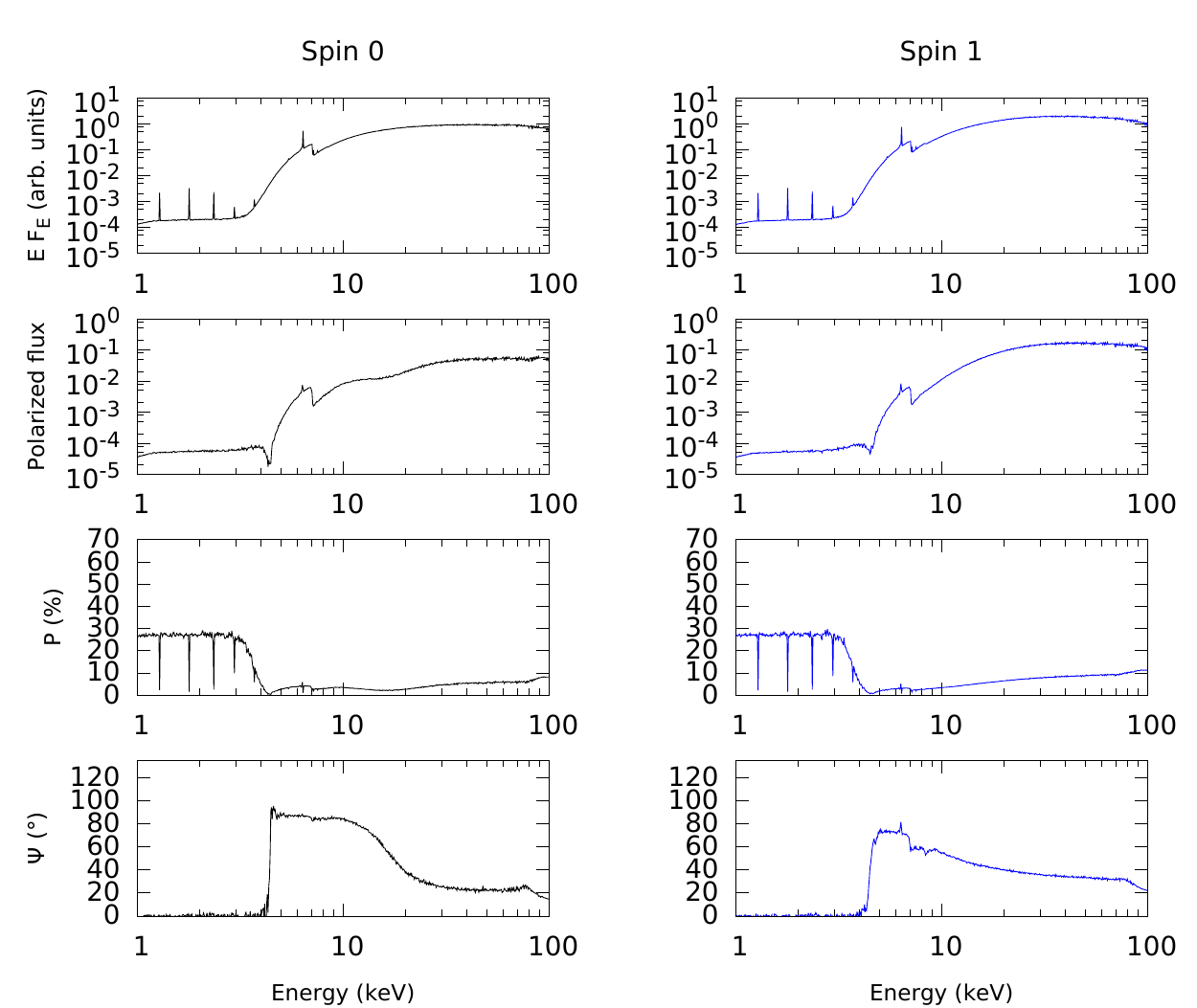}
    \includegraphics[trim = 0mm 0mm 0mm 0mm, clip, width=8.5cm]{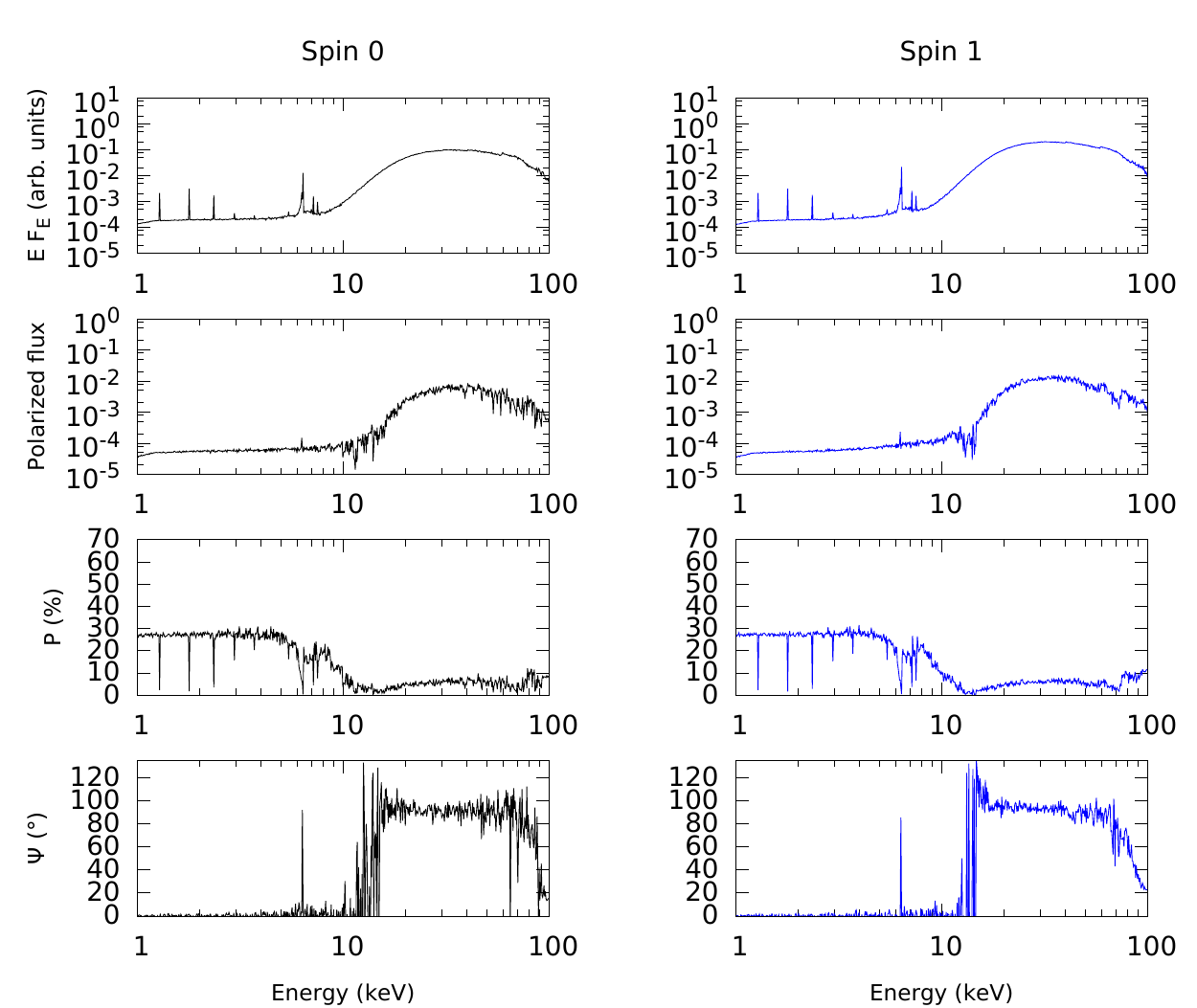}    
    \caption{X-ray flux (F$_{\rm E}$ is energy flux at energy E), polarized flux, polarization 
	    degree and polarization position angle for a type-2 AGN with Compton-thin 
	    absorbing polar winds (n$_{\rm H_{wind}}$ = 10$^{21}$~at.cm$^{-2}$).
	    Top-left: n$_{\rm H_{torus}}$ = 10$^{23}$~at.cm$^{-2}$; top-right:
	    10$^{24}$~at.cm$^{-2}$; bottom: 10$^{25}$~at.cm$^{-2}$. See text 
	    for additional details about the model components. The input 
	    spectrum is polarized (2\% parallel polarization) and GR 
	    effects are included (left column: non-spinning Schwarzschild black 
	    hole; right column: maximally spinning Kerr black hole).}
    \label{Fig:Polarized_primary_GR_perp_Absorbing_nh21_wind}
\end{figure*}

\begin{figure*}
    \includegraphics[trim = 0mm 0mm 0mm 0mm, clip, width=8.5cm]{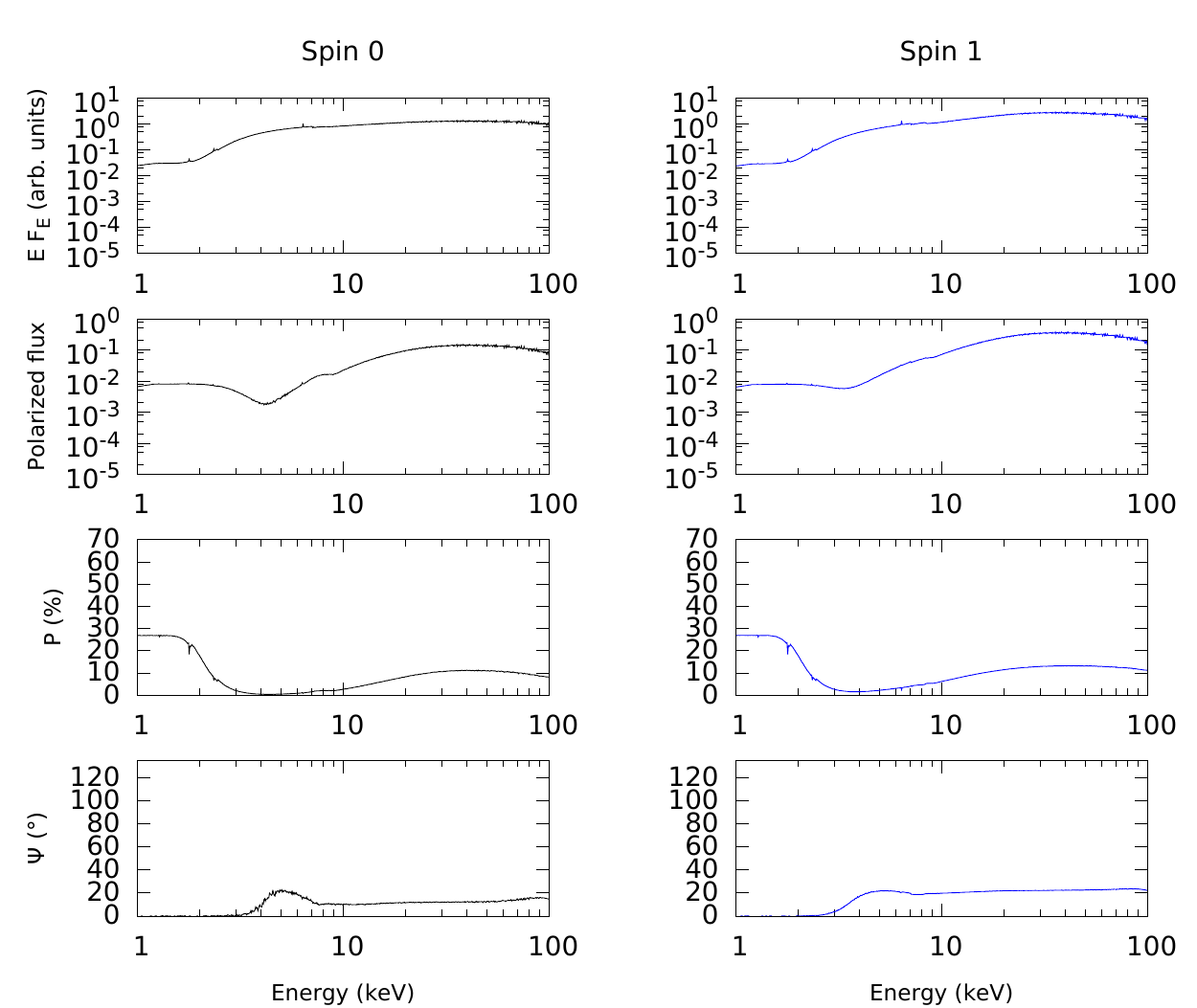}
    \hspace{5pt}\vrule\hspace{5pt}%
    \includegraphics[trim = 0mm 0mm 0mm 0mm, clip, width=8.5cm]{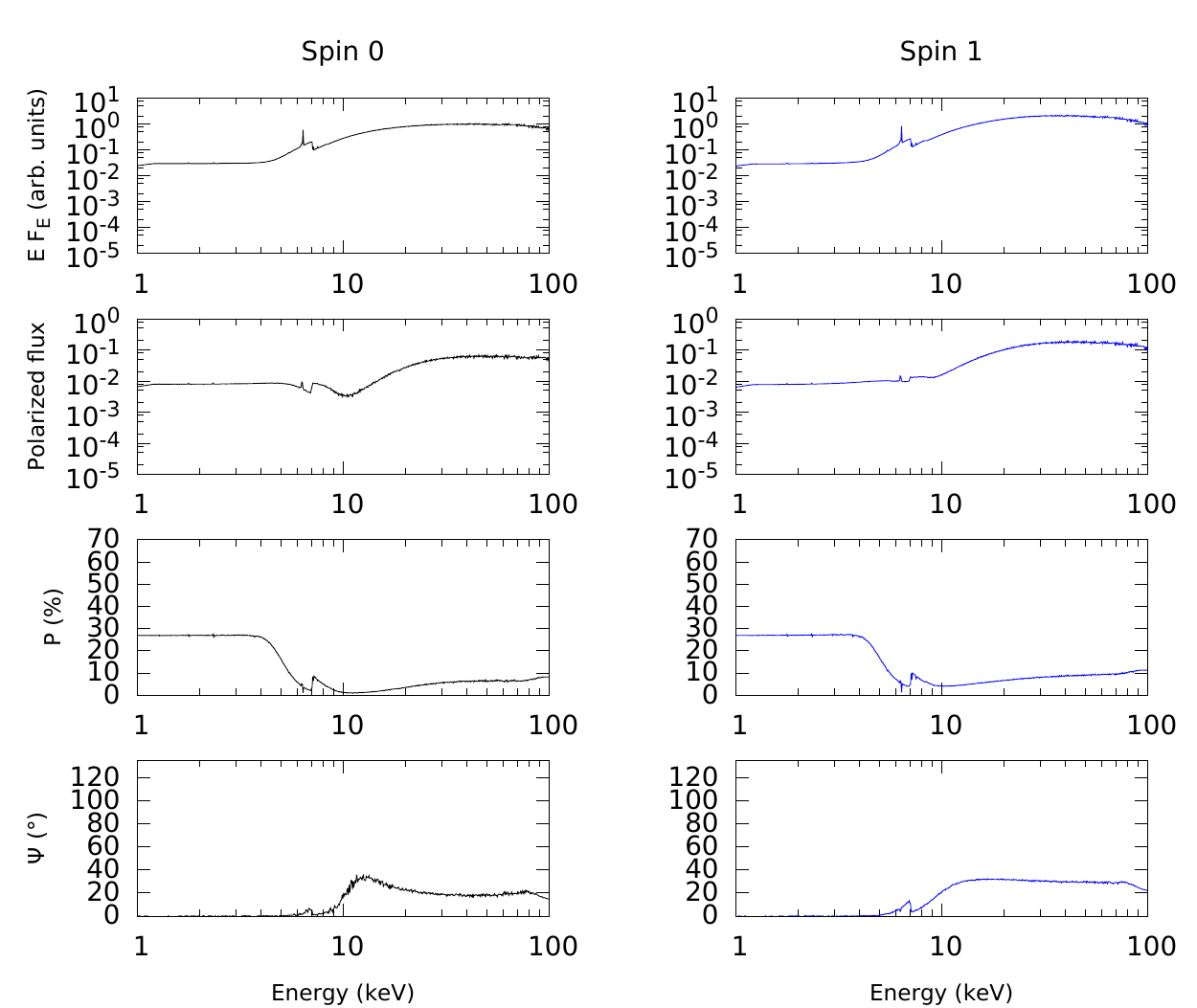}
    \includegraphics[trim = 0mm 0mm 0mm 0mm, clip, width=8.5cm]{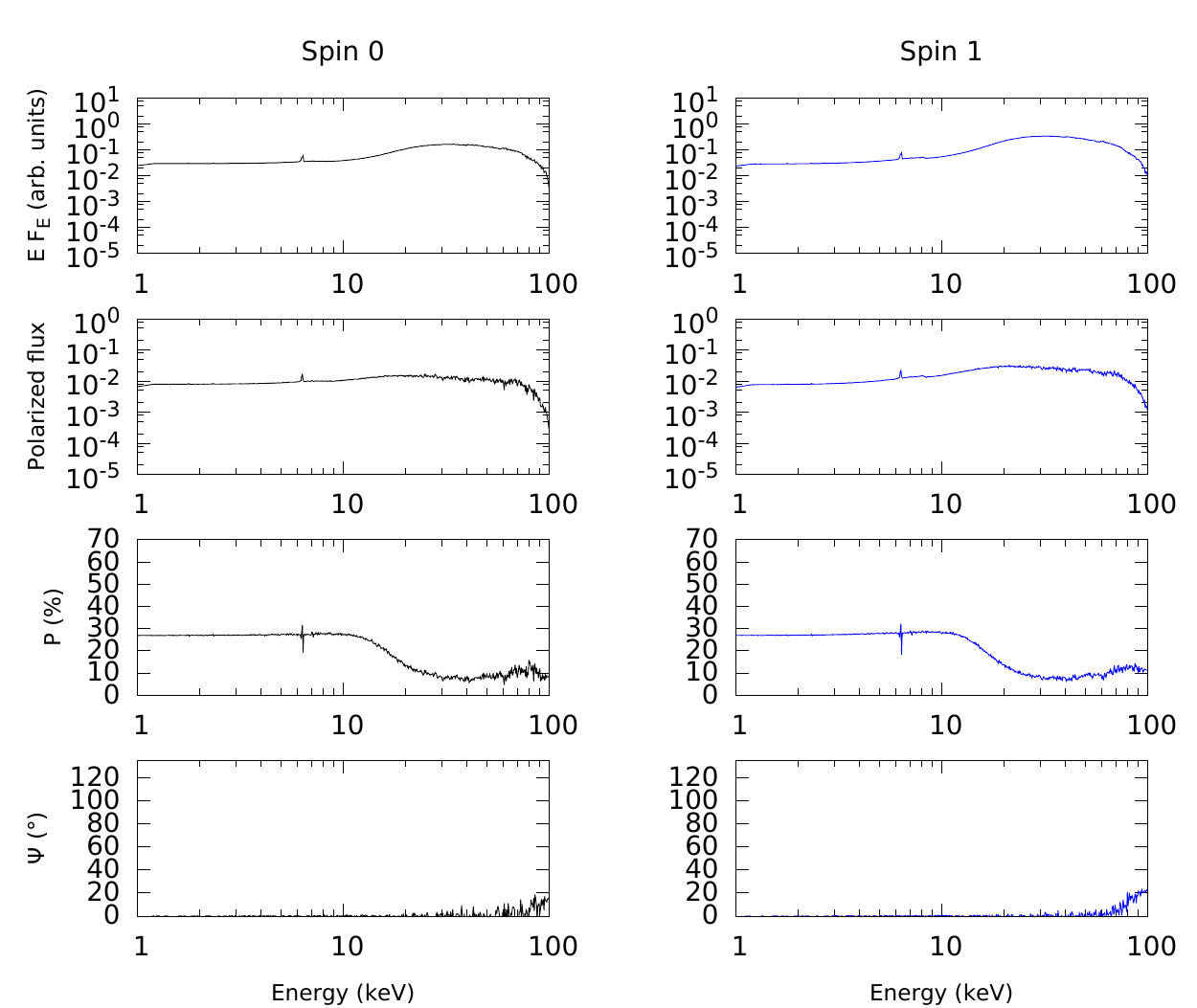}    
    \caption{X-ray flux (F$_{\rm E}$ is energy flux at energy E), polarized flux, polarization 
	    degree and polarization position angle for a type-2 AGN with fully-ionized 
	    polar winds. Top-left: n$_{\rm H_{torus}}$ = 10$^{23}$~at.cm$^{-2}$; top-right:
	    10$^{24}$~at.cm$^{-2}$; bottom: 10$^{25}$~at.cm$^{-2}$. See text 
	    for additional details about the model components. The input 
	    spectrum is polarized (2\% parallel polarization) and GR 
	    effects are included (left column: non-spinning Schwarzschild black 
	    hole; right column: maximally spinning Kerr black hole).}
    \label{Fig:Polarized_primary_GR_perp_Ionized_wind}
\end{figure*}

\begin{figure*}
    \includegraphics[trim = 0mm 0mm 0mm 0mm, clip, width=8.5cm]{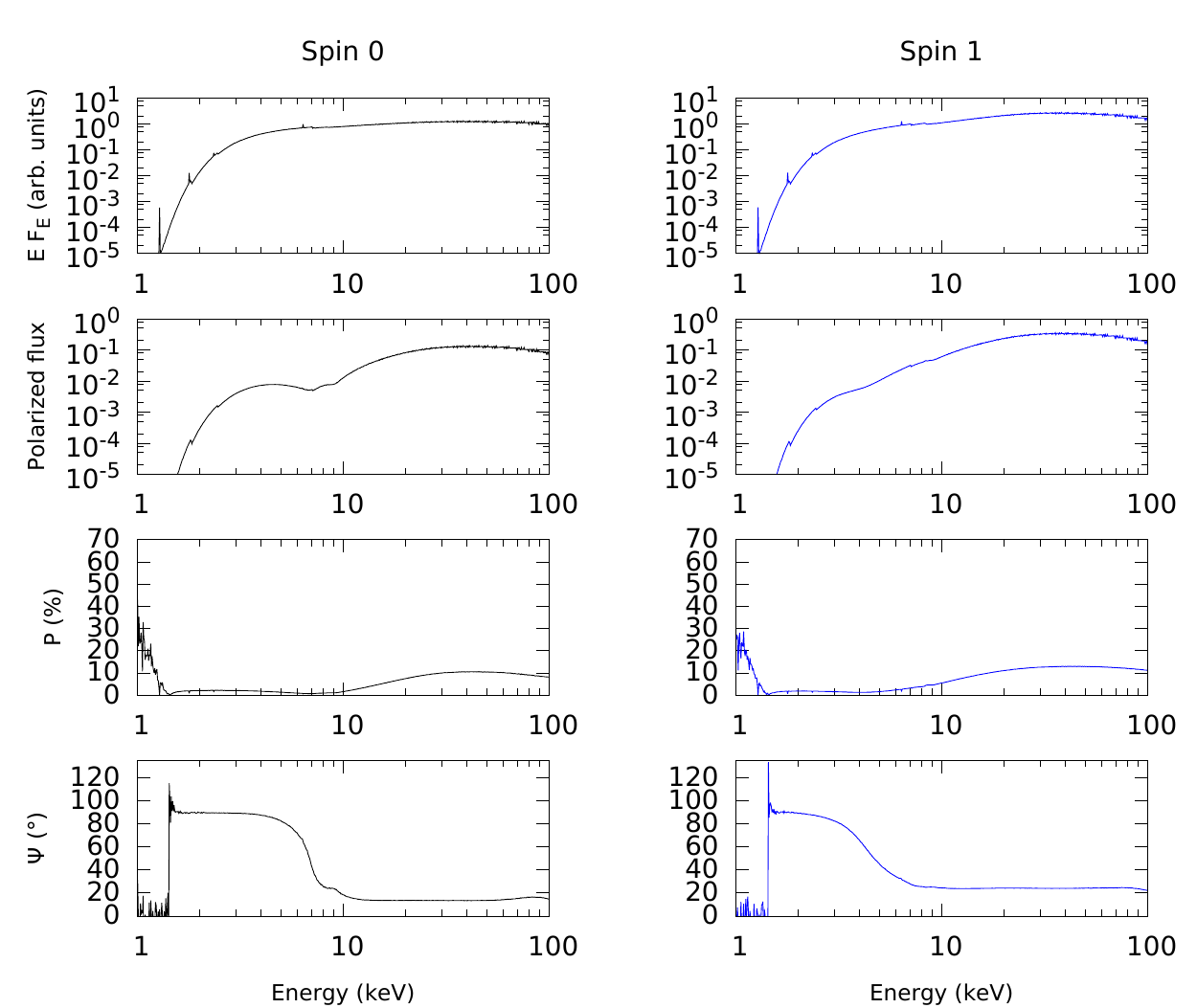}
    \hspace{5pt}\vrule\hspace{5pt}%
    \includegraphics[trim = 0mm 0mm 0mm 0mm, clip, width=8.5cm]{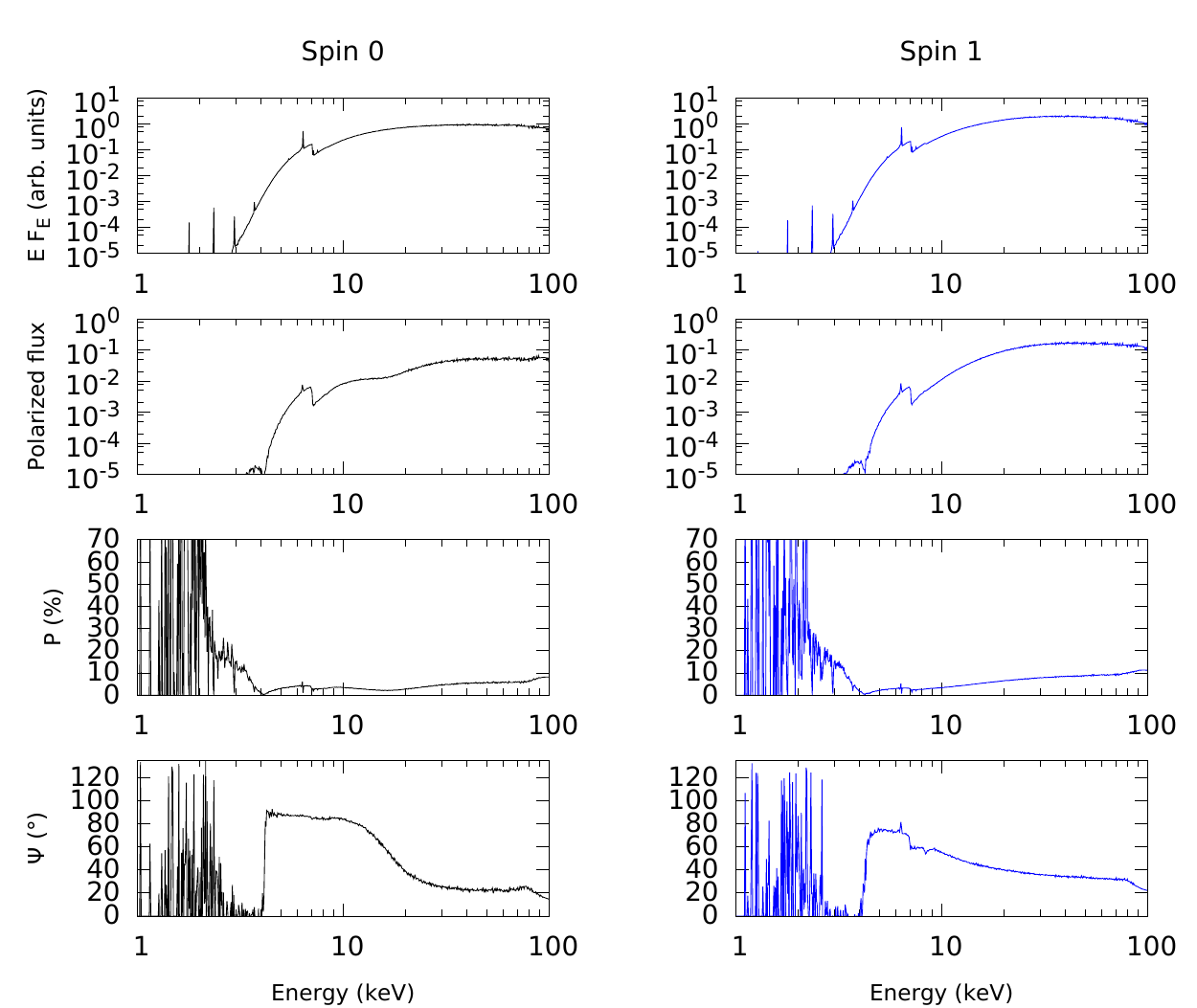}
    \includegraphics[trim = 0mm 0mm 0mm 0mm, clip, width=8.5cm]{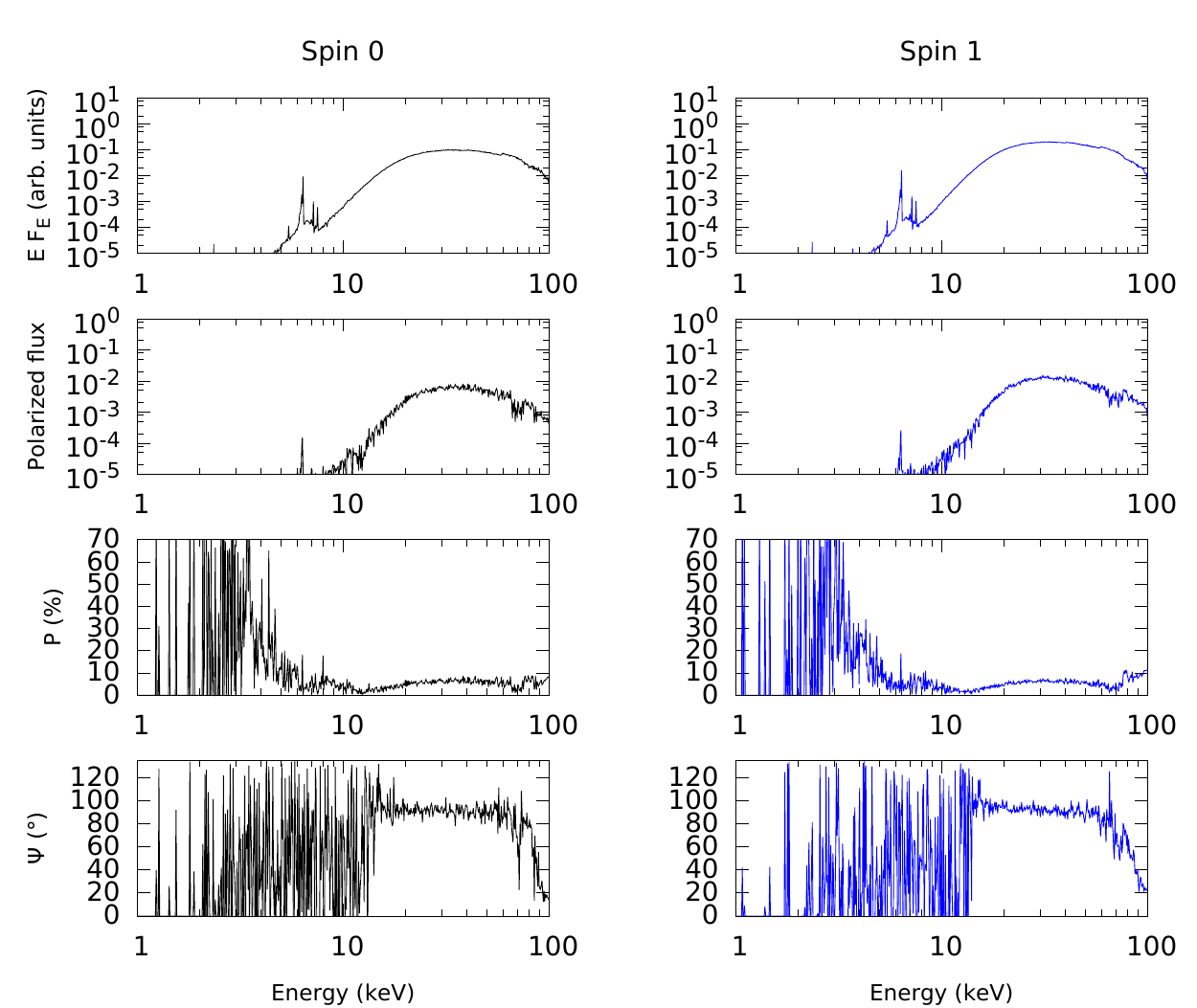}    
    \caption{X-ray flux (F$_{\rm E}$ is energy flux at energy E), polarized flux, polarization 
	    degree and polarization position angle for a type-2 AGN without polar winds.
	    Top-left: n$_{\rm H_{torus}}$ = 10$^{23}$~at.cm$^{-2}$; top-right:
	    10$^{24}$~at.cm$^{-2}$; bottom: 10$^{25}$~at.cm$^{-2}$. See text 
	    for additional details about the model components. The input 
	    spectrum is polarized (2\% parallel polarization) and GR 
	    effects are included (left column: non-spinning Schwarzschild black 
	    hole; right column: maximally spinning Kerr black hole).}
    \label{Fig:Polarized_primary_GR_perp_No_wind}
\end{figure*}

\clearpage

\setcounter{figure}{0}
\renewcommand{\thefigure}{F\arabic{figure}}

\begin{figure*}
    \includegraphics[trim = 0mm 0mm 0mm 0mm, clip, width=8.5cm]{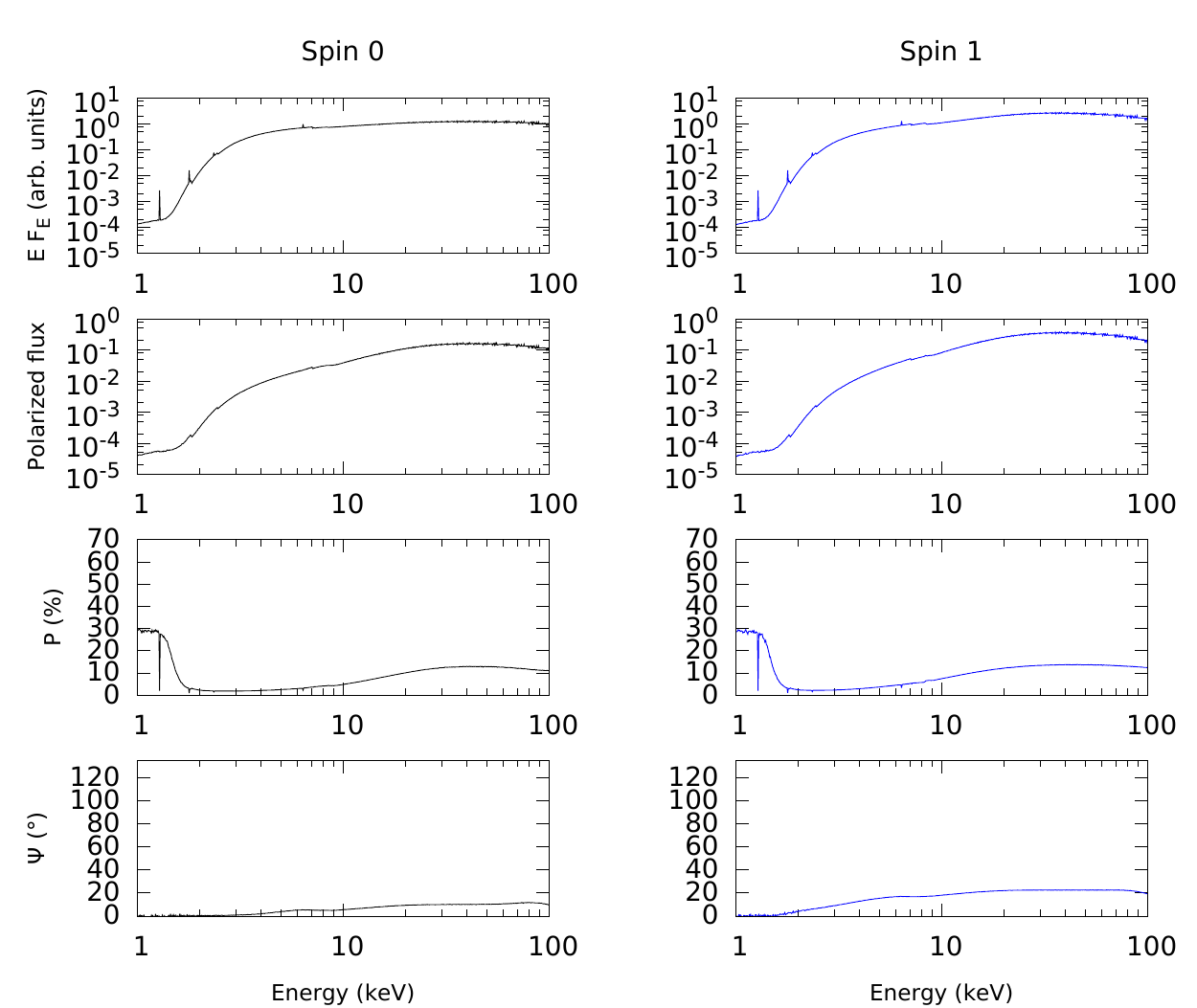}
    \hspace{5pt}\vrule\hspace{5pt}%
    \includegraphics[trim = 0mm 0mm 0mm 0mm, clip, width=8.5cm]{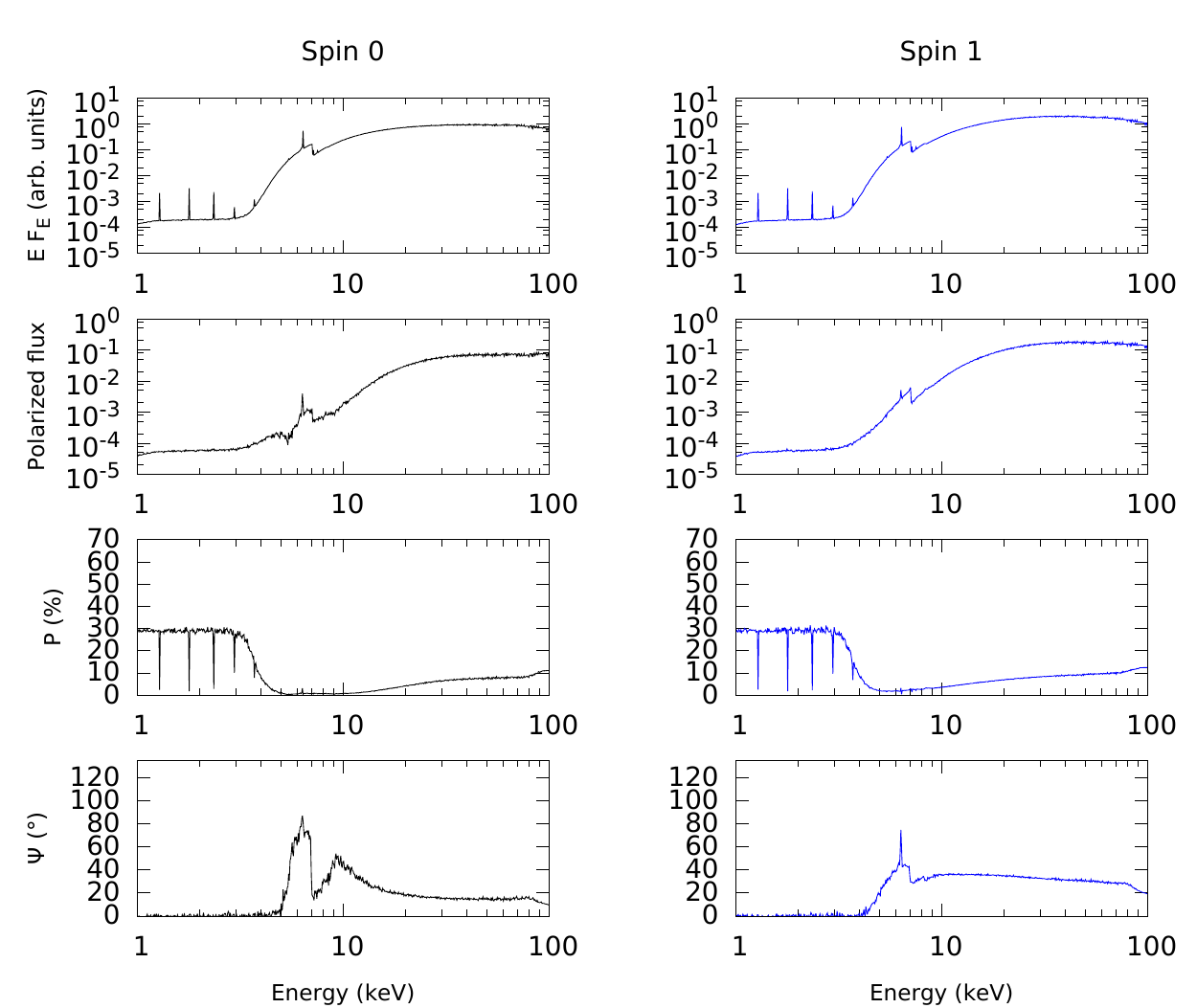}
    \includegraphics[trim = 0mm 0mm 0mm 0mm, clip, width=8.5cm]{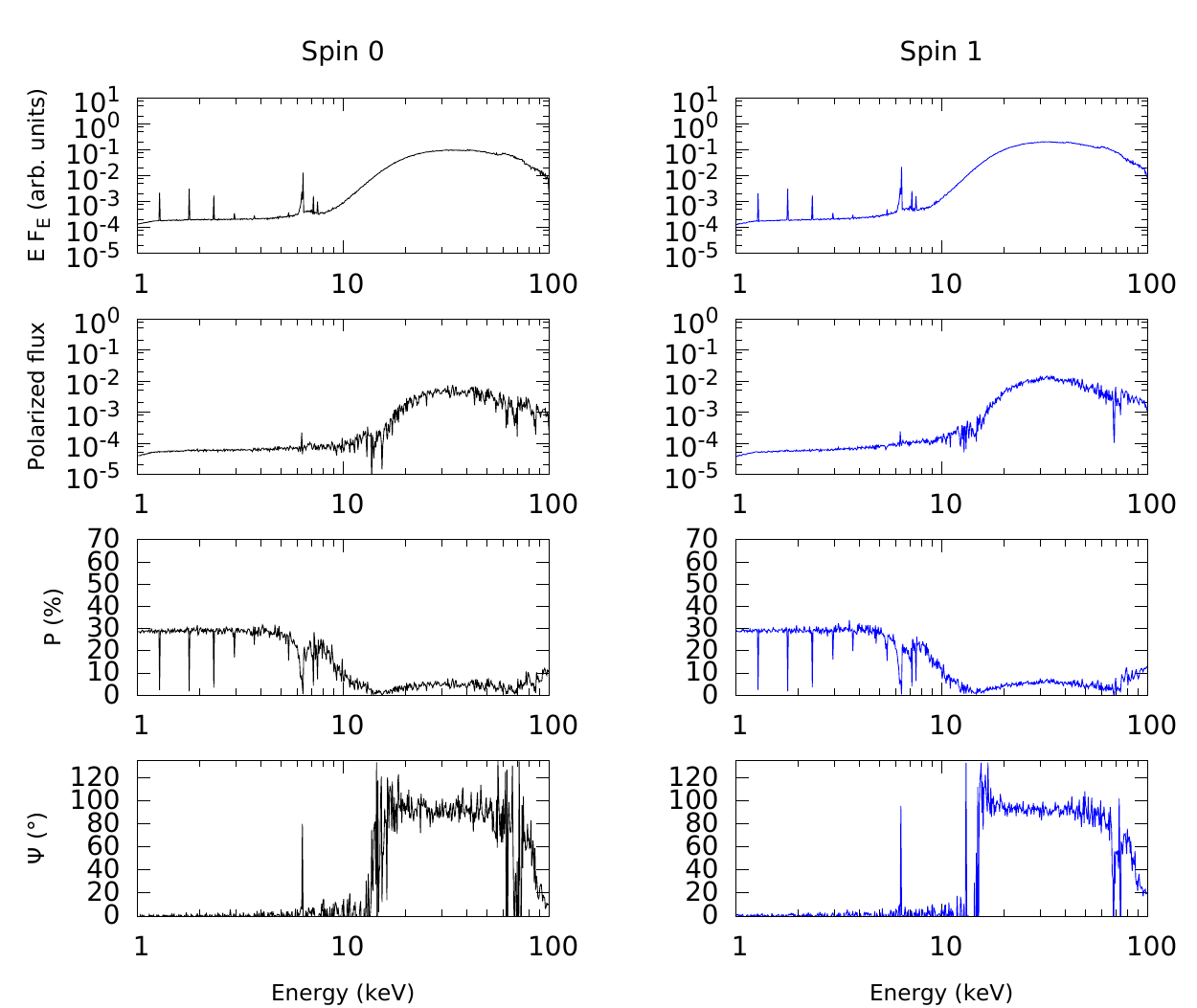}    
    \caption{X-ray flux (F$_{\rm E}$ is energy flux at energy E), polarized flux, polarization 
	    degree and polarization position angle for a type-2 AGN with Compton-thin 
	    absorbing polar winds (n$_{\rm H_{wind}}$ = 10$^{21}$~at.cm$^{-2}$).
	    Top-left: n$_{\rm H_{torus}}$ = 10$^{23}$~at.cm$^{-2}$; top-right:
	    10$^{24}$~at.cm$^{-2}$; bottom: 10$^{25}$~at.cm$^{-2}$. See text 
	    for additional details about the model components. The input 
	    spectrum is polarized (2\% perpendicular polarization) and GR 
	    effects are included (left column: non-spinning Schwarzschild black 
	    hole; right column: maximally spinning Kerr black hole).}
    \label{Fig:Polarized_primary_GR_para_Absorbing_nh21_wind}   
\end{figure*}

\begin{figure*}
    \includegraphics[trim = 0mm 0mm 0mm 0mm, clip, width=8.5cm]{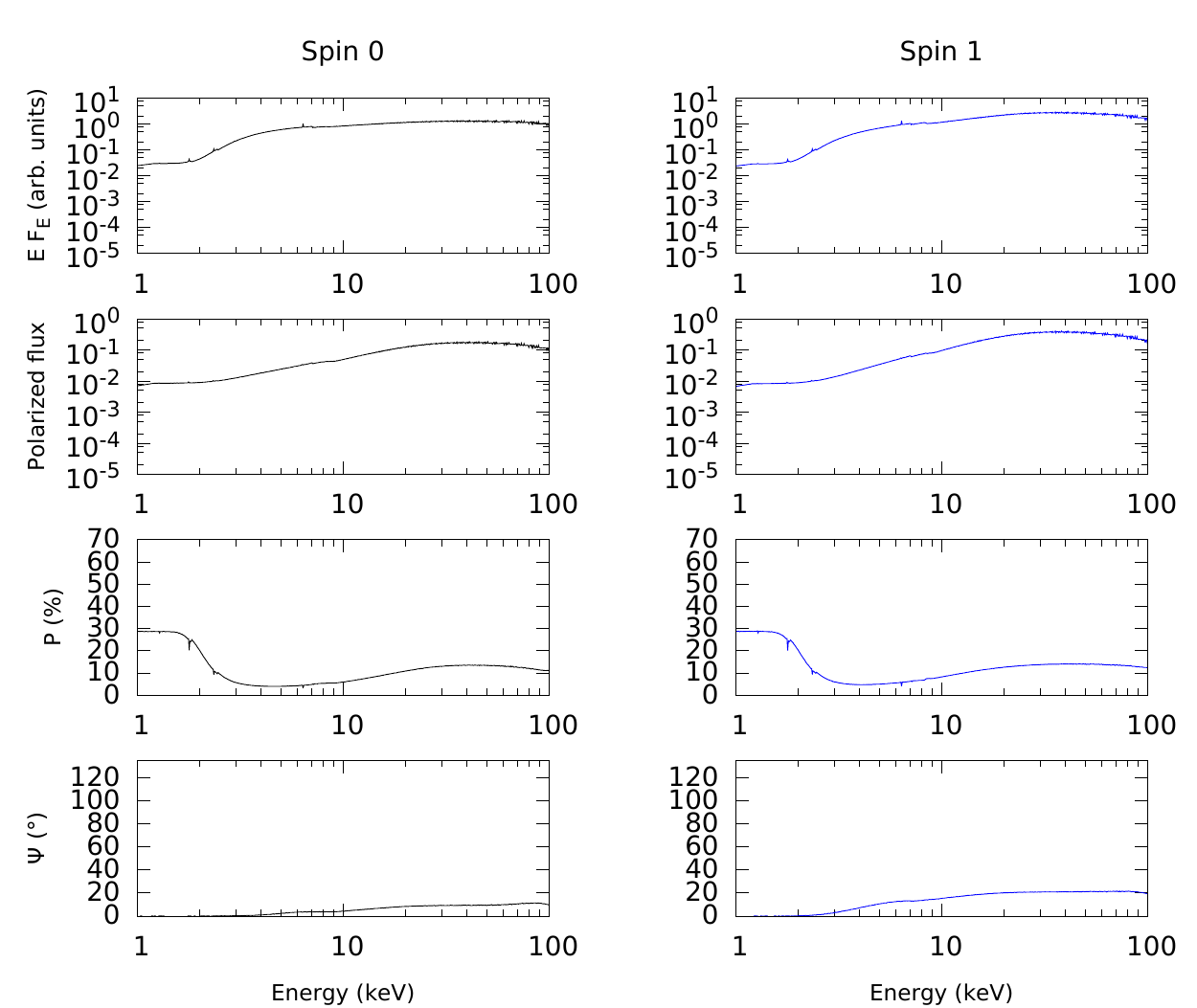}
    \hspace{5pt}\vrule\hspace{5pt}%
    \includegraphics[trim = 0mm 0mm 0mm 0mm, clip, width=8.5cm]{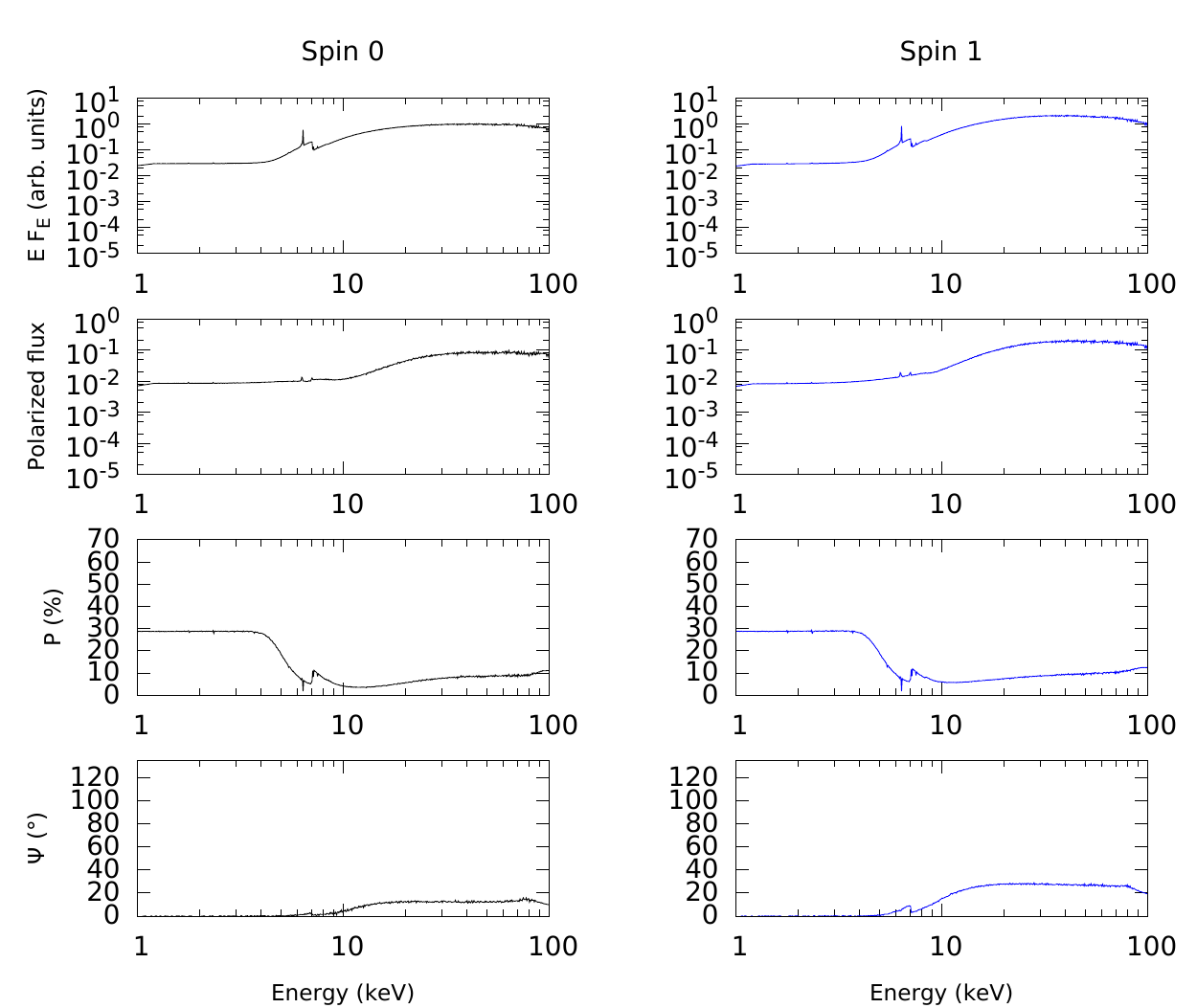}
    \includegraphics[trim = 0mm 0mm 0mm 0mm, clip, width=8.5cm]{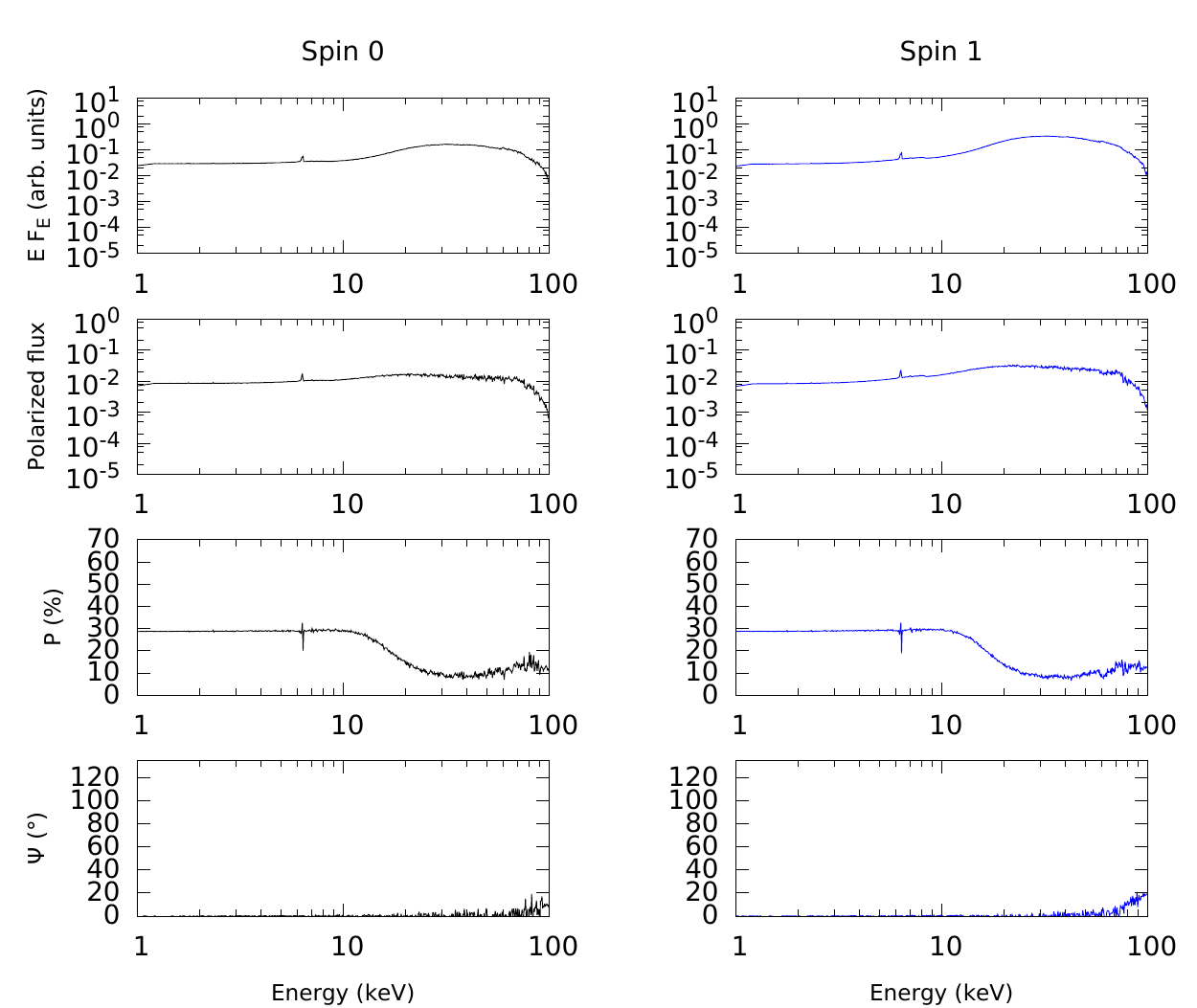}    
    \caption{X-ray flux (F$_{\rm E}$ is energy flux at energy E), polarized flux, polarization 
	    degree and polarization position angle for a type-2 AGN with fully-ionized 
	    polar winds. Top-left: n$_{\rm H_{torus}}$ = 10$^{23}$~at.cm$^{-2}$; top-right:
	    10$^{24}$~at.cm$^{-2}$; bottom: 10$^{25}$~at.cm$^{-2}$. See text 
	    for additional details about the model components. The input 
	    spectrum is polarized (2\% perpendicular polarization) and GR 
	    effects are included (left column: non-spinning Schwarzschild black 
	    hole; right column: maximally spinning Kerr black hole).}
    \label{Fig:Polarized_primary_GR_para_Ionized_wind}
\end{figure*}

\begin{figure*}
    \includegraphics[trim = 0mm 0mm 0mm 0mm, clip, width=8.5cm]{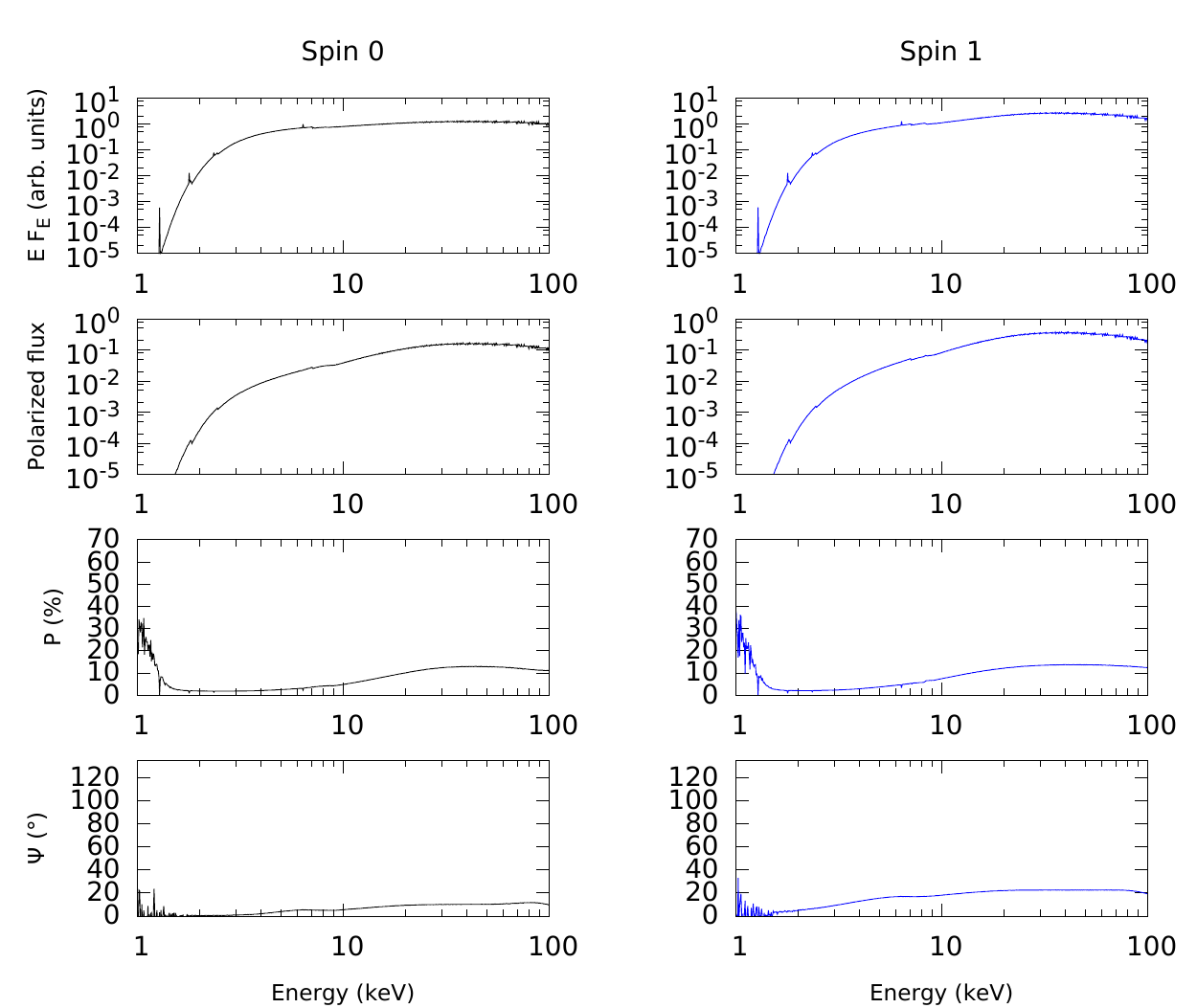}
    \hspace{5pt}\vrule\hspace{5pt}%
    \includegraphics[trim = 0mm 0mm 0mm 0mm, clip, width=8.5cm]{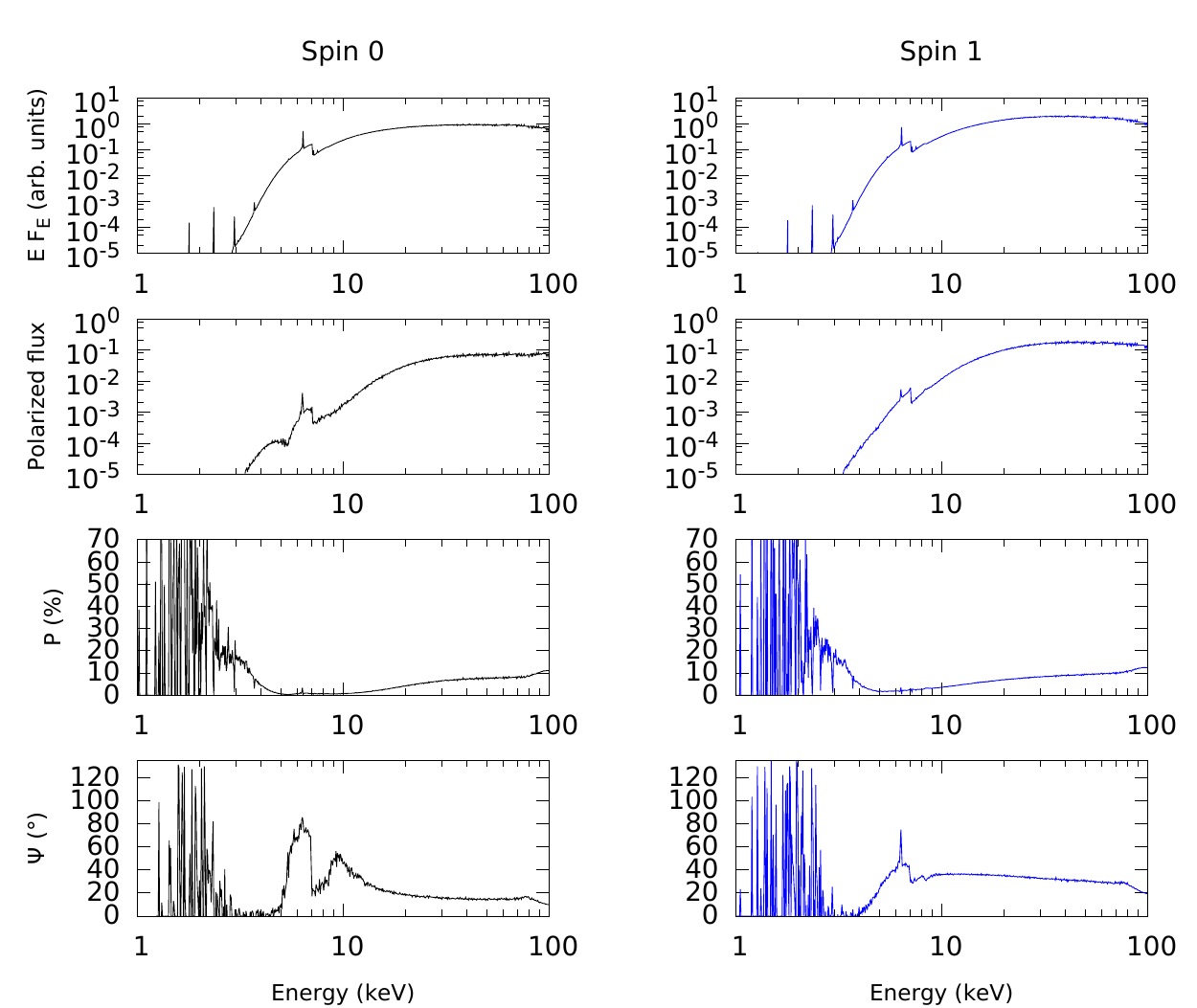}
    \includegraphics[trim = 0mm 0mm 0mm 0mm, clip, width=8.5cm]{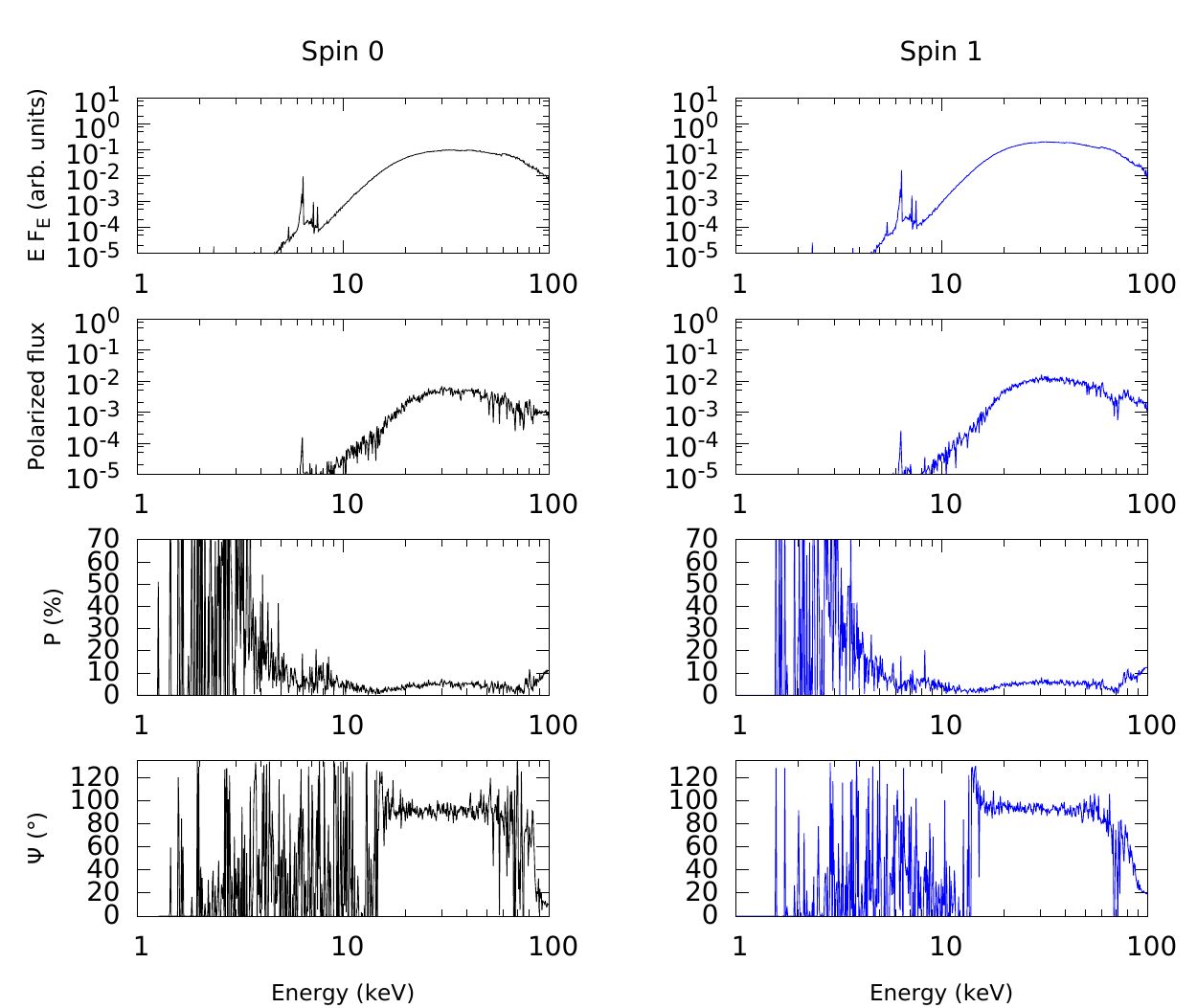}    
    \caption{X-ray flux (F$_{\rm E}$ is energy flux at energy E), polarized flux, polarization 
	    degree and polarization position angle for a type-2 AGN without polar 
	    winds. Top-left: n$_{\rm H_{torus}}$ = 10$^{23}$~at.cm$^{-2}$; top-right:
	    10$^{24}$~at.cm$^{-2}$; bottom: 10$^{25}$~at.cm$^{-2}$. See text 
	    for additional details about the model components. The input 
	    spectrum is polarized (2\% perpendicular polarization) and GR 
	    effects are included (left column: non-spinning Schwarzschild black 
	    hole; right column: maximally spinning Kerr black hole).}
    \label{Fig:Polarized_primary_GR_para_No_wind}
\end{figure*}

\end{document}